# The role of the bile salt surfactant sodium deoxycholate in aqueous two-phase separation of single-wall carbon nanotubes revealed by systematic parameter variations

*Joeri Defillet[1, †], Marina Avramenko[1,†], Miles Martinati[1], Miguel Ángel López Carrillo[1],*

*Domien Van der Elst[1], Wim Wenseleers[1*] and Sofie Cambré[1*]*

[1]Nanostructured and Organic Optical and Electronic Materials, Physics Department, University of Antwerp, Belgium

[†] These authors contributed equally to this work

[*] Corresponding authors: Wim.Wenseleers@uantwerp.be, sofie.cambre@uantwerpen.be

**Abstract**

Aqueous two-phase (ATP) extraction has been demonstrated as a fast, scalable, and effective separation technique to sort single-wall carbon nanotubes (SWCNTs) according to their diameter and chiral structure. The exact mechanism behind the chirality-dependent migration of SWCNTs between the two phases is however not completely understood, and depends on many parameters (*e.g.*, choice of surfactants and their concentration, pH, temperature, …), making it difficult to optimize the multivariable parameter space. In this work, we present a systematic study of the choice and concentration of specific surfactants on the ATP sorting, by performing a series of single-step ATP separations in which each time only one parameter is systematically varied, while monitoring the structure-specific migration of every SWCNT chirality between both phases with detailed wavelength-dependent spectroscopy. These systematic studies reveal that the diameter-dependent stacking of a discrete number of sodium deoxycholate molecules fitting around the SWCNT circumference determines the separation order in the form of a periodically modulated pattern as a function of SWCNT diameter. Addition of cosurfactants can be used to compete with the bile salt surfactant to enhance the separation yields, but does not affect the sorting order. The results are afterwards directly applied to predict the parameters required to separate specific chiral structures in just two ATP steps.

Graphical Abstract:

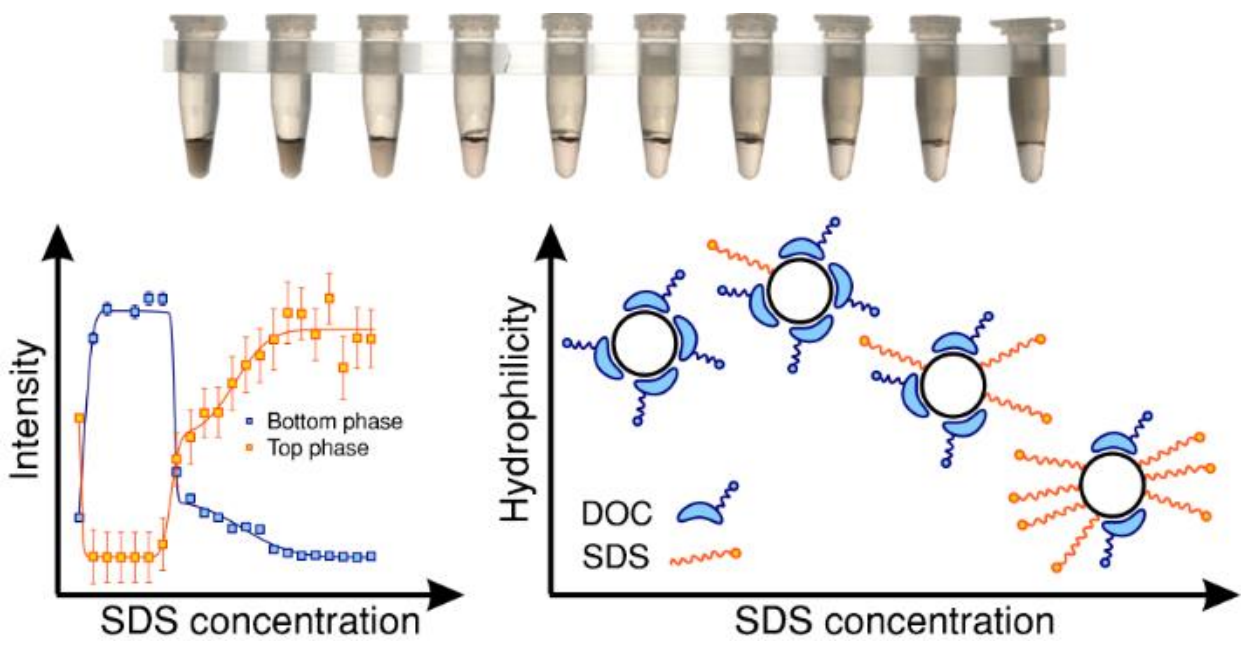

## 1. Introduction

Since their discovery in 1993 [1] the interest in single-wall carbon nanotubes (SWCNTs) has grown exponentially. Their well-defined structure-dependent electronic and optical properties [2] combined with unique mechanical and thermal properties, make SWCNTs fascinating materials for an extensive set of applications in nanoelectronics, photonics, photovoltaics and biochemical sensing [3]. However, a longstanding issue for their implementation in actual devices remains that synthesis methods invariably produce mixtures of different diameters, chiral structures, and metallic and semiconducting tubes, each with their specific electronic and optical properties. While significant progress is being made to synthesize extremely narrow diameter distributions [4–8], most synthesis methods are not yet sufficiently selective and versatile to produce any chirality on demand, and thus an appropriate versatile, scalable, cheap and fast separation technique is crucial to create an industrial breakthrough. A variety of separation techniques have been introduced in the field and were recently reviewed in references [8,9], such as dielectrophoresis [10], density gradient ultracentrifugation (DGU) [11,12], selective polymer wrapping [13–15], DNA-[16,17] or surfactant-[18,19] solubilization followed by size-exclusion chromatography and more recently aqueous two-phase extraction (ATP) [20]. ATP in particular shows promise to become a fast, highly scalable and easily tunable separation method. This technique, developed in the 1980s for biochemical separations [21–23], has more recently been successfully applied to separate SWCNTs [24–38]. In ATP, two water-soluble, yet immiscible polymers (*e.g.*, polyethylene glycol [PEG] and dextran) are mixed together at sufficiently high concentration, after which they spontaneously form two phases. The bottom, dextran-rich phase, is slightly more hydrophilic compared to the top, PEG-rich phase, resulting in the spontaneous redistribution of SWCNTs across both phases [33]. Previous studies show that by choosing a specific DNA sequence to wrap the SWCNTs [25], by changing the surfactants and their concentration [24,26,34,38], by adding salts [34,35], by adding hydration modulators that change the surfactant wrapping [36] or by changing the pH [29,31] SWCNTs can be sorted by diameter and chirality, electronic structure (*i.e.*, metallic or semiconducting) [28,30] and even handedness [29,35]. For example, enantiomers of large diameter SWCNTs (*i.e.*, $d$ = 1.4 nm) have recently been isolated for the first time, using a mixture of 3 surfactants and modulating the pH to induce separation [31].

Despite the impressive results obtained so far, the underlying separation mechanisms are not quite understood, and ATP still has significant room for improvement to increase simplicity, reproducibility, yield and in particular predictability of the separations [33]. In a typical ATP separation, SWCNTs are first dispersed in a sodium deoxycholate (DOC) aqueous solution. This bile salt surfactant was previously found to very efficiently individualize SWCNTs in an aqueous suspension, even without using sonication [39]. Afterwards, a small fraction of the DOC-dispersed SWCNTs is added to a PEG/dextran mixture, typically resulting in a very low DOC concentration (typically ~0.05 % wt/V, *i.e.*, well below the critical micelle concentration (CMC) of DOC in water ~ 0.1-0.3 % wt/V [40]), and sodium dodecyl sulfate (SDS) is used as a cosurfactant to change the effective composition of the surfactant layer adsorbed on the SWCNT walls. In ATP sorting of biomolecules, the separation can occur due to a difference in surface area (size-dependent separation), a difference in hydrophilicity (since both phases have different hydrophilicity), a different electrochemical interaction or a specific affinity for one of the two phases (*e.g.*, specific binding to one of the two polymers) [41]. For SWCNT separations, it is proposed that the difference in hydrophilicity governs the separation, and that depending on the specific surfactant layer adsorbed on the SWCNT walls, the SWCNTs are partitioning into a different phase [34]. Note that redox chemistry [28], as well as changes of the pH [29] and modifying the hydration [29,31,36] have also been employed to alter the separation outcomes, each having their specific influence on the stacking of the surfactants on the SWCNT walls.

While in most cases ATP separations are performed iteratively, by replacing one of the phases by a mimicking phase and slightly adjusting the surfactant composition and other parameters such as pH, temperature and salts in each consecutive step (typically 6-10) [33], it was shown that in principle, after optimization of the sorting parameters, a two-step separation is sufficient to isolate a specific chirality [29,31,34,38], drastically improving the reproducibility and simplicity of the separations. Optimization of the separation parameters can be very cumbersome, because of the many parameters involved, such as the choice of polymers, concentration and choice of the surfactants, addition of other components such as salts and oxidants, changes of the pH and temperature that each have their specific influence on the sorting outcomes. In recent works, Sims *et al.* [32,42] proposed to therefore perform surfactant-exchange PL experiments inside the polymer matrices typically used in ATP, and spectroscopically detectable shifts in PL were assigned to possible transition points in ATP, thus providing a faster tool to estimate the relevant surfactant concentrations for ATP separation. More recently, this was confirmed by Podlesny *et al.*, for the optimization of other surfactant systems [43].

In this work, we circumvent this optimization of parameters for each different chiral structure, by performing a series of single-step ATP separations in which each time only one single sorting parameter is systematically varied. By employing detailed multi-excitation-wavelength optical spectroscopic characterization of the resulting bottom and top phases combined with extensive analysis of these spectroscopic data, one systematic variation experiment directly yields information on the migration of a wide range of SWCNT diameters and chiralities. In particular, we simultaneously obtain information on more than 30 different chiral structures in the diameter range of 0.6 – 1.24 nm, allowing to study the effect of diameter and chiral structure on the sorting outcomes. These studies reveal that chirality-dependent interactions of the bile salt surfactant sodium deoxycholate with the SWCNT walls determine the specific chirality-dependent separation order in the form of a periodically modulated pattern as a function of diameter, indicating stacking of a discrete number of DOC molecules around the SWCNT circumference. Finally, the obtained results are then directly applied to develop new (optimized) sorting protocols for specific chiral structures.

## 2. Materials and methods

**2.1 Materials**: Polyethylene glycol (PEG, Alfa Aesar, MW 6 kDa) and dextran (Tokyo Chemical Industry Co., MW 40 kDa) were used as received. Sodium deoxycholate (DOC, 99%), sodium dodecyl sulfate (SDS, 99%), sodium dodecylbenzenesulfonate (SDBS, 88%), Triton X-100, Tween 60, Tween 80 and cetyltrimethylammonium bromide (CTAB, 99%) were acquired from Acros Organics. Sodium dioctyl sulfosuccinate (DIOCT, 99%) was obtained from Sigma-Aldrich. $D_2O$ was obtained from Cortecnet (99.8 atom%D). HiPco SWCNTs were obtained from NoPo Nanotechnologies India Inc. (batch number-2015.820).

**2.2 SWCNT solubilization:** Raw SWCNT powder (40 mg) was solubilized in a solution of 40 mg of DOC in 4 mL of $D_2O$. $D_2O$ is used as solvent instead of $H_2O$ due to its higher transparency in the IR, important for the spectroscopic analysis. The solution is stirred for at least 1 month, using a small bar magnet inside the solution and putting the vial on a magnetic stirring plate at room temperature. To enhance the concentration of isolated SWCNTs in the sample, a brief sonication (15 min) was applied in a bath sonicator the first 3 days of the stirring procedure (BRANSONIC, 1510E-MTH). Afterwards, the samples were centrifuged in a tabletop centrifuge (Sigma 2-16KCH) at 16215 g (14000 rpm) for 4 h using a swing-out rotor, to remove undissolved material. The supernatant was collected and used for the separations, and if needed for the subsequent separation, the SWCNT concentration was increased by concentrating the samples in an Ultra centrifugal filter. The final DOC concentration in the samples was each time checked by absorption spectroscopy of the SWCNT solution, in which the DOC surfactant has a distinguishable absorption band at 2300 nm (see Fig. S1 in the Supporting Information). This final check was required to improve reproducibility, as the concentration step sometimes results in slightly altered DOC concentrations.

**2.3 Preparation of stock solutions and systematic ATP separations:** To be able to perform the systematic ATP separations, different stock solutions of various concentrations were prepared such that these stock solutions can be mixed together with the SWCNT samples to obtain a systematic variation of the concentrations. Examples of such experiments are provided in Tables S1 – S4 in the Supporting Information, yielding the different concentrations and volumes that are mixed together to get the desired surfactant concentration for sorting. For simplicity and for easy conversion of the concentrations from $D_2O$ to $H_2O$, we chose to define the concentrations of each of these stock solutions as mass per added volume: for example, a 20% wt/V PEG solution thus corresponds to mixing 200 mg of PEG with 1 mL of $D_2O$. For the variation of SDS concentration at a fixed DOC concentration, a stock solution of 20% wt/V PEG, 20% wt/V dextran, and 10 and 20% wt/V SDS were prepared. In each of the separations, the total volume was not changed (assuming no volume change upon mixing), such that the phase separation remains the same. The final DOC and SDS concentrations, and their ratio, was then determined by assuming a volumetric dilution over both phases. Note that such a volumetric dilution is not the reality, as absorption spectroscopy highlights that the surfactants do not distribute evenly over the two phases (Fig. S1), which is important when designing a multi-step separation approach. Nevertheless, for a single-step separation like presented in this work, the used surfactant concentrations at the start of the experiment can be accurately determined.

**2.4 Spectroscopic techniques:** The absorption spectra were measured with a Varian Cary 5000 UV-VIS-nIR spectrometer in the range of 175-2500 nm. Before measuring, each bottom and top phase was diluted by a factor of 2 with a 4% wt/V DOC in $D_2O$ solution to raise the final DOC concentration above 2% wt/V. As such, changes in Raman cross sections and PL quantum efficiencies due to a different surfactant surrounding can be avoided. Note for example that the Raman cross-section of the (5,3) SWCNT is largely quenched at low DOC concentrations and also the PL efficiency as well as the peak positions of the SWCNT electronic transitions are strongly dependent on the surrounding surfactant layer, whereas dilution to 2% wt/V DOC results in similar peak positions in each of the samples. The measurements were all performed in 60 μL micro-cuvettes with an optical path length of 3 mm.

The PLE experiments were performed with a dedicated, in-house developed setup comprising a high-power pulsed Xenon lamp (Edinburgh Instruments, custom adapted Xe900/XP920) for the excitation and a liquid-nitrogen cooled extended InGaAs diode array detector for the detection (Princeton instruments OMA V:1024/LN-2.2, sensitive up to 2200 nm). The PLE maps were collected in a 90-degree geometry, inside the above-mentioned 3 mm / 60 μL microcells. Gratings and slits of the excitation and emission spectrometers (Acton Spectrapro 2355 and 2156, respectively) were chosen to provide for an average resolution in emission of 10 nm and 8 nm in excitation. Spectra were calibrated for spectral sensitivity of the detector, and spectral and temporal variations of the lamp intensity. Additionally, PLE maps were corrected for the inner-filter effect of both excitation and emission paths. With highly concentrated bottom and top phases, the samples were additionally diluted with a 2% wt/V DOC stock solution, to allow for such an inner-filter effect correction.

The Raman spectra were measured with a Dilor XY800 triple spectrometer with liquid nitrogen cooled CCD detection and a series of laser excitation wavelengths coming from an $Ar^+$ (457 and 502 nm) and $Kr^+$-ion laser (647 and 676 nm), a tunable Ti:sapphire laser (710, 725, 785 and 824 nm) and a tunable rhodamine 6G dye laser (570 nm).

**2.5 Fitting Models**: The 2D RRS and PLE data are fitted simultaneously for all samples within one systematic separation series by using previously developed models for single-wavelength Raman spectra [44–46] and wavelength-dependent PLE spectra [31,47–50] of SWCNTs.

In case of Raman spectroscopy, the fitting model is composed of a sum of Lorentzians of which the peak positions and linewidths are shared for all the Raman spectra within a systematic separation series. While the peak positions and linewidths of these Lorentzians are obtained through a numerical least-squares fitting algorithm, the relative amplitudes of each of the peaks in each of the different samples are calculated analytically by linear regression. As such, the amplitudes are not numerically optimized fitting parameters, and all RRS spectra are thus fitted with two parameters for each Lorentzian. We include both Lorentzians for empty and water-filled SWCNTs in resonance with the different laser excitation wavelengths. This approach results first of all in a much better determination of the Raman frequencies and linewidths of the different Lorentzians, since they are now defined by taking into account all the Raman spectra in which they are observed and since their relative amplitudes vary in between spectra due to the obtained chirality sorting. In addition, such a simultaneous fitting procedure allows us to determine with the highest precision the amplitudes of each chirality even in spectra with very low intensities.

For PLE spectroscopy, a 2D wavelength-dependent fitting model was previously developed, comprising an emission and excitation line shape. The emission profile is fitted using a Voigt line shape with the emission energy ($E_{11}$) and its linewidth as fitting parameters. The excitation profile is fitted using both excitonic and band-to-band transitions, as well as a phonon side band in excitation. The number of fit parameters is reduced, by making assumptions on the position and relative amplitudes of these phonon sidebands, for which more details can be found in references [31,47–50]. Importantly, all PLE maps are then fitted simultaneously, with shared peak positions and linewidths that are optimized numerically, but similarly as in the Raman maps, with varying amplitudes that are obtained through linear regression and thus not optimized numerically. Again, this approach allows for fitting the PLE peaks of each chiral structure more accurately, and for determining the amplitudes of each chiral structure in each PLE map to the highest precision.

Moreover, to account for imperfections of both RRS and PLE fit models, which could lead to non-normally distributed residuals around zero, we reduced the effective number of degrees of freedom used in the calculation of the fit errors to the number of zero-crossings in the residuals, which is typically much lower. This effectively results in larger but more accurate error bars on the fit parameters and amplitudes.

### 2.6 Theoretical modelling

For constructing the molecular models, first the geometry of an isolated DOC molecule was optimized at the semi-empirical level using the AM1 Hamiltonian in MOPAC 2016 [51]. Then, these DOC molecules were placed one by one around the circumference of a SWCNT and each time partly relaxed at the molecular mechanics OPLS level in HyperChem 7.52 [52] to ensure that the cholesterol groups stacked at van der Waals distances from each other and from the nanotube. For the geometry of the nanotube, standard carbon-carbon distances of 1.421Å were used. Atomic charges of the DOC molecules used in the OPLS calculation were derived from the AM1 calculation. The process of adding DOC molecules was repeated until the entire circumference of the SWCNT was filled. The final geometry was partly relaxed at the OPLS level to a root-mean-square gradient of <0.0001 kcal/(Å mol). The geometry of the flexible polar tails of the DOC molecules was not optimized in the OPLS calculation (but it was in the preceding higher level AM1 calculation). The geometry of the SWCNT was also kept fixed.

## 3. Results and discussion

For the systematic studies performed in this work, we start by preparing a stock solution of SWCNTs synthesized by the high-pressure CO-conversion method (HiPco) by dispersing the raw SWCNTs in a 1% wt/V DOC solution in $D_2O$ through bath sonication and magnetic stirring, followed by a centrifugation to obtain a solution of well isolated SWCNTs (see methods section for complete details). PEG and dextran stock solutions were prepared by adding 200mg of PEG (resp. dextran) to 1mL of $D_2O$, here referred to as a 20% wt/V PEG and dextran solution. Similarly, DOC and SDS stock solutions in $D_2O$ were prepared at different concentrations, *e.g.*, 1% wt/V DOC, 5% wt/V DOC, 10% wt/V SDS and 20% wt/V SDS etc. Note that for simplicity, all concentrations in this work are calculated with respect to the added volume of the liquid, and not with respect to the total volume (nor weight) of the resulting solution, the main reason being the easy exchange of surfactant concentrations when doing separations in $H_2O$ or $D_2O$. Here, deuterated water was chosen instead of $H_2O$ for its much better optical transparency in the IR, allowing the full characterization of the resulting samples by optical spectroscopy.

For each experiment, at least 20 different Eppendorf centrifuge tubes were filled with a two-phase mixture of PEG and dextran (*i.e.*, 700 µL of the PEG and 300 µL of the dextran stock solutions). Based on previous ATP sorting results [24,26,34,38], in the first experiment we vary the concentration of SDS, while keeping the DOC concentration constant at 0.0507% wt/V, closely matching the DOC concentration typically used in literature for single-chirality sorting of SWCNTs [24,26,34,38]. To this end, varying volumes of $D_2O$ and the SDS surfactant stock solutions were added to the Eppendorf tube (keeping the total volume constant) and mixed with PEG and dextran using a vortex mixer. Finally, the same volume of the SWCNT stock solution was added to each Eppendorf tube (see Table S1 in the SI) and everything was again mixed with a vortex mixer. Tables S1 – S4 in the Supporting Information (SI) present the variation of parameters and resulting surfactant concentrations. After mixing, the samples were centrifuged for 10 min at 5000 g (Eppendorf microcentrifuge with fixed angle rotor) to speed up the phase separation resulting in Fig. 1(a). The resulting bottom and top phases were collected manually using a syringe. Each of these fractions was further characterized using a combination of optical absorption, resonant Raman scattering (RRS) with multiple excitation wavelengths and wavelength-dependent IR fluorescence-excitation (PLE) spectroscopy to determine the concentration of a wide range of SWCNT chiralities in the resulting fractions.

### 3.1 Systematic ATP separation with 0.0507% wt/V DOC and a varying SDS concentration:

Fig. 1(a) presents a photograph of a discrete set of phase-separated samples where the R# marks the approximate ratio of the SDS versus DOC concentrations in each separated sample (*e.g.,* R10 corresponds to a SDS/DOC ratio of 9.87). Starting from sample R10 up to R42, it can be seen that the bottom phase becomes lighter in color, indicating that SWCNTs migrate to the top phase. For sample R0 however (*i.e.*, no SDS added), it seems that most SWCNTs are separated into the top phase. To verify this visual observation, we can look at the absorption spectra of the different bottom and top phases, where the SDS concentration is varied from 0 up to 2.1% wt/V in steps of 0.1% wt/V while keeping the DOC concentration fixed at 0.0507% wt/V. Fig. 1(b-c) present the absorption spectra of the bottom phases and Fig. 1(d-e) of the top phases. Without adding SDS (black curves in Fig. 1(b,d)) most SWCNTs indeed separate into the top phase. As soon as very small aliquots of SDS are added to the two-phase mixture and up to an SDS/DOC ratio of approximately 8, SWCNTs migrate to the bottom phase, and thus disappear from the top phase (Fig. 1(b, d)). For higher SDS concentrations, SWCNTs start to migrate to the top phase again with some chiralities moving at lower SDS concentrations than others (Fig. 1(c, e)). Although not obtaining detailed chirality-dependent information from these absorption spectra, several observations can thus already be made. First of all, without adding SDS, and using a DOC concentration of 0.0507% wt/V, most SWCNTs separate into the less hydrophilic PEG-rich top phase indicating that the surrounding surfactant layer is not sufficiently covering the SWCNTs thereby the entire hybrid structure is less hydrophilic. Note that DOC, being a bile salt surfactant, by itself shows a complex aggregation mechanism, where the critical micelle formation does not correspond to a discrete concentration but occurs over a certain concentration range with first the formation of primary micelles with low aggregation numbers ($N$ = 2), followed by secondary micelles, the latter forming tubular structures at sufficiently high concentration [40]. A DOC concentration of 0.0507% wt/V is well below the primary micelle formation of DOC in water at room temperature (typically occurring at ~0.1% wt/V) and even more below the formation of secondary micelles (typically occurring at ~0.3% wt/V). This shows that at the concentrations used, no DOC micelles are present in the solution. On the other hand, SDS has a critical micelle concentration of 0.2% wt/V [53], hence for SDS/DOC ratios above 4, SDS micelles will be present in the solution.

When adding SDS, we can observe two different regimes. At sufficiently low SDS concentrations, SWCNT-surfactant systems become more hydrophilic when adding SDS, *i.e.*, migrate to the bottom phase, while at much higher SDS concentrations they become less hydrophilic and hence separate back into the top phase. The fact that SDS-covered SWCNTs are less hydrophilic compared to DOC-covered SWCNTs was also previously confirmed by Subbaiyan *et al*, who found SDS-solubilized SWCNTs to separate always in the top phase, whatever the concentration of SDS in the two-phase mixture [34]. To understand this better, we characterize all samples in more detail with PLE and RRS, monitoring the specific chirality composition in each of the different phases (*vide infra*).

To describe the partitioning of a specific chirality with chiral indices $(n,m)$ in the two phases, we determine in each fraction the relative partition coefficient $K_{(n,m)}^{top}$ (resp. $K_{(n,m)}^{bottom}$) and define it as the intensity of the $(n,m)$ chirality in the top (resp. bottom) phase divided over the maximum intensity in that phase over the measured surfactant concentration range. These relative PL and Raman intensities are proportional to the varying SWCNT concentration in the different samples, as long as the PL efficiencies and Raman cross sections do not change between the different samples. Before characterizing the samples by optical spectroscopy, each bottom and top phase was therefore diluted by a factor of 2 with a 4% wt/V DOC solution to raise the final DOC concentration above 2% wt/V. As such, changes in Raman cross sections and PL quantum efficiencies due to a different surfactant surrounding can be avoided, because these high DOC concentrations replace all the SDS molecules on the SWCNT wall, evidenced by the fact that we do not observe any changes in peak positions and linewidths throughout the different bottom and top phases.

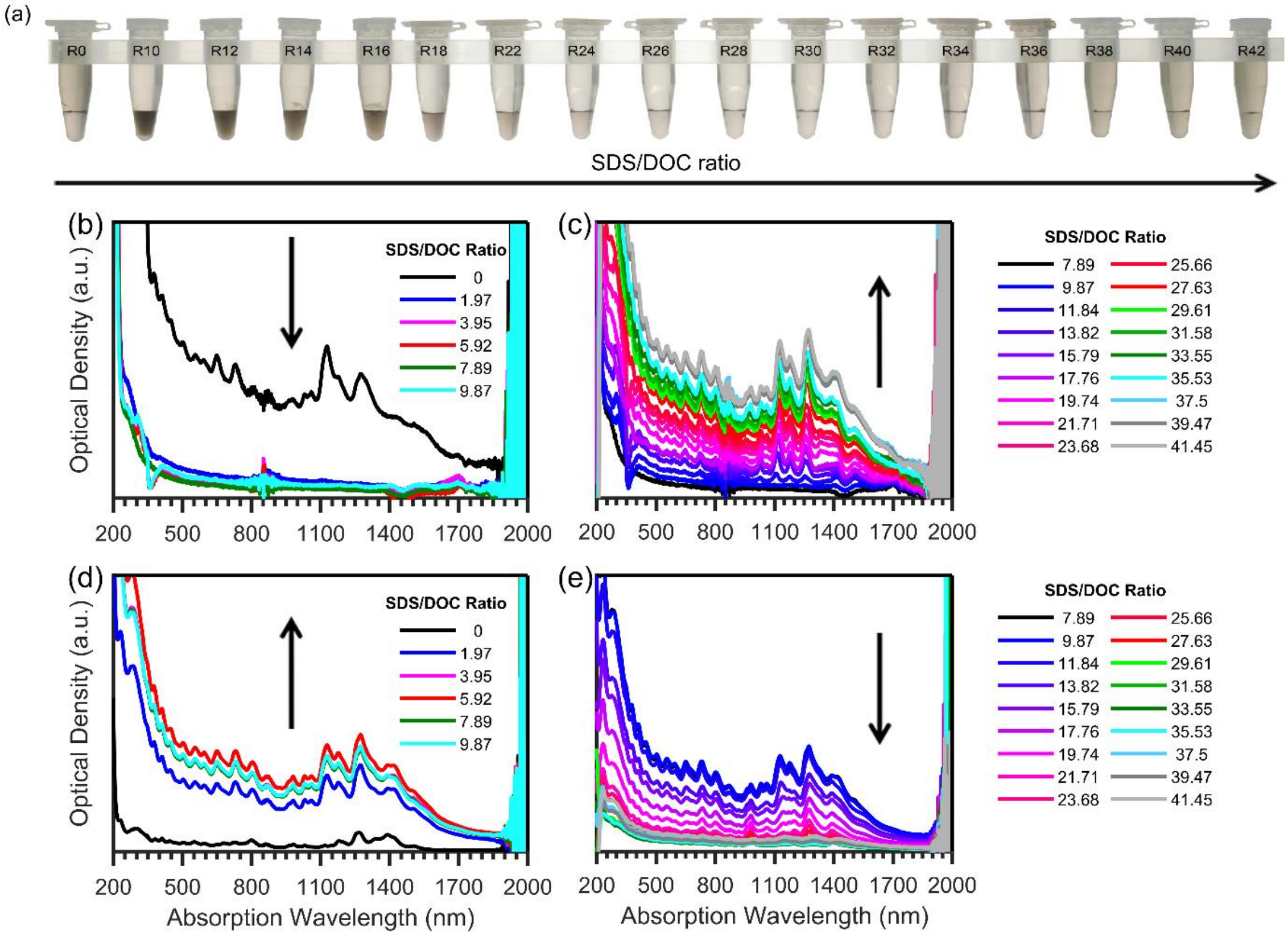

Figure 1: Systematic ATP separation by varying the SDS concentration at a fixed DOC concentration of 0.0507% wt/V. (a) Selection of photographs of different SDS/DOC ratios showing the redistribution of the CNTs among the two phases. The specific SDS/DOC ratio is indicated by the R# on the centrifuge tubes (*e.g.*, R10 corresponding to an SDS/DOC ratio of 9.87). (b,c) Absorption spectra of the top phases for varying SDS/DOC ratios showing first a decrease of concentration (panel b) and afterwards an increasing concentration (panel c) with increasing SDS/DOC ratio. (d,e) Analogous absorption spectra for the corresponding bottom phases, showing first an increase and afterwards a decrease of SWCNT concentrations.

Note also that throughout the entire series, we monitor that no drastic change in the phase separation itself occurs (*e.g.*, volumetric changes, and composition of the phases), which might be the case, for example, at too high surfactant concentrations. Since PEG, dextran and the different surfactants have characteristic absorption peaks in the IR range of the spectrum (see Fig. S1 in the SI) these absorption peaks can be used to assess the relative concentration of all components in the different samples and as such monitor that the PEG/dextran ratio does not change *e.g.*, when adding more surfactants. Note that the absorption spectra indicate that also the surfactants do not distribute evenly between the two phases, and *e.g.* the SDS concentration in the bottom phase can be up to a factor of 6 lower compared to the assumed equal distribution of surfactants amongst phases (see Fig. S1), which is very important to realize when performing multi-step separations in which each time a mimicking new top phase is added to extract other chiralities, as it is difficult to keep track of the actual surfactant concentrations [24,29,31,33].

Fig. 2 presents a set of selected PLE maps of bottom and top phases for different SDS/DOC ratios (the complete series is presented in Figs. S2 and S3 in the SI). Similarly, as in the absorption spectra, in the absence of SDS, some chiralities separate in the top phase and some in the bottom phase. When adding slight amounts of SDS, all SWCNTs migrate to the bottom phase. At higher SDS/DOC ratios, SWCNTs migrate to the top phase in a non-monotonous manner with diameter. To analyze the migration of each chirality in the different samples, we can extract the PL intensities of each chirality by integrating the PLE maps over a discrete emission and excitation wavelength range, as previously done in reference [44]. However, more accurate intensities can be obtained when fitting the 2D PLE maps using our empirical 2D PLE fitting model [31,48–50]. This fitting model is based on an accurate, empirical description of the complicated excitation line shape associated with both excitonic and band-to-band excitations as well as phonon side bands for the SWCNTs and allows for fitting the exact peak positions, linewidths and intensities for each of the chiral species present in a particular PLE map. For fitting the series of PLE maps, a simultaneous fit of all PLE maps from the bottom and top phases was performed, in which SWCNT peak positions and linewidths are optimized but shared for all the PLE maps, and only intensities were allowed to vary between the different PLE maps. As such, much more accurate peak positions and linewidths for each of the chiralities can be obtained, and amplitudes for each chirality in each of the different PLE maps can be readily extracted. Typical fits of a few PLE maps can be found in Fig. S4 in the SI, showing excellent agreement between fit and experiment. After the peak positions and linewidths are accurately determined, the intensity (amplitudes of the PLE fits) of each SWCNT chirality is determined by a simple linear regression and can be plotted for both the bottom and top phase as a function of added SDS concentration (*vide infra* Fig. 4).

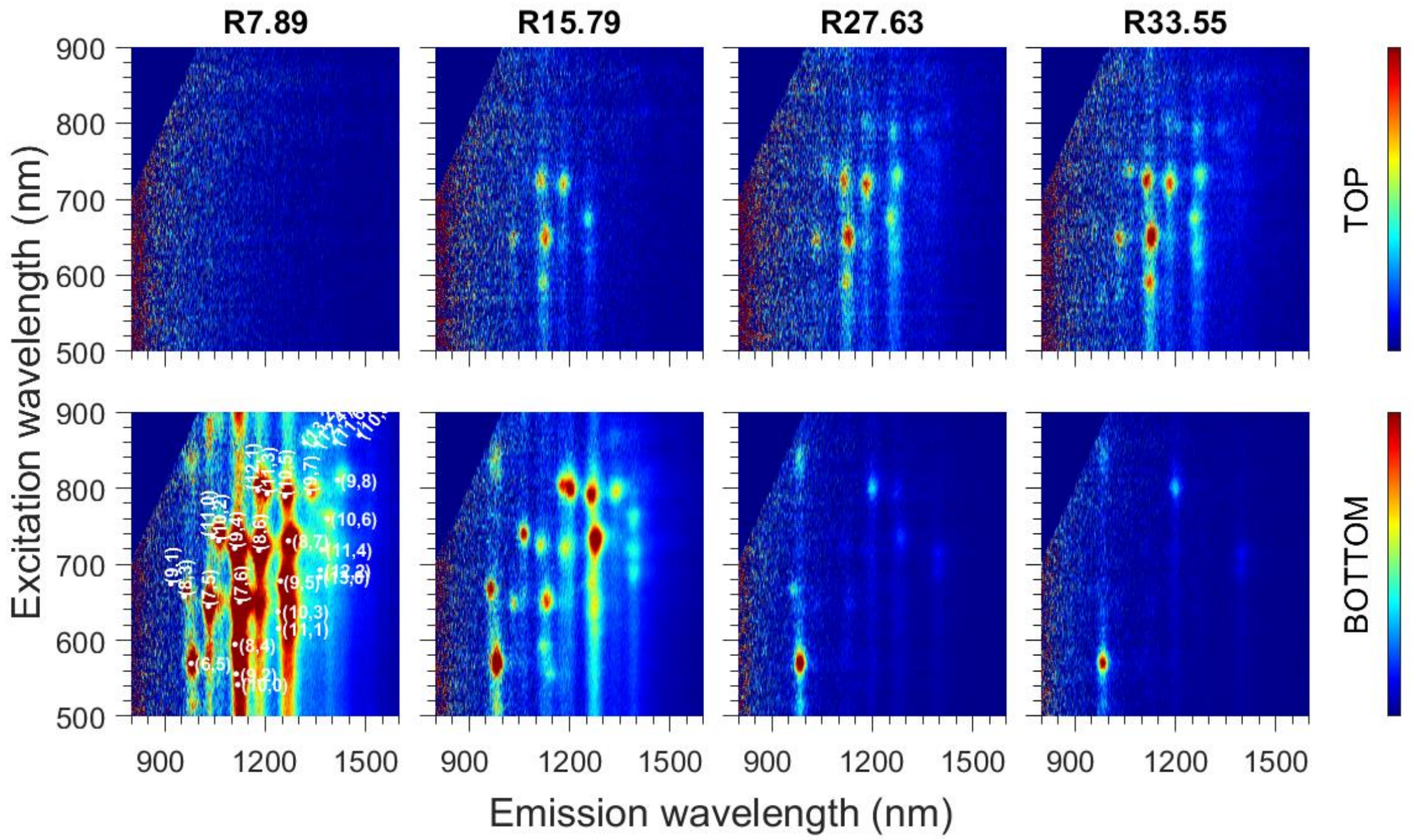

Figure 2: Selection of PLE maps of 4 bottom and 4 corresponding top phases at different SDS/DOC ratios (indicated by the R# in the titles). Absolute intensities (represented by the color scale) of bottom (resp. top) phases of the different SDS/DOC ratios can be directly compared (but not between bottom and top phases due to the different dilution factors). All other PLE maps can be found in the SI.

PLE spectroscopy can unfortunately not be used to access the smallest (and largest) diameter semiconducting SWCNTs as well as the metallic SWCNTs present in the sample due to the too low (or absent) emission from these chiralities. Therefore, in addition to the PLE analysis, we also performed resonant Raman measurements at different excitation wavelengths to extract data for 31 different SWCNT chiralities (see Table S5 in the SI), focusing in particular on the smallest diameters present in the samples (*e.g.,* (5,3) and (7,2)) as well as on specific armchair SWCNTs ((6,6) and (7,7)) and other metallic SWCNTs (*e.g.*, (7,4), (8,5) and (9,3)). In addition, we also measured several SWCNT chiralities that overlap with those obtained from the PLE analysis, to cross-check the relative abundance changes based on the PLE and RRS intensities.

Fig. 3 presents an example of Raman spectra of the bottom and top phases excited at 725 nm, in resonance with (11,4), (8,7), (8,6) (9,4) and (10,2) SWCNTs. Intensities for each of the chiralities as a function of increasing SDS concentration were obtained by fitting the Raman spectra with a sum of Lorentzians, with shared peak positions and linewidths for all the spectra but varying intensities (see methods for more details on the fitting model). All other Raman spectra obtained at other excitation wavelengths (with other SWCNTs in resonance) are presented in the SI Figs. S5-S19. Note that the Raman spectra even allow us to also extract the migration of empty (closed) and water-filled (opened) SWCNTs in the two-phase mixture, due to their slightly shifted radial breathing mode (RBM) vibrational frequencies [45,46,54]. Nevertheless, we found that the difference between empty and water-filled SWCNT migration is extremely small in these separations, and thus that filling does not seem to have a significant impact on the separations in this work. Moreover, for some chiralities the RBMs of empty and water-filled SWCNTs are not sufficiently resolved to accurately determine the fraction of empty and water-filled tubes in each sample, and slight variations in the spectrometer calibration can strongly change these relative intensities. Therefore, a RRS partition coefficient curve of a particular chirality represents the much more accurate sum of the corresponding curves of empty and water-filled SWCNTs.

Fig. 4 then presents the resulting intensity variations obtained from both RRS (circles) and PLE (squares) for a selected set of SWCNT chiralities, *i.e.*, related to the relative partition coefficients $K_{(n,m)}^{top}$ and $K_{(n,m)}^{bottom}$, as a function of increasing SDS/DOC ratio, for the SWCNTs in the bottom phase (shown in blue) and top phase (denoted in orange). The full set of partition coefficient curves is presented in the SI, Fig. S20. For each of the chiralities, the partition coefficient curves cross each other at about half the intensity, proving that when a SWCNT disappears in the bottom phase, it directly appears in the top phase. Note also that the small scatter on the data points of subsequent ATP separations, *i.e.*, each data point in Fig. 4 corresponding to a different single-step ATP separation, shows the very high accuracy and reproducibility of these single-step ATP separations. While the scatter between data points gives us information on the reproducibility of the single-step ATP separations, the error bars plotted on the individual data points give us information on the accuracy of the intensity determination from the different fitting routines for RRS and PLE (see methods section). Our observations in absorption spectroscopy that without SDS, some chiralities are already in the top phase, then move to the bottom phase when SDS is added, and then move again to the top phase when even more SDS is added, can be nicely seen for each of the examples presented in Fig. 4. Indeed, the first data point in each of the panels (*i.e.*, 0% wt/V SDS) shows a significant intensity in the top phase, while adding only 0.2% wt/V of SDS immediately shifts all intensity to the bottom phase.

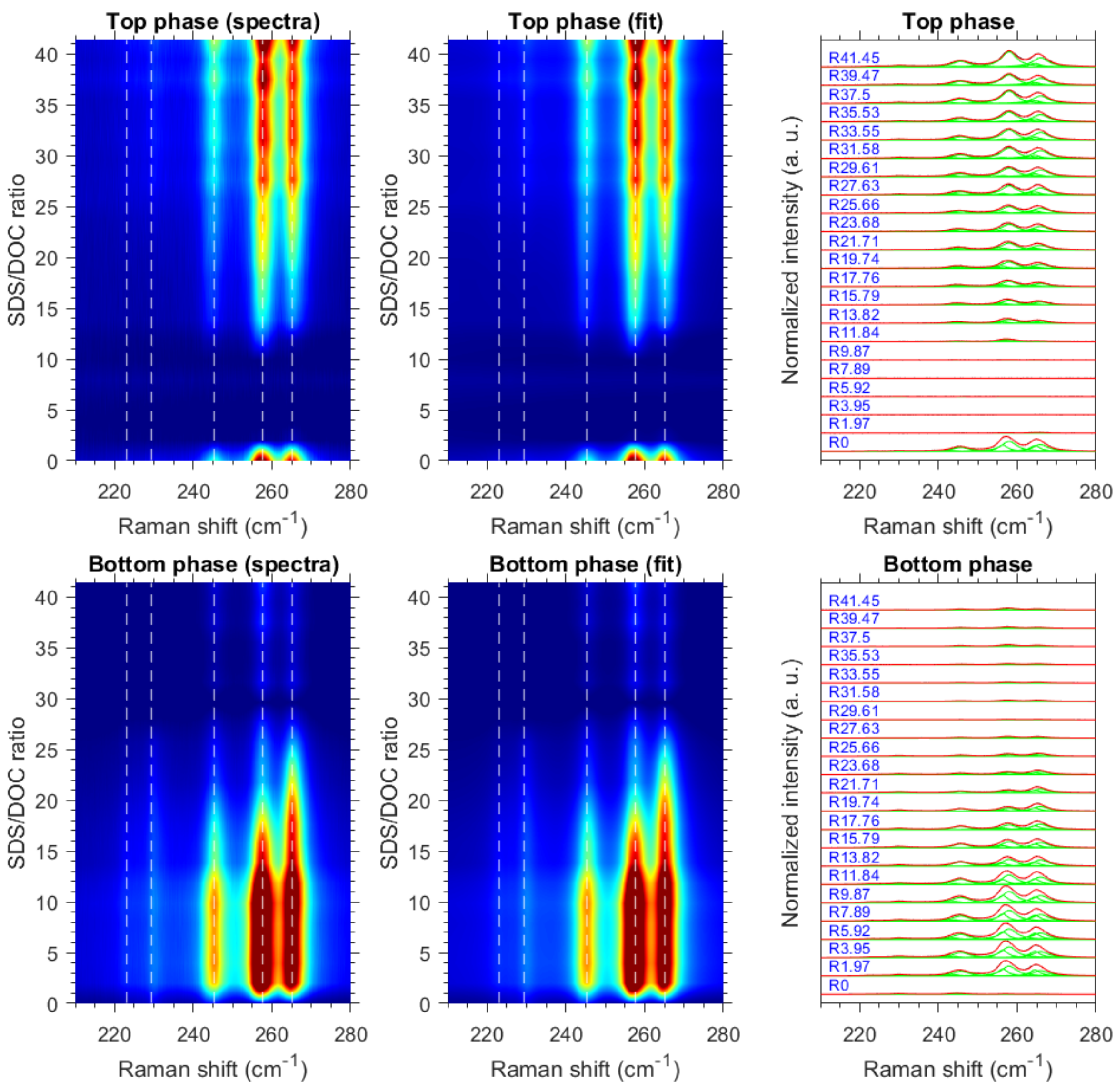

Figure 3: Resonant Raman scattering of bottom and top phases measured as a function of SDS/DOC ratio excited at 725 nm displaying from left to right the (11,4), (8,7), (8,6) (9,4) and (10,2) SWCNTs (denoted with dashed white lines). Experimental spectra are shown either as interpolated color maps (left panels) or individual and vertically shifted spectra (black curves in the right panels). The resulting fits are shown also either as interpolated color maps (middle panels), or superimposed on the Raman spectra (right panels) with individual components shown in green and total fits shown in red.

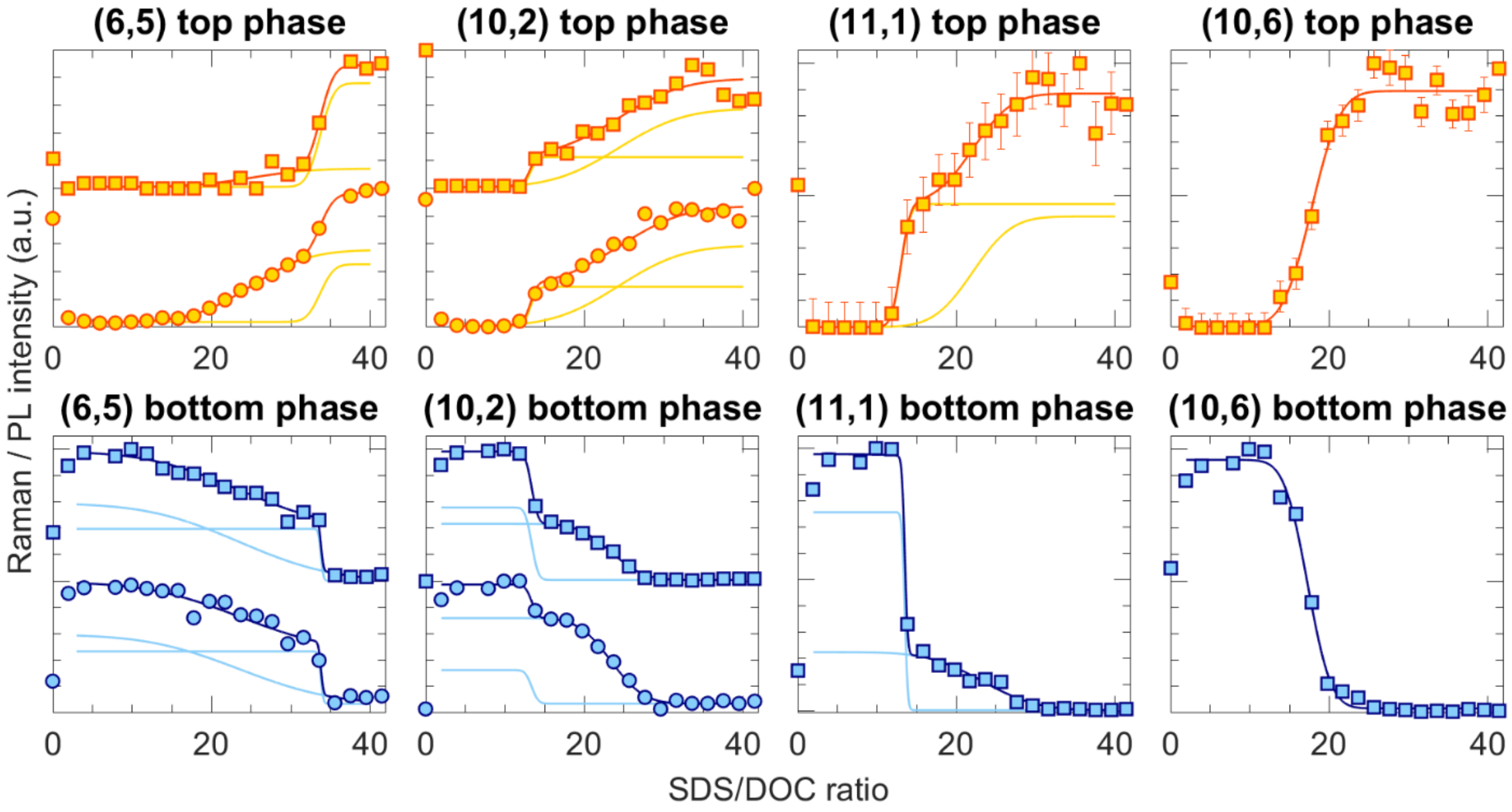

Figure 4: Normalized PL (squares) and Raman (circles) intensities (*i.e.*, equal to $K^{top}_{(n,m)}$ and $K^{bottom}_{(n,m)}$) for a selected set of SWCNT chiralities as a function of increasing SDS/DOC ratio (while keeping the DOC concentration constant at 0.0507 % wt/V), in the top (orange) and bottom (blue) phases. The fit components (see main text) for the bottom and top phases are shown with blue and yellow solid lines, respectively; their sums are shown in darker colors. Note that error bars on the data points are obtained directly from the fit of the Raman/PLE maps but are typically smaller than the size of the marker symbols. The full set of data for 37 studied chiralities is presented in Fig. S20.

At much higher SDS concentrations, the SDS molecules will start to compete with the DOC molecules, partly replacing them and thereby making the hybrid structure less hydrophilic, resulting in a migration to the top phase. Note that previously it was indeed found that SDS-covered SWCNTs always separate into the top phase, and at sufficiently high concentrations DOC-covered SWCNTs separate into the bottom phase [38], further supporting the above hypothesis that once SDS molecules start to replace DOC molecules, the SWCNTs move to the top phase. Most importantly, the exchange of DOC with SDS molecules on the SWCNT surface, occurs in a diameter-dependent manner, and thereby allows for a diameter-dependent separation of SWCNTs with ATP. Similar competition between another bile salt surfactant sodium cholate and SDS was studied in the references [55,56]: it was shown that at high SDS concentrations bile salt micelles are penetrated by SDS molecules, changing the SWCNT environment.

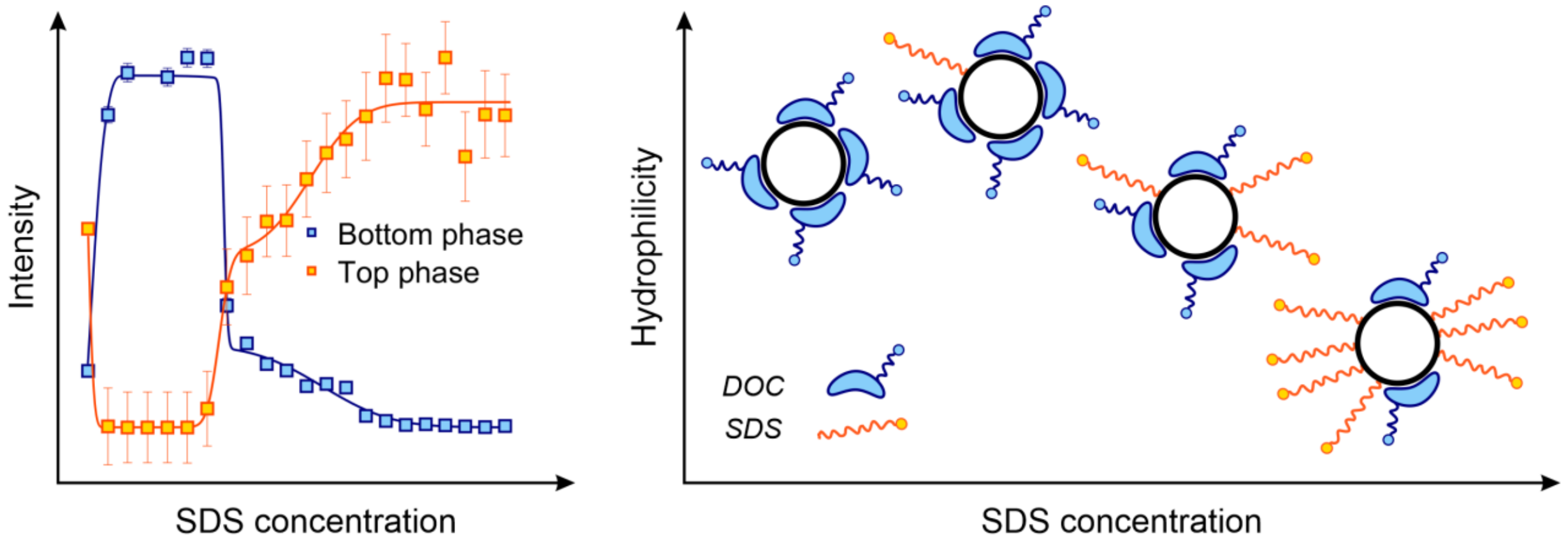

Figure 5. Schematic representation of the sorting mechanism when starting from a sufficiently low DOC concentration and systematically adding higher SDS concentrations. As an example, the migration curves for the (11,1) chirality are shown on the left.

Since this first separation step at very low SDS concentrations, moving all chiralities from top to bottom phase, seems to occur for all the chiralities at a very low SDS concentration (*i.e.*, in our experiments occurring in the first SDS concentration step), this first step is obviously not of interest for us from a chirality-sorting point of view. Therefore, in the following we mainly focus on the chirality-dependent phase changes at higher SDS concentration, where SWCNTs move from bottom to top phase with increasing SDS concentration. The resulting concentration variations in both phases, as presented in Fig. 4, were therefore fitted (neglecting the data point at 0% wt/V SDS) with an error function or a complementary error function to obtain the transition points, as well as the steepness of the transition, defined as the standard deviation or width $\sigma$ of the Gaussian distribution corresponding to the error function (*i.e.*, meaning that 68% of the transition occurs within this concentration range). If available, RRS and PL partition coefficient curves were fitted simultaneously for the same phase in order to obtain the best values for the transition points for each chirality. In some cases, if no experimental data point is present within the transition, the position and, in particular, the width of the transition is difficult to determine by means of a fit as the fit is underdetermined, resulting in vast errors on the position and the width where such a transition takes place. In those cases, we estimated the width as the widest visible boundaries of the transition curve (*i.e.*, the first points at a lower and a higher concentration than the transition).

It is worth noting that approximately two-thirds of all considered chiralities show only a one-step transition (like (10,6) in Fig. 4) while the partition coefficient curves for the other chiralities show more than one transition point, typically two (*e.g.*, see (6,5), (11,1) and (10,2) in Fig. 4). For 3 chiralities, we even observe 3 steps ((12,1), (12,6) and (13,4) in Fig. S20). Observing two steps for some chiralities is not surprising, given that similar surfactant combinations have been found to allow for sorting different enantiomers from each other [27,31–33,57]. Indeed, the bile salt surfactant used is a natural chiral surfactant, therefore only one of the two enantiomers of the surfactant is present, that could wrap differently around the two different SWCNT enantiomers, leading to a 'better' or 'worse' wrapping by DOC and thus, in our above model, a more difficult or more easy replacement by SDS, respectively. Interestingly, in most cases the two steps represent about 50% of the total intensity, which is to be expected for a racemic mixture of both enantiomers. A deviation from this 50% intensity could on the other hand be originating from a different Raman or PL cross-section for both (surfactant-wrapped) enantiomers and doesn't mean that the specific SWCNT sample has a different concentration of the two enantiomers.

While the origin of the two steps can be evidently ascribed to the two enantiomers, the observation of a third step is less obvious. Notably, this third step, if present, can always be found at an SDS/DOC ratio higher than 23 (1.16% wt/V SDS concentration), it is broad and low in its relative intensity. We assign it to a transition of (small) bundles because the third step is clearly seen only in RRS partition coefficient curves and (almost) absent in the PL ones, in line with the expected quenching of PL for bundles. Therefore, we only add this transition to our fitting model to fit the other transitions better, but they are not taken into account further on in the text. Besides, it sometimes happens that, for a particular chirality, in the PL partition coefficient curve only one transition is present while the RRS curve shows two transitions. In that case, the second transition that is clearly visible in RRS but (almost) absent in PL is also attributed to bundles for the same reason described above (see, for example, the second transitions of (7,5) and (7,6) in Fig. S20). Naturally, such a transition is also used only for fitting purposes.

Furthermore, it is important to note that in all cases, relative PL and RRS intensities of some transitions may be different for the same chirality (like for (6,5) in Fig. 4, top phase) which can be a sign of different PL and Raman cross-sections of the two enantiomers in the different polymer environments.

Coming back to the two-step transitions, we notice that one step is always steeper than the other (see, for example, the first broad and second sharp step of (6,5) and the first sharp and second broad step of (11,1) in Fig. 4). The difference in the widths of the two transitions for the same chirality can also be explained by the fact that DOC is a chiral and natural surfactant, with a semi-rigid cholesterol building block as the apolar part that stacks in a very ordered way on the SWCNT surface of the ‘best-fitting’ handedness, leading to steeper transitions, while more irregular structures logically lead to more inhomogeneous stacking configurations and thus broader transitions for the enantiomers with opposite handedness. Unfortunately, it is not always possible to resolve both transitions well, and that is why for approximately half of the studied chiralities we observe only one transition. Note that the steeper a transition is, the more interesting it is for the separation purpose (*vide infra*), although using only one transition out of two does not allow to sort 100% of SWCNTs of a particular chirality, drastically reducing the separation yield. Interestingly, (6,4) and (12,2) SWCNTs (Fig. S20) clearly show a non-zero intensity in the bottom phase at the highest SDS concentrations used, and the second transition is therefore not yet reached, because the SDS concentration cannot be further increased without making the SWCNT solubilization unstable [58]. However, since their transition is far from those of other chiralities, one can easily separate them from the other chiralities by moving all other chiralities to the top phase, and keeping these chiralities in the bottom phase, as previously demonstrated in reference [34].

Finally, the results for transitions as a function of SWCNT diameter (based on the bottom phase data) are summarized in Fig. 6(a) (transitions based on the top phase data are presented in Fig. S21, and numerical values of the transition points are presented in Tables S6 and S7). To highlight both the accuracy of the determination of the transition point and the width of the transition, each data point has two error bars associated to it. The colored, solid line error bars represent the accuracy of the determination of the transition point, which is obtained from the fit taking into account the experimentally obtained error on the individual data points presented in Fig. 4. The grey dashed error bars, on the other hand, represent the widths of the transitions, determined as the standard deviations or linewidth of the corresponding Gaussians. The steepest transitions (or the only one for those chiralities showing only a single separation step) are shown with blue circles and are the most important ones for developing future sorting methodologies. All other transitions (if present) are denoted with red squares. Surprisingly, an intriguing trend towards periodic modulation (shown with a dashed green line as a guide to the eye) with two maxima at approximately 0.71 and 1.03 nm SWCNT diameter can be clearly seen, which will be further investigated in the next sections. In addition, Fig. S22 demonstrates that there is no prominent dependence of partition coefficients on SWCNT chiral angle that is in correspondence with previous gel chromatography studies where it was found that DOC-based separations also occur there in a diameter-dependent manner [59].

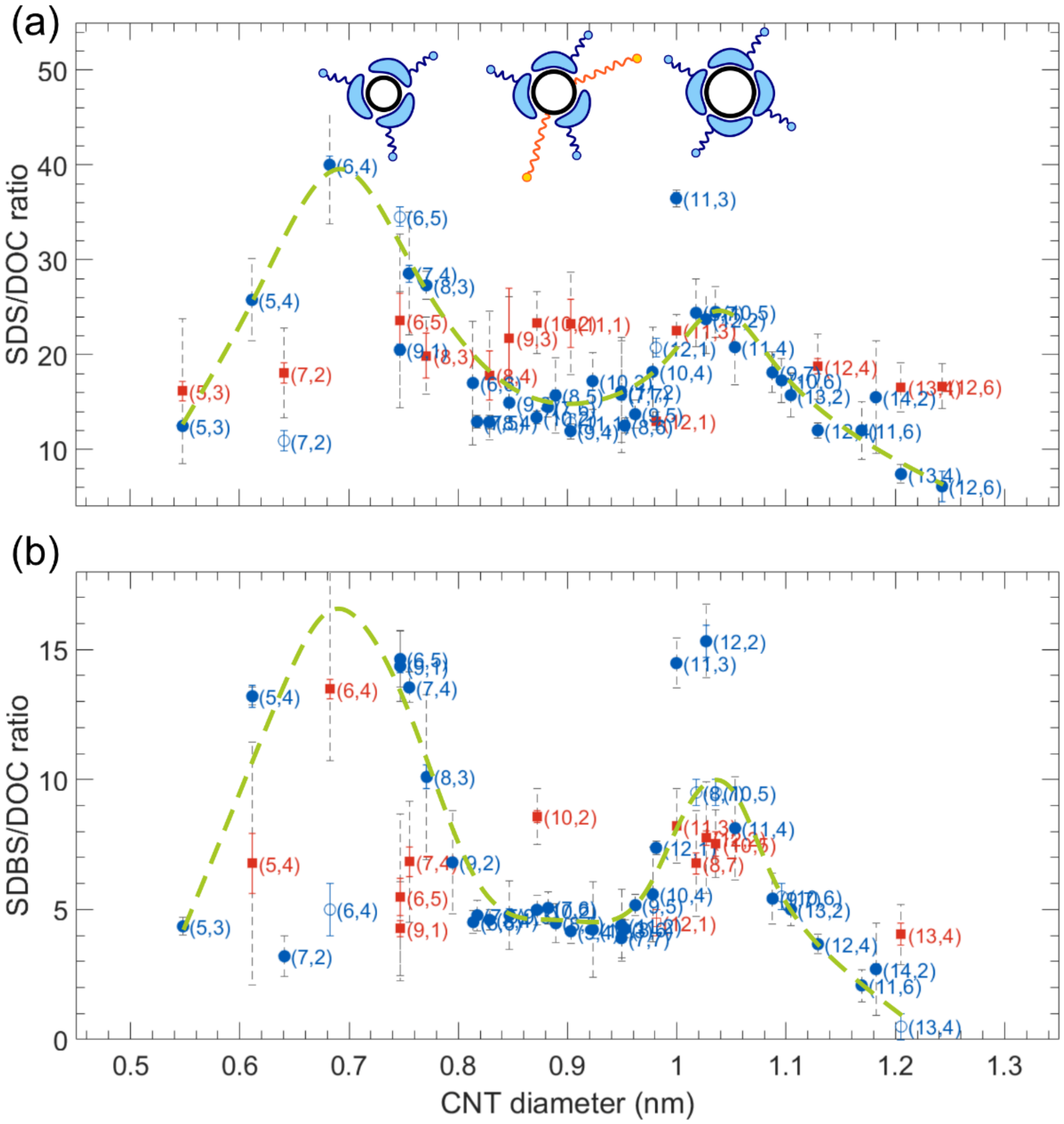

Figure 6: Transition points from bottom to top phase with increasing SDS concentration as determined from the bottom phase data for SWCNTs in SDS/DOC (at a fixed DOC concentration of 0.0507% wt/V) (a) and with increasing sodium dodecylbenzensulfonate (SDBS) concentration as determined from the bottom phase data for SWCNTs in SDBS/DOC (at a fixed DOC concentration of 0.1% wt/V) (b) as a function of SWCNT diameter, obtained by fitting the PL and RRS intensities of the bottom phase with a complementary error function (as presented in Fig. 4). Blue circles correspond to the transitions that are either the only one (for all the chiralities that show one transition) or the steepest one (for all the chiralities that have two- or three-step transition curves). The second transition, if present, is shown with red squares. Blue and red solid-line error bars are $1\sigma$ errors of the peak position fit. Dashed grey error bars are the transition linewidths obtained from the fit procedure (defined as the standard deviation or width ($\sigma$) of the corresponding Gaussians). If available, RRS and PL partition coefficient curves were fitted simultaneously for the same phase to obtain the best values for the transition points. If it was difficult to determine parameters of a transition (position, width) by means of fitting, *e.g.*, in case no experimental data point was present in the transition and the fit gave vast errors due to an underdetermined fit, we estimated the width of the transition from the nearest neighbor data points from the transition at lower and higher SDS/DOC ratio (corresponding points are shown with open symbols). The green dashed curve is a line to guide the eye, showing a trend towards periodic modulation, while schematic illustrations in the upper part of the figure illustrate an explanation of this trend. Note that for SDS/DOC the (6,4) and (12,2) SWCNTs have not fully transferred yet to the top phase at the highest SDS concentrations (respectively 50% and 20% of the maximum intensity left, see also Fig. S20). Similarly, for SDBS/DOC, the (6,4) SWCNT has not yet fully transferred (approx. 60% at the highest SDBS concentrations, see Fig. S26). We therefore take the lowest possible concentration at which the transition takes place as the data point and extend the grey dashed lines to the highest concentrations without showing an upper error bar.

### 3.2 Variation of cosurfactants

Previously it was postulated that mainly the ratio of both surfactant concentrations (*i.e.*, SDS/DOC ratio) determines the separation outcomes, and consequently, that in principle the DOC concentration could be changed when similarly changing the SDS concentration to keep the ratio constant [26,34], but typically this was difficult because of the above-mentioned upper limit for SDS SWCNT solutions to be stable [58]. Indeed, first attempts by Fagan *et al.* [26] showed that increasing the DOC concentration to ~0.1% wt/V DOC and thereby also increasing the SDS concentrations could result in a significant improvement in diameter resolution, in particular for larger diameter SWCNTs. Increasing the DOC concentration would also have the advantage that the SWCNT concentrations in the separations can be increased, as typically SWCNTs are first dispersed in 1% wt/V DOC and then diluted in the subsequent ATP separations, which would drastically affect separation yields. Note that 0.05% wt/V DOC is the lowest concentration at which SWCNT solutions are stable [40,60]. In the framework of our proposed model that is presented in Fig. 5, we started looking for other cosurfactants than SDS that can better compete with the DOC wrapping and thereby more quickly replace the DOC molecules and induce phase transitions at much lower added cosurfactant concentrations, which is useful as the needed low DOC concentration and the corresponding high SDS concentrations are limiting the yield of the separations.

The choice of this cosurfactant is based on two of its properties. First, the surfactant needs to be capable of solubilizing the CNTs, where we based ourselves on the extensive set of surfactants tested in reference [39], and secondly, it needs to separate the CNTs in the top phase in a one-surfactant ATP separation, so that it can compete with the bile salt surfactant DOC and move CNTs from the bottom phase to the top phase based on the partial exchange of the DOC surfactant layer.

In principle, one could expect that the better a cosurfactant competes with DOC, the more DOC (and therefore CNT solution) can be added to the two-phase system, as a result increasing the separation yield. To this end, six other surfactants were found to obey the above criterion that they separate CNTs into the top phase, and were then compared to SDS, namely sodium dodecylbenzensulfonate (SDBS), polyethylene glycol sorbitan monostearate (TWEEN60), polyethylene glycol sorbitan monooleate (TWEEN80), Polyoxyethylene octyl phenyl ether (TRITON X100), Cetyltrimethylammonium bromide (CTAB) and sodium dioctyl sulfosuccinate (DIOCT). We performed seven identical ATP separations with a cosurfactant/DOC ratio of 10 and a DOC concentration of 0.05% wt/V. Fig. 7 shows the PLE maps of the bottom and top phases for the different cosurfactants. While for SDS barely any of the SWCNT chiralities have migrated to the top phase at this cosurfactant/DOC ratio (see also Fig. 5), we find that for SDBS, TWEEN60 and TRITON X100 all SWCNTs have migrated to the top phase, showing these surfactants to be better competitors for DOC than SDS. However, while the same concentrations of SWCNTs were used in all separations, the absorption spectra of separations with TWEEN60 indicate a lack of SWCNT transitions in top and bottom phase, indicating that the SWCNTs have aggregated. Secondly, the PL intensity of SWCNTs with CTAB and TRITON X100 is severely quenched with respect to the PL intensity for the SDBS separations (note that in this case, we did not dilute to a 2% wt/V DOC concentration for conducting the PLE experiments, to show this effect). Also, by comparing the absorption spectra of these top and bottom phases (see SI Fig. S23), the reason for the observed quenching of the PL is most likely the interaction with the environment for TRITON X100 and CTAB, as the concentration of isolated tubes is very similar to the SDBS separations. Both CTAB, DIOCT and TWEEN80 give intermediate competition with DOC, and thus SDBS was chosen as a cosurfactant for further experiments (possibly TRITON X100 would have been a similarly suitable co-surfactant (based on absorption) but was not tried further in this work because of the quenched PL). Note that the exceptional competitive properties of SDBS have been previously studied in combinations with other surfactants in reference [61].

We carried out two sets of systematic single-step ATP experiments for the SDBS/DOC combination with varying SDBS concentrations, first using the same 0.05% wt/V concentration of DOC as with SDS, and afterwards one where we deliberately increase the DOC concentration to 0.1% wt/V, the concentration where primary DOC micelles are starting to form [40]. Similar as previously, we took PLE maps and Raman spectra of each of the fractions, fitted these experimental data to extract intensities as a function of SDBS concentration in the different phases and defined the transition points using the above-described methodology. The difference between these two DOC concentration cases is illustrated in Fig. 8, by plotting the partitioning in the bottom phase for a few representative chiralities, while the overview of the transition points is summarized in Fig. 6(b) and Fig. S24(a), as well as in Tables S8 and S9 for 0.1% wt/V and 0.05% wt/V DOC concentration, respectively. For 0.05% wt/V DOC concentration, we observe that SDBS competes so strongly with DOC that even adding the smallest aliquots of SDBS (SDBS/DOC ratios lower than two) result in the first/only transition to already take place, in agreement with the results in Fig. 7 that at a ratio of 10 all CNTs have already moved to the top phase. The transitions are usually extremely steep, and occur between two measurement points, which makes them difficult to be accurately fitted. The second, less steep transitions are present only for one-third of the examined chiralities. Clearly, 0.05% wt/V DOC concentration in combination with SDBS is not a suitable choice for separations as it would require the concentrations of SDBS to be extremely accurately pipetted for the separations to match a specific transition point. Interestingly however, as it can be seen in Fig. S24(a), the order of the chirality separation does show the same trend as the one observed for SDS/DOC, with a very similar periodic modulation depending on SWCNT diameter. Fortunately, when increasing the DOC concentration to 0.1% wt/V, all transitions shift to higher SDBS concentrations, while remaining highly resolved (*i.e.*, steep transitions as required for good separations) (Fig. 8). Interestingly again, the trend towards periodic modulation is seen again (Fig. 6(b)) also for this higher SDBS concentration, with approximately the same position of the maxima (0.72 and 1.02 nm), pointing to DOC being the origin of this periodic modulation.

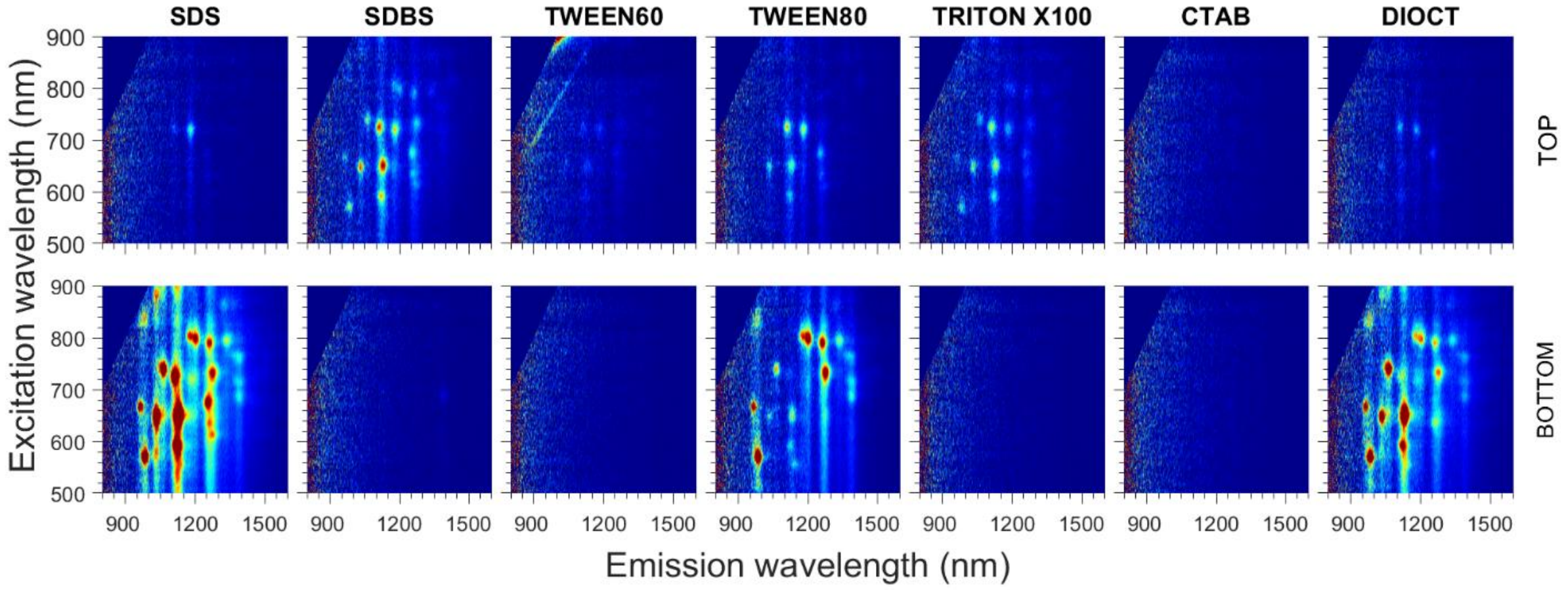

Figure 7: PLE maps of the bottom (lower row) and top (upper row) phases for the different cosurfactants (SDS, SDBS, TWEEN60, TWEEN80, TRITON X100, CTAB, DIOCT). The PL intensities in all bottom (top) phases can be directly compared to each other, while the PL intensities between bottom and top phases are different due to the different volumetric dilutions in the two phases (the top phase having a larger volume than the bottom phase after phase separation).

To further clarify the role of DOC in the periodic modulation, we also performed one additional systematic single-step ATP experiment, in which only DOC without any cosurfactants was used. Its results are summarized in Fig. S24(b) and Table S10, while different types of partition coefficient curves are shown in Fig. S25. However, in this case an additional complication appears. Half of the studied chiralities demonstrate their first, steep transition at DOC concentrations below 0.04% wt/V (see (6,5), (7,5), (12,6) in Fig. S25). Then, for approximately one-third of the considered chiralities, bundling due to too low DOC concentration occurs before moving to another phase that results in 'unfinished' partition coefficient curves like the ones for (13,2) and (10,6) (Fig. S25) and very wide upper limits for transitions, *i.e.*, their exact positions cannot be determined since a full transition has not occurred yet at the lowest DOC concentrations. Note that the 0.05% wt/V concentration at which most ATP separations so far have been conducted, is at the limit of having most of the SWCNT chiralities residing in the bottom phase, and thus increasing the DOC concentration for future experiments, by using another cosurfactant, will drastically help in keeping the SWCNT chiralities stable during the separations. Due to these complications, it is very difficult to observe the diameter-dependent modulation in these separations for DOC itself in Fig. S24(b). Note that in this case, the 'peaks' in the modulation curves (green dashed lines in Fig. 6) should correspond to 'dips' as the separation is now different, *i.e.* in the previous case (with cosurfactant/DOC combinations), the better-wrapped SWCNT chiralities move to the top phase at higher cosurfactant concentrations while here (with DOC alone) the SWCNT chiralities that are better covered by the DOC molecules, will be the last ones to separate to the top phase when reducing the DOC concentration. Indeed, when looking to the red data points in Fig. S24(b), we observe a minimum around 0.72 nm, similarly as for the SDS/DOC (00507% wt/V) and SDBS/DOC (0.1% wt/V) separations presented in Fig. 6. The second 'dip' at about 1.02 nm is more difficult to observe, because all of the chiralities above 0.96 nm show very wide and unfinished transitions with DOC, thereby their exact transition points could not be determined.

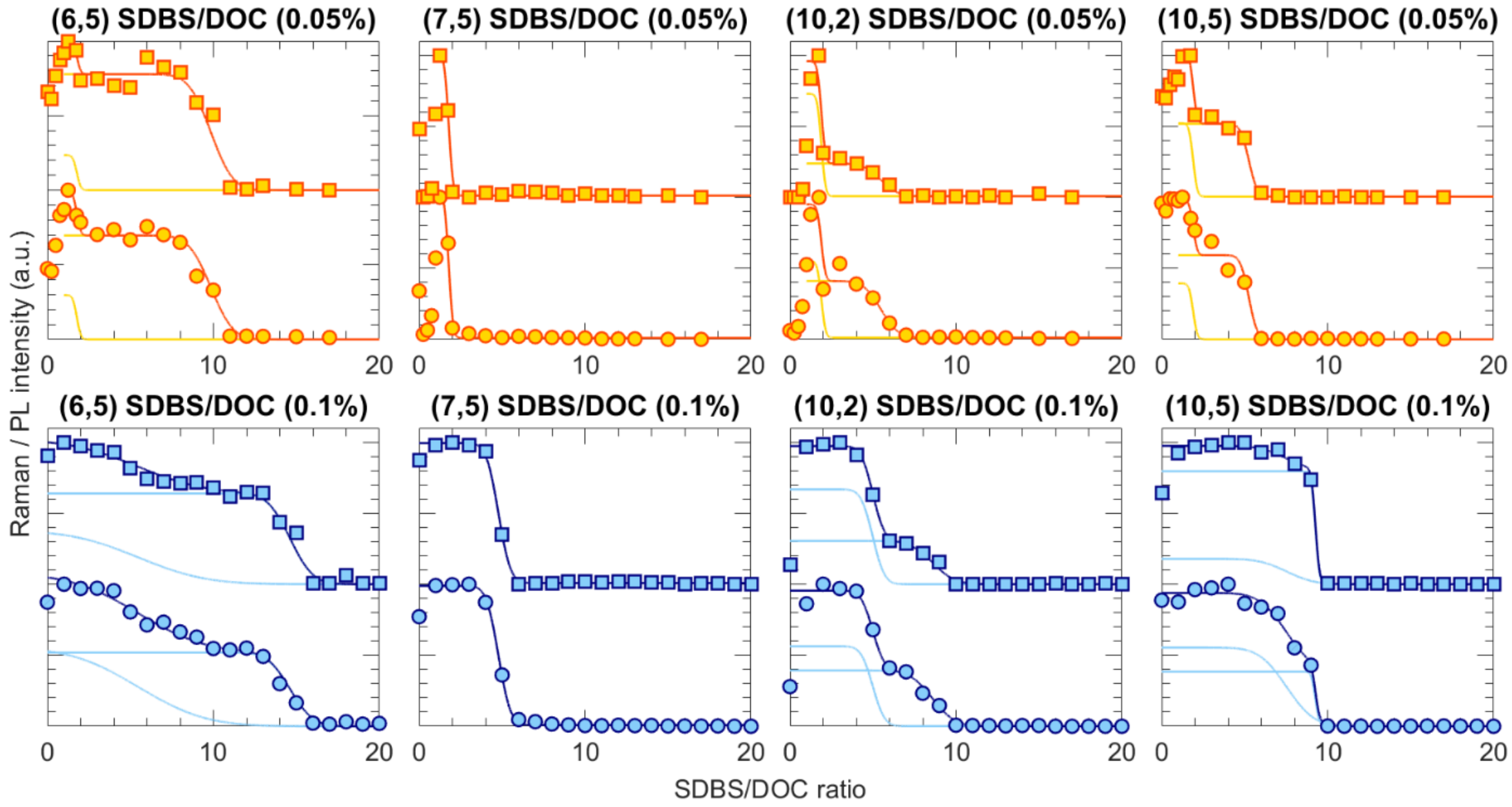

Figure 8: Normalized PL (squares) and Raman (circles) intensities (*i.e.*, $K_{(n,m)}^{top}$ and $K_{(n,m)}^{bottom}$) for a selected set of SWCNT chiralities as a function of increasing SDBS/DOC ratio (bottom phases). The individual fit components for the cases of 0.05% wt/V and 0.1% wt/V DOC concentrations are shown with yellow and blue solid lines, respectively; their sums (overall fits) are shown in darker blue and orange respectively. Error bars on the Raman and PL intensities are not visible because they are typically smaller than the data symbols.

### 3.3 Origin of the periodic modulation

The above systematic characterizations with different cosurfactants and without cosurfactants, point at DOC and its chirality-dependent stacking on the SWCNT walls to determine the separation order, while the competition with cosurfactants can be used to enhance the differences between the different transition points, *i.e.*, ‘rescaling’ the observed periodic modulation pattern. The latter statement is clearly illustrated by comparing the two (most representative) data sets from this work in Fig. 6.

We have also compared our results with SDS/DOC transition points reported by J. Fagan *et al.* [26,32], which were derived empirically from subsequent ATP steps in a multi-step separation methodology [26] and by surfactant exchange PL experiments [32] (see Fig. S27). In that work [26], a monotonous, 1/diameter relation was postulated with increasing SDS concentration, with a few chiralities disobeying this 1/diameter trend. Although this seems contradictory to our findings, when plotting the literature transition points on top of our experimental data, they are in fact in excellent agreement with our results. The main difference why the periodic modulation was not found in previous work is the much broader range of chiralities studied here, including in particular smaller diameter chiralities, possible through the proposed systematic studies, pointing at a more complicated, periodically modulated pattern.

So, the question that could be raised is why for SWCNTs with a diameter of approximately 0.72 nm and 1.02 nm, the DOC molecules wrap the SWCNTs more effectively, such that a higher cosurfactant concentration is needed to separate those chiralities to the top phase. To this end we devised a molecular model, consisting of a (6,4) and (12,2) chirality, with approximately those diameters, surrounded by DOC molecules, which is presented in Fig. 9. First, the geometry of the DOC molecules was optimized using the OPLS force field [62,63] and atomic charges from a semi-empirical calculation with the AM1 Hamiltonian [64] (see section 2.6 for the details). Then, these optimized DOC molecules were placed around the SWCNT, and the geometry partly relaxed using the OPLS force field, so that also the distance between the molecules and the SWCNTs, and between the different molecules converged to their equilibrium van der Waals distances.

The model shows that 7 DOC molecules precisely fit around the circumference of a (6,4) SWCNT, while 8 DOC molecules fit tightly around the (12,2) chirality, showing indeed that the two maxima found in the transition points with diameter (Fig. 6) correspond to those SWCNT diameters that correspond to a discrete number of DOC molecules fitting exactly around their circumference. Intermediate diameters result in holes in the DOC layer, and thus in a more hydrophobic structure. These holes will be easily filled up by the (less hydrophilic) cosurfactant, and the less perfect DOC stacking will more easily allow for DOC to be replaced further by the cosurfactant, resulting in SWCNT migration at lower added cosurfactant concentrations.

Interestingly, note that similar diameter-dependent modulations have been observed before in the density of sodium-cholate-coated SWCNTs in density gradient ultracentrifugation experiments [65,66]. However, those results cannot be directly compared with the present work, because of the different surfactant and much higher surfactant concentrations, and different diameter range considered in those studies.

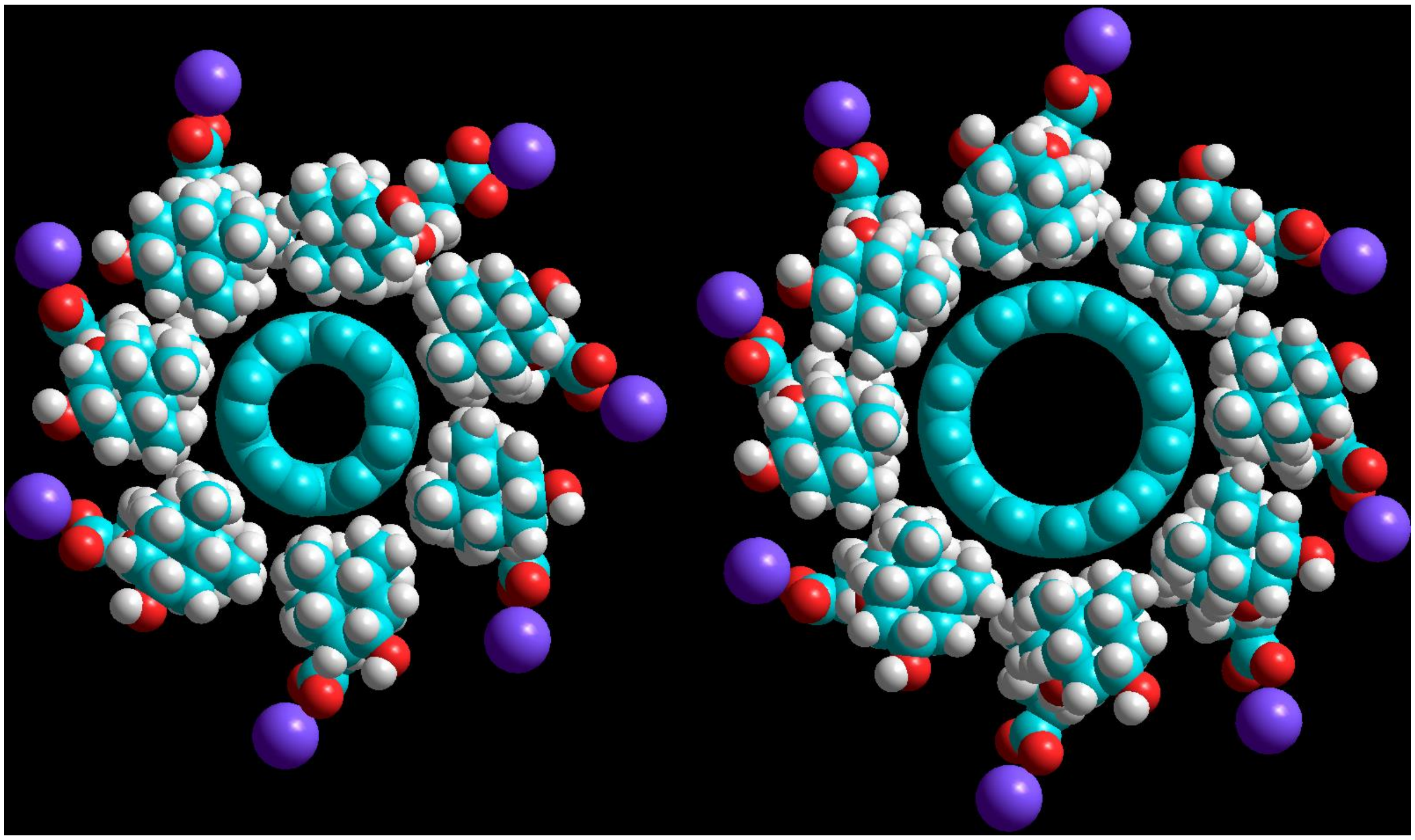

Figure 9: Molecular model of the stacking of DOC around (6,4) (left) and (12,2) (right) SWCNT chiralities, demonstrating a tightly fitting surfactant layer, with 7 and 8 DOC molecules fitting around the circumference, respectively.

### 3.4 Application of these results to develop new sorting methodologies.

To develop new sorting methodologies based on the above results, the following procedure can be used. Typically, one can start with a first separation where the chirality of interest is just separated into the bottom phase, *i.e.*, by selecting an SDS/SDBS concentration just below the transition point. The bottom phase can then be collected. Then, in a second step, a mimicking top phase can be added, of which the composition can be obtained by measuring the top phase of the first step in absorption spectroscopy and determining the concentrations of each of the components (PEG, dextran and surfactants as presented in Fig. S1 in the SI), while the SDS/SDBS concentration can be increased just above the transition point, thereby moving only this one chirality to the top phase.

These results thus demonstrate that in order to separate a specific chirality to the highest purity, its transition point needs to be far from those of other chiralities and should preferably also be very steep. The periodic modulation observed is thus essential for the success of the separation, but also identifies only a few chiralities that can be sorted to the highest precision (those with diameters closely matching the 'peaks' of the periodic modulation). In particular, our results explain why in previous work the (6,5) and (6,4) SWCNTs could be particularly well separated [24]. Indeed, our results indicate that this is possible due to the periodic diameter modulation induced by DOC, which results in the transition point of (6,4)/(6,5) chiralities at the highest cosurfactant concentrations, sufficiently distinct from those of other chiralities. Thus, when taking the above results into account, we can develop a two-step sorting protocol that results in the purification of (6,5) SWCNTs and compare the effect of using different cosurfactants (SDS or SDBS) on the separation purity and yield.

We first compare the results for SDS/DOC (0.05% wt/V DOC) and SDBS/DOC (0.1% wt/V DOC) when adding the same quantity of SWCNT solution. When looking in Fig. 6(a), an SDS/DOC ratio of 29 results in most of the other chiralities separating in the top phase, while (6,4), (6,5) and (11,3) separate in the bottom phase, and the ideal ratio for SDBS/DOC (0.1% wt/V DOC) is found to be 12. The exact quantities for the different separations can be found in Table S11. First, all the surfactants and polymers are added together and mixed, then the SWCNTs are added, and the suspension is remixed after which a low-speed centrifugation is applied to accelerate the phase separation. Afterwards, the bottom phase is collected, and an absorption spectrum of the top phase is acquired to estimate the concentration of PEG and dextran in this top phase. For both separations, the top phase was found to be composed of 13.66% wt/V PEG and 2.41% wt/V dextran.

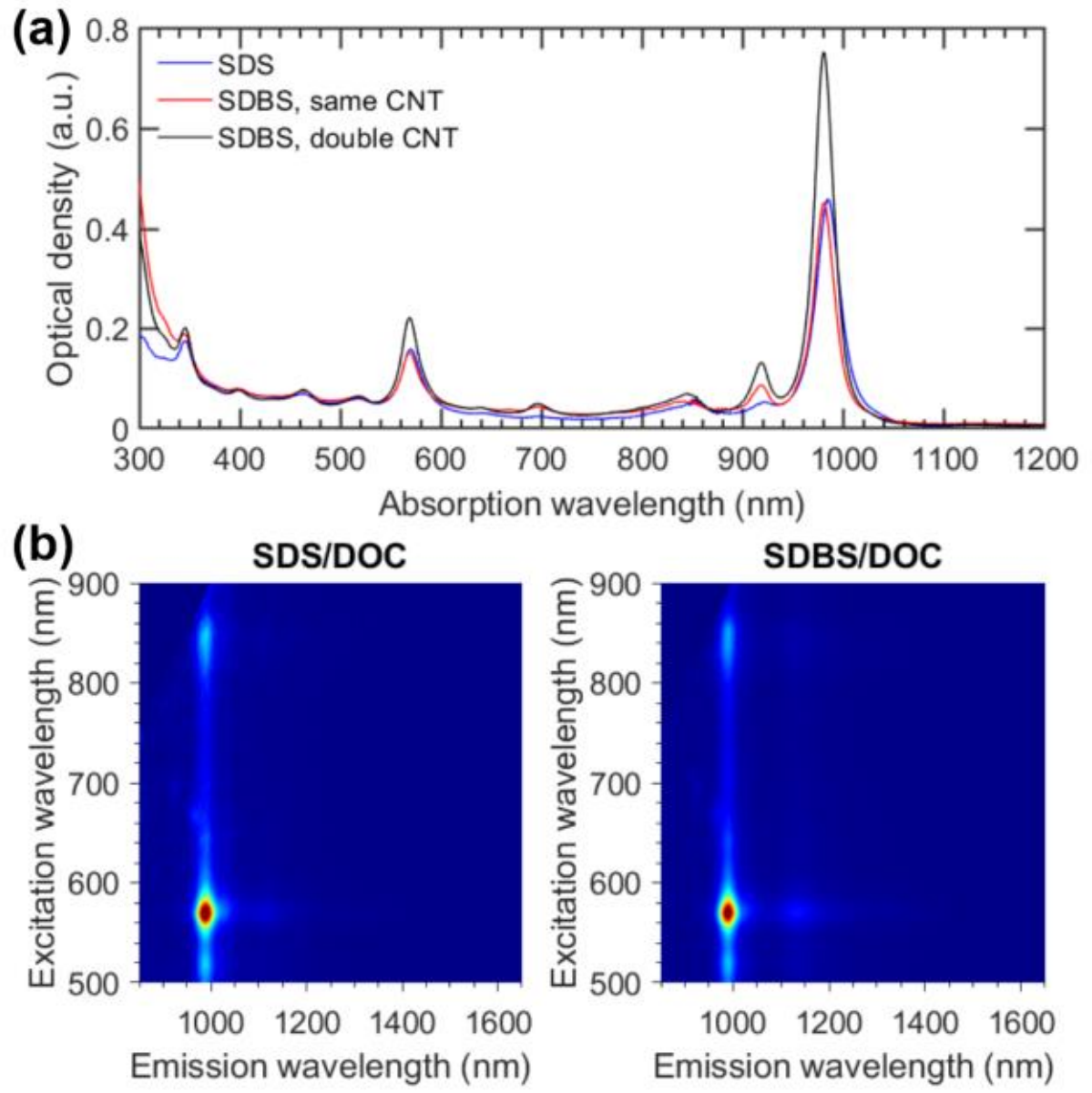

Figure 10: (a) Absorption of the two-step (6,5) separations using SDS/DOC (blue curve), SDBS/DOC using the same quantity of SWCNTs (red curve), and SDBS/DOC with twice the amount of SWCNTs added (black curve). (b) PLE maps of the SDS/DOC and SDBS/DOC separations showing the high and very similar purity of the samples.

Since the absorption spectrum of DOC is not sufficiently distinctive for the low concentrations used, it is difficult to obtain the exact concentration and therefore we assume an equal distribution of DOC among bottom and top phase. On the other hand, it is possible to estimate an upper limit for the concentration of SDS/SDBS in the bottom phase after the first step, being 1.3% wt/V for SDS and 0.8% wt/V for SDBS (with typical error of the estimation of approximately 0.1% wt/V). To estimate this upper limit, we start from the surfactant concentrations used in the separation assuming equal distribution of the cosurfactants across both phases (*i.e.,* 1.45% wt/V and 1.2% wt/V for SDS and SDBS respectively) and subtract the surfactant absorption spectra (with the concentration as a scaling factor) from the absorption spectrum of the bottom phase, typically resulting in negative absorption peaks in the bottom phase corresponding to SDS/SDBS. Then, we reduce the SDS/SDBS concentration until the negative absorption peaks are no longer observed, representing the above-mentioned upper limit of the concentration. Accordingly, the concentration of SDS/SDBS needs to be increased in the mimicking top phase to reach the desired separation concentration in the second step. The exact composition for the mimicking top phase used in the second step is also provided in Table S11, assuming a ratio of 175 µL from the mimicking top phase and 75 µL of the bottom phase in the second step. For the SDS/DOC separation, the top mimicking phase is made at an SDS/DOC ratio of 43.6, so that after mixing and before phase separation an SDS/DOC ratio of approximately 38 is reached, such that the (6,5) chirality moves to the top phase, and the (6,4) chirality remains in the bottom phase. Similarly, for SDBS/DOC, the top mimicking phase is made at an SDBS/DOC ratio of 18, so that after mixing and before phase separation the SDBS/DOC ratio amounts to approximately 15. Note that the data point for (6,4) marked in Fig. 6 is the lowest estimation of the transition point (since (6,4) did not fully migrate to the top phase within the measured cosurfactant concentration range, see Fig. S26) such that at this SDBS/DOC ratio the (6,5) tube moves completely to the top phase and a significant fraction of the (6,4) chirality remains in the bottom phase. Both two-step separations result in highly pure (6,5) SWCNT samples with similar purity and yield, as exemplified in Fig. 10(a) (absorption spectra, red and blue curves) and Fig. 10(b) (PLE maps).

However, as the DOC concentration in the SDBS/DOC separation is a factor of 2 higher, twice the amount of SWCNTs can be added to the same 2-phase polymer mixture, thereby allowing for enhancing the overall yield/volume, without changing the purity of the samples. The absorption spectrum of such a separation is also shown in Fig. 10(a) (black curve).

## 4. Conclusions

To conclude, we present a new systematic methodology to investigate the effect of various parameters on the sorting of SWCNTs by ATP. The methodology consists of performing a series of single-step ATP separations in which surfactant concentrations are systematically varied one by one, while monitoring the migration of each SWCNT chirality between the two phases using a combination of absorption, resonant Raman and PLE spectroscopy. Importantly, in this manner, sorting information is obtained from more than 30 different chiral structures spanning a diameter range from 0.6 to 1.24 nm. We find a periodic modulation of the transition points as a function of diameter and relate it to the structure-specific and diameter-dependent stacking of sodium deoxycholate, depending on how an integer number of molecules fit around the circumference of the SWCNTs. For example, we found that the diameter of (6,4) SWCNTs exactly matches the stacking of 7 DOC molecules, while 8 DOC molecules fit around a (12,2) chirality. Addition of cosurfactants aids in separating the different transition points and as such allowing for a separation of individual chiralities. The methodology is demonstrated for the separation of (6,5) SWCNTs, resulting in higher yields per volume of the polymers when using SDBS instead of SDS as a cosurfactant, where SDBS outperforms SDS in view of competition with DOC.

Since these results indicate that the size of the bile salt surfactant molecules determines which SWCNT diameters show a distinct transition point and, as a result, can be efficiently separated, future work thus requires this bile salt surfactant to be altered to allow for other chiralities to be purified to the highest purities in a two-step procedure. Also, the influence of other parameters, such as addition of salts, temperature and pH, can be studied in the same systematic manner to allow for a future predictive sorting of specific SWCNT chiralities. Since the specific stacking of surfactants around the SWCNT circumference will also influence the density of the SWCNT-surfactant hybrids, and the interaction of the SWCNT-surfactant hybrid with its environment, we expect the results of this work to be broadly applicable to other sorting methodologies as well, in particular, density gradient ultracentrifugation [66] and gel chromatography [67].

**Declaration of competing interest**

The authors declare that they have no conflicts of interest.

**CRediT authorship contribution statement**

**Joeri Defillet:** Investigation; Data curation; Methodology; Formal analysis; Writing - review & editing. **Marina Avramenko:** Data curation; Validation; Formal analysis; Visualization; Writing - original draft. **Miles Martinati:** Investigation; Formal analysis; Writing - review & editing. **Miguel Angel Lopez Carrillo:** Formal analysis; Writing - review & editing. **Domien van der Elst:** Investigation; Writing - review & editing. **Wim Wenseleers:** Methodology; Investigation; Software; Resources; Visualization; Supervision; Funding acquisition; Writing - review & editing. **Sofie Cambré:** Conceptualization; Investigation; Software; Data curation; Methodology; Resources; Visualization; Supervision; Funding acquisition; Writing - original draft.

**Joeri Defillet** and **Marina Avramenko** contributed equally to this work.

**Acknowledgment**

JD, MA, MM and SC acknowledge funding from the European Research Council through an ERC Starting Grant No. 679841 (ORDERin1D) which was granted to SC and partially funded the research of JD, MM and MA. MM and MALC acknowledge the University of Antwerp Research Fund (BOF-DOCPRO4) that provided them with a PhD grant. This research was furthermore also supported by the Fund for Scientific Research Flanders (FWO) through projects G040011N, G02112N, G035918N and G036618N and the EOS CHARMING project G0G6218N [EOS-ID 30467715].

## Supporting information

Supporting Information is available online at ….. and includes (1) Examples of systematic parameter variations, (2) Measuring the composition of the phases by absorption spectroscopy, (3) PLE maps of bottom and top phases for SDS/DOC variation, (4) Examples of simultaneous 2D PLE fits of the PLE maps, (5) RRS spectra and fits at different excitation wavelengths, (6), fits of partition coefficients for SDS/DOC, (7) overview figures for other surfactant combinations (8) tables of transition points for all surfactant combinations, (9) details on the separation parameters for (6,5) SWCNTs in two steps and (10) Comparison with literature data.

## References

[1] S. Iijima, T. Ichihashi, Single-shell carbon nanotubes of 1-nm diameter, Nature. 363 (1993) 603–605. http://dx.doi.org/10.1038/363603a0.

[2] H. Kataura, Y. Kumazawa, Y. Maniwa, I. Umezu, S. Suzuki, Y. Ohtsuka, Y. Achiba, Optical properties of single-wall carbon nanotubes, Synth. Met. 103 (1999) 2555–2558. https://doi.org/http://dx.doi.org/10.1016/S0379-6779(98)00278-1.

[3] A. Jorio, M.S. Dresselhaus, G. Dresselhaus, Carbon nanotubes: advanced topics in the synthesis, structure, properties and applications, Springer, 2008.

[4] F. Yang, X. Wang, D. Zhang, J. Yang, LuoDa, Z. Xu, J. Wei, J.-Q. Wang, Z. Xu, F. Peng, X. Li, R. Li, Y. Li, M. Li, X. Bai, F. Ding, Y. Li, Chirality-specific growth of single-walled carbon nanotubes on solid alloy catalysts, Nature. 510 (2014) 522–524. https://doi.org/10.1038/nature13434 http://www.nature.com/nature/journal/v510/n7506/abs/nature13434.html#supplementary-information.

[5] F. Yang, X. Wang, M. Li, X. Liu, X. Zhao, D. Zhang, Y. Zhang, J. Yang, Y. Li, Templated Synthesis of Single-Walled Carbon Nanotubes with Specific Structure, Acc. Chem. Res. 49 (2016) 606–615. https://doi.org/10.1021/acs.accounts.5b00485.

[6] H. An, A. Kumamoto, H. Takezaki, S. Ohyama, Y. Qian, T. Inoue, Y. Ikuhara, S. Chiashi, R. Xiang, S. Maruyama, Chirality specific and spatially uniform synthesis of single-walled carbon nanotubes from a sputtered Co-W bimetallic catalyst, Nanoscale. 8 (2016) 14523–14529. https://doi.org/10.1039/c6nr02749k.

[7] F. Yang, X. Wang, J. Si, X. Zhao, K. Qi, C. Jin, Z. Zhang, M. Li, D. Zhang, J. Yang, Z. Zhang, Z. Xu, L.-M. Peng, X. Bai, Y. Li, Water-Assisted Preparation of High-Purity

Semiconducting (14,4) Carbon Nanotubes, ACS Nano. 11 (2016) 186–193. https://doi.org/10.1021/ACSNANO.6B06890.

[8] F. Yang, M. Wang, D. Zhang, J. Yang, M. Zheng, Y. Li, Chirality Pure Carbon Nanotubes: Growth, Sorting, and Characterization, Chemical Reviews. 120 (2020) 2693–2758. https://doi.org/10.1021/ACS.CHEMREV.9B00835.

[9] D. Janas, Towards monochiral carbon nanotubes: a review of progress in the sorting of single-walled carbon nanotubes, Materials Chemistry Frontiers. 2 (2017) 36–63. https://doi.org/10.1039/C7QM00427C.

[10] R. Krupke, F. Hennrich, H. v Löhneysen, M.M. Kappes, Separation of Metallic from Semiconducting Single-Walled Carbon Nanotubes, Science. 301 (2003) 344. http://science.sciencemag.org/content/301/5631/344.abstract.

[11] M.S. Arnold, A.A. Green, J.F. Hulvat, S.I. Stupp, M.C. Hersam, Sorting carbon nanotubes by electronic structure using density differentiation, Nature Nanotechn. 1 (2006) 60. https://doi.org/http://www.nature.com/nnano/journal/v1/n1/suppinfo/nnano.2006.52_S1.html.

[12] S. Ghosh, S.M. Bachilo, R.B. Weisman, Advanced sorting of single-walled carbon nanotubes by nonlinear density-gradient ultracentrifugation, Nature Nanotechn. 5 (2010) 443. https://doi.org/http://www.nature.com/nnano/journal/v5/n6/abs/nnano.2010.68.html#supplementary-information.

[13] N. Stürzl, F. Hennrich, S. Lebedkin, M.M. Kappes, Near Monochiral Single-Walled Carbon Nanotube Dispersions in Organic Solvents, The Journal of Physical Chemistry C. 113 (2009) 14628–14632. https://doi.org/10.1021/jp902788y.

[14] K. Akazaki, F. Toshimitsu, H. Ozawa, T. Fujigaya, N. Nakashima, Recognition and One-Pot Extraction of Right- and Left-Handed Semiconducting Single-Walled Carbon Nanotube Enantiomers Using Fluorene-Binaphthol Chiral Copolymers, Journal of the American Chemical Society. 134 (2012) 12700–12707. https://doi.org/10.1021/ja304244g.

[15] F. Lemasson, N. Berton, J. Tittmann, F. Hennrich, M.M. Kappes, M. Mayor, Polymer Library Comprising Fluorene and Carbazole Homo- and Copolymers for Selective Single-Walled Carbon Nanotubes Extraction, Macromolecules. 45 (2012) 713–722. https://doi.org/10.1021/ma201890g.

[16] M. Zheng, A. Jagota, M.S. Strano, A.P. Santos, P. Barone, S.G. Chou, B.A. Diner, M.S. Dresselhaus, R.S. McLean, G.B. Onoa, G.G. Samsonidze, E.D. Semke, M. Usrey, D.J. Walls, Structure-Based Carbon Nanotube Sorting by Sequence-Dependent DNA Assembly, Science. 302 (2003) 1545. http://science.sciencemag.org/content/302/5650/1545.abstract.

[17] X. Tu, S. Manohar, A. Jagota, M. Zheng, DNA sequence motifs for structure-specific recognition and separation of carbon nanotubes, Nature. 460 (2009) 250.

https://doi.org/10.1038/nature08116 https://www.nature.com/articles/nature08116#supplementary-information.

[18] H. Liu, D. Nishide, T. Tanaka, H. Kataura, Large-scale single-chirality separation of single-wall carbon nanotubes by simple gel chromatography, Nature Commun. 2 (2011) 309. https://doi.org/10.1038/ncomms1313 http://www.nature.com/articles/ncomms1313#supplementary-information.

[19] C. Blum, N. Stürzl, F. Hennrich, S. Lebedkin, S. Heeg, H. Dumlich, S. Reich, M.M. Kappes, Selective Bundling of Zigzag Single-Walled Carbon Nanotubes, ACS Nano. 5 (2011) 2847–2854. https://doi.org/10.1021/nn1033746.

[20] C.Y. Khripin, J.A. Fagan, M. Zheng, Spontaneous Partition of Carbon Nanotubes in Polymer-Modified Aqueous Phases, JACS. 135 (2013) 6822. https://doi.org/10.1021/ja402762e.

[21] J.G. Pryde, J.H. Phillips, FRACTIONATION OF MEMBRANE-PROTEINS BY TEMPERATURE-INDUCED PHASE-SEPARATION IN TRITON X-114 - APPLICATION TO SUBCELLULAR-FRACTIONS OF THE ADRENAL-MEDULLA, Biochemical Journal. 233 (1986) 525–533. <Go to ISI>://WOS:A1986AXZ8600029.

[22] C. Bordier, Phase separation of integral membrane proteins in triton x114 solution, THE JOURNAL OF BIOLOGICAL CHEMISTRY. 256 (1981) 1604–1607.

[23] G.F. Cecilia Holm and Per Belfrage, Demonstration of the amphiphilic character of hormone-sensitive lipase by temperature-induced phase separation in Triton X-114 and charge-shift electrophoresis, THE JOURNAL OF BIOLOGICAL CHEMISTRY. 261 (1986) 15659–15661.

[24] J.A. Fagan, C.Y. Khripin, C.A. Silvera Batista, J.R. Simpson, E.H. Hároz, A.R. Hight Walker, M. Zheng, Isolation of Specfic Small Diameter Single-Wall Carbon Nanotube Species via Aqueous Two-Phase Extraction, Adv. Mater. 26 (2014) 2800. https://doi.org/10.1002/adma201304873.

[25] G. Ao, C.Y. Khripin, M. Zheng, DNA-Controlled Partition of Carbon Nanotubes in Polymer Aqueous Two-Phase Systems, Journal of the American Chemical Society. 136 (2014) 10383–10392. https://doi.org/10.1021/ja504078b.

[26] J.A. Fagan, E.H. Hároz, R. Ihly, H. Gui, J.L. Blackburn, J.R. Simpson, S. Lam, A.R. Hight Walker, S.K. Doorn, M. Zheng, Isolation of >1 nm Diameter Single-Wall Carbon Nanotube Species Using Aqueous Two-Phase Extraction, ACS Nano. 9 (2015) 5377–5390. https://doi.org/10.1021/acsnano.5b01123.

[27] M. Zhang, C.Y. Khripin, J.A. Fagan, P. McPhie, Y. Ito, M. Zheng, Single-Step Total Fractionation of Single-Wall Carbon Nanotubes by Countercurrent Chromatography, Analytical Chemistry. 86 (2014) 3980–3984. https://doi.org/10.1021/ac5003189.

[28] H. Gui, J.K. Streit, J.A. Fagan, A.R. Hight Walker, C. Zhou, M. Zheng, Redox Sorting of Carbon Nanotubes, Nano Letters. 15 (2015) 1642–1646. https://doi.org/10.1021/nl504189p.

[29] H. Li, G. Gordeev, O. Garrity, S. Reich, B.S. Flavel, Separation of Small-Diameter Single-Walled Carbon Nanotubes in One to Three Steps with Aqueous Two-Phase Extraction, ACS Nano. 13 (2019) acsnano.8b09579. https://doi.org/10.1021/acsnano.8b09579.

[30] L. Wei, B.S. Flavel, W. Li, R. Krupke, Y. Chen, Exploring the upper limit of single-walled carbon nanotube purity by multiple-cycle aqueous two-phase separation, Nanoscale. 9 (2017) 11640–11646. https://doi.org/10.1039/C7NR03302H.

[31] H. Li, G. Gordeev, O. Garrity, N.A. Peyyety, P.B. Selvasundaram, S. Dehm, R. Krupke, S. Cambré, W. Wenseleers, S. Reich, M. Zheng, J.A. Fagan, B.S. Flavel, Separation of Specific Single-Enantiomer Single-Wall Carbon Nanotubes in the Large-Diameter Regime, ACS Nano. 14 (2019) 948–963. https://doi.org/10.1021/ACSNANO.9B08244.

[32] C.M. Sims, J.A. Fagan, Near-infrared fluorescence as a method for determining single-wall carbon nanotube extraction conditions in aqueous two polymer phase extraction, Carbon. 165 (2020) 196–203. https://doi.org/10.1016/J.CARBON.2020.04.044.

[33] J.A. Fagan, Aqueous two-polymer phase extraction of single-wall carbon nanotubes using surfactants, Nanoscale Advances. 1 (2019) 3307–3324. https://doi.org/10.1039/C9NA00280D.

[34] N.K. Subbaiyan, S. Cambré, A.N.G. Parra-Vasquez, E.H. Hároz, S.K. Doorn, J.G. Duque, Role of surfactants and salt in aqueous two-phase separation of carbon nanotubes toward simple chirality isolation, ACS Nano. 8 (2014) 1619–1628. https://doi.org/10.1021/nn405934y.

[35] G. Ao, J.K. Streit, J.A. Fagan, M. Zheng, Differentiating Left- and Right-Handed Carbon Nanotubes by DNA, Journal of the American Chemical Society. 138 (2016) 16677–16685. https://doi.org/10.1021/jacs.6b09135.

[36] E. Turek, T. Shiraki, T. Shiraishi, T. Shiga, T. Fujigaya, D. Janas, Single-step isolation of carbon nanotubes with narrow-band light emission characteristics, Scientific Reports 2019 9:1. 9 (2019) 1–8. https://doi.org/10.1038/s41598-018-37675-4.

[37] B. Podlesny, T. Shiraki, D. Janas, One-step sorting of single-walled carbon nanotubes using aqueous two-phase extraction in the presence of basic salts, Scientific Reports 2020 10:1. 10 (2020) 1–9. https://doi.org/10.1038/s41598-020-66264-7.

[38] N.K. Subbaiyan, A.N.G. Parra-Vasquez, S. Cambré, M.A.S.S. Cordoba, S.E. Yalcin, C.E. Hamilton, N.H. Mack, J.L. Blackburn, S.K. Doorn, J.G. Duque, Bench-top aqueous two-phase extraction of isolated individual single-walled carbon nanotubes, Nano Research. 8 (2015) 1755–1769. https://doi.org/10.1007/s12274-014-0680-z.

[39] W. Wenseleers, I.I. Vlasov, E. Goovaerts, E.D. Obraztsova, A.S. Lobach, A. Bouwen, Efficient Isolation and Solubilization of Pristine Single-Walled Nanotubes in Bile Salt

Micelles, Advanced Functional Materials. 14 (2004) 1105–1112. https://doi.org/10.1002/adfm.200400130.

[40] R. Ninomiya, K. Matsuoka, Y. Moroi, Micelle formation of sodium chenodeoxycholate and solubilization into the micelles: comparison with other unconjugated bile salts, Biochimica et Biophysica Acta (BBA) - Molecular and Cell Biology of Lipids. 1634 (2003) 116–125. https://doi.org/10.1016/J.BBALIP.2003.09.003.

[41] A.L. Grilo, M. Raquel Aires-Barros, A.M. Azevedo, Partitioning in Aqueous Two-Phase Systems: Fundamentals, Applications and Trends, Separation & Purification Reviews. 45 (2016) 68–80. https://doi.org/10.1080/15422119.2014.983128.

[42] C.M. Sims, J.A. Fagan, Surfactant chemistry and polymer choice affect single-wall carbon nanotube extraction conditions in aqueous two-polymer phase extraction, Carbon. 191 (2022) 215–226. https://doi.org/10.1016/J.CARBON.2022.01.062.

[43] B. Podlesny, B. Olszewska, Z. Yaari, P. v. Jena, G. Ghahramani, R. Feiner, D.A. Heller, D. Janas, En route to single-step, two-phase purification of carbon nanotubes facilitated by high-throughput spectroscopy, Scientific Reports. 11 (2021). https://doi.org/10.1038/s41598-021-89839-4.

[44] S. Cambré, P. Muyshondt, R. Federicci, W. Wenseleers, Chirality-dependent densities of carbon nanotubes by in situ 2D fluorescence-excitation and Raman characterisation in a density gradient after ultracentrifugation, Nanoscale. 7 (2015) 20015–20024. https://doi.org/10.1039/C5NR06020F.

[45] S. Cambré, B. Schoeters, S. Luyckx, E. Goovaerts, W. Wenseleers, Experimental Observation of Single-File Water Filling of Thin Single-Wall Carbon Nanotubes Down to Chiral Index (5,3), Physical Review Letters. 104 (2010) 207401. https://doi.org/10.1103/PhysRevLett.104.207401.

[46] W. Wenseleers, S. Cambré, J. Čulin, A. Bouwen, E. Goovaerts, Effect of Water Filling on the Electronic and Vibrational Resonances of Carbon Nanotubes: Characterizing Tube Opening by Raman Spectroscopy, Advanced Materials. 19 (2007) 2274–2278. https://doi.org/10.1002/adma.200700773.

[47] J. Campo, S. Cambré, B. Botka, J. Obrzut, W. Wenseleers, J.A. Fagan, Optical Property Tuning of Single-Wall Carbon Nanotubes by Endohedral Encapsulation of a Wide Variety of Dielectric Molecules, ACS Nano. 15 (2020) 2301–2317. https://doi.org/10.1021/ACSNANO.0C08352.

[48] S. Cambré, J. Campo, C. Beirnaert, C. Verlackt, P. Cool, W. Wenseleers, Asymmetric dyes align inside carbon nanotubes to yield a large nonlinear optical response, Nature Nanotechnology. 10 (2015) 248–252. https://doi.org/10.1038/nnano.2015.1.

[49] A. Castan, S. Forel, F. Fossard, J. Defillet, A. Ghedjatti, D. Levshov, W. Wenseleers, S. Cambré, A. Loiseau, Assessing the reliability of the Raman peak counting method for the

characterization of SWCNT diameter distributions: a cross characterization with TEM, Carbon. 171 (2021) 968–979. https://doi.org/10.1016/J.CARBON.2020.09.012.

[50] S. van Bezouw, D.H. Arias, R. Ihly, S. Cambré, A.J. Ferguson, J. Campo, J.C. Johnson, J. Defillet, W. Wenseleers, J.L. Blackburn, Diameter-Dependent Optical Absorption and Excitation Energy Transfer from Encapsulated Dye Molecules toward Single-Walled Carbon Nanotubes, ACS Nano. 12 (2018) 6881–6894. https://doi.org/10.1021/ACSNANO.8B02213.

[51] James J. P. Stewart, MOPAC2016, (2016). http://openmopac.net/.

[52] HyperChem(TM) Professional 7.52, (n.d.).

[53] P. Kékicheff, C. Grabielle-Madelmont, M. Ollivon, Phase diagram of sodium dodecyl sulfate-water system: 1. A calorimetric study, Journal of Colloid and Interface Science. 131 (1989) 112–132. https://doi.org/10.1016/0021-9797(89)90151-3.

[54] S. Cambré, W. Wenseleers, Separation and Diameter-Sorting of Empty (End-Capped) and Water-Filled (Open) Carbon Nanotubes by Density Gradient Ultracentrifugation, Angew. Chem. Int. Ed. 50 (2011) 2764. https://doi.org/DOI 10.1002/anie.201007324.

[55] R.M. Jain, M. Ben-Naim, M.P. Landry, M.S. Strano, Competitive Binding in Mixed Surfactant Systems for Single-Walled Carbon Nanotube Separation, Journal of Physical Chemistry C. 119 (2015) 22737–22745. https://doi.org/10.1021/ACS.JPCC.5B07947.

[56] D. Yang, L. Li, X. Wei, Y. Wang, W. Zhou, H. Kataura, S. Xie, H. Liu, Submilligram-scale separation of near-zigzag single-chirality carbon nanotubes by temperature controlling a binary surfactant system, Science Advances. 7 (2021) 84–101. https://doi.org/10.1126/SCIADV.ABE0084.

[57] J.K. Streit, J.A. Fagan, M. Zheng, A Low Energy Route to DNA-Wrapped Carbon Nanotubes via Replacement of Bile Salt Surfactants, Analytical Chemistry. 89 (2017) 10496–10503. https://doi.org/10.1021/ACS.ANALCHEM.7B02637.

[58] P. Poulin, B. Vigolo, P. Launois, Films and fibers of oriented single wall nanotubes, Carbon. 40 (2002) 1741–1749. https://doi.org/10.1016/S0008-6223(02)00042-8.

[59] X. Zeng, D. Yang, H. Liu, N. Zhou, Y. Wang, W. Zhou, S. Xie, H. Kataura, Detecting and Tuning the Interactions between Surfactants and Carbon Nanotubes for Their High-Efficiency Structure Separation, Advanced Materials Interfaces. 5 (2018) 1700727. https://doi.org/10.1002/ADMI.201700727.

[60] A. Hirano, T. Tanaka, Y. Urabe, H. Kataura, Purification of Single-Wall Carbon Nanotubes by Controlling the Adsorbability onto Agarose Gels Using Deoxycholate, Journal of Physical Chemistry C. 116 (2012) 9816–9823. https://doi.org/10.1021/JP301380S.

[61] M. Park, J. Park, J. Lee, S.Y. Ju, Scaling of binding affinities and cooperativities of surfactants on carbon nanotubes, Carbon. 139 (2018) 427–436. https://doi.org/10.1016/J.CARBON.2018.07.003.

[62] W.L. Jorgensen, D.S. Maxwell, J. Tirado-Rives, Development and testing of the OPLS all-atom force field on conformational energetics and properties of organic liquids, Journal of the American Chemical Society. 118 (1996) 11225–11236. https://doi.org/10.1021/JA9621760/SUPPL_FILE/JA11225.PDF.

[63] W.L. Jorgensen, J. Tirado-Rives, The OPLS [optimized potentials for liquid simulations] potential functions for proteins, energy minimizations for crystals of cyclic peptides and crambin, Journal of the American Chemical Society. 110 (1988) 1657–1666. https://doi.org/10.1021/JA00214A001.

[64] M.J.S. Dewar, E.G. Zoebisch, E.F. Healy, J.J.P. Stewart, Development and use of quantum mechanical molecular models. 76. AM1: a new general purpose quantum mechanical molecular model, Journal of the American Chemical Society. 107 (2002) 3902–3909. https://doi.org/10.1021/JA00299A024.

[65] S. Cambré, W. Wenseleers, Separation and Diameter-Sorting of Empty (End-Capped) and Water-Filled (Open) Carbon Nanotubes by Density Gradient Ultracentrifugation, Angewandte Chemie International Edition. 50 (2011) 2764–2768. https://doi.org/10.1002/anie.201007324.

[66] S. Cambré, P. Muyshondt, R. Federicci, W. Wenseleers, Chirality-dependent densities of carbon nanotubes by in situ 2D fluorescence-excitation and Raman characterisation in a density gradient after ultracentrifugation, Nanoscale. 7 (2015) 20015–20024. https://doi.org/10.1039/c5nr06020f.

[67] H. Liu, D. Nishide, T. Tanaka, H. Kataura, Large-scale single-chirality separation of single-wall carbon nanotubes by simple gel chromatography, Nature Communications 2011 2:1. 2 (2011) 1–8. https://doi.org/10.1038/ncomms1313.

# Supporting Information for:

# The role of bile salt surfactants in aqueous two-phase separation of single-wall carbon nanotubes revealed by systematic parameter variations

*Joeri Defillet[1, †], Marina Avramenko[1,†], Miles Martinati[1], Miguel Ángel López Carrillo[1],*

*Domien Van der Elst[1], Wim Wenseleers[1] and Sofie Cambré[1]*

[1]Nanostructured and Organic Optical and Electronic Materials, Physics Department, University of Antwerp, Belgium

[†] These authors contributed equally to this work

**Table of Contents:**

## Section 1. Examples of systematic parameter variations

**Table S1.** Quantities of stock solutions used for the variation of the SDS concentration at a fixed DOC concentration. Stock solutions were prepared with following concentrations: 20%wt/V PEG, 20% wt/V dextran, 10% wt/V SDS, 20% wt/V SDS, all in $D_2O$. The SWCNT solution had a starting DOC concentration of 1.6%wt/V, as verified by absorption spectroscopy. The total volume is kept constant, assuming no change upon mixing.

| | **PEG 20% (µL)** | **Dextran 20% (µL)** | **SWCNT (1.6% DOC) (µL)** | **$D_2O$ (µL)** | **SDS 10% (µL)** | **SDS 20% (µL)** | **total volume (µL)** | **DOC conc (% wt/V)** | **SDS conc (% wt/V)** | **SDS/DOC ratio** |
|---|---|---|---|---|---|---|---|---|---|---|
| 1 | 700 | 300 | 38 | 162 | 0 | 0 | 1200 | 0.0507 | 0 | 0 |
| 2 | 700 | 300 | 38 | 150 | 12 | 0 | 1200 | 0.0507 | 0.1 | 1.97 |
| 3 | 700 | 300 | 38 | 138 | 24 | 0 | 1200 | 0.0507 | 0.2 | 3.95 |
| 4 | 700 | 300 | 38 | 126 | 36 | 0 | 1200 | 0.0507 | 0.3 | 5.92 |
| 5 | 700 | 300 | 38 | 114 | 48 | 0 | 1200 | 0.0507 | 0.4 | 7.89 |
| 6 | 700 | 300 | 38 | 102 | 60 | 0 | 1200 | 0.0507 | 0.5 | 9.87 |
| 7 | 700 | 300 | 38 | 90 | 72 | 0 | 1200 | 0.0507 | 0.6 | 11.84 |
| 8 | 700 | 300 | 38 | 78 | 84 | 0 | 1200 | 0.0507 | 0.7 | 13.82 |
| 9 | 700 | 300 | 38 | 66 | 96 | 0 | 1200 | 0.0507 | 0.8 | 15.79 |
| 10 | 700 | 300 | 38 | 54 | 108 | 0 | 1200 | 0.0507 | 0.9 | 17.76 |
| 11 | 700 | 300 | 38 | 42 | 120 | 0 | 1200 | 0.0507 | 1.0 | 19.74 |
| 12 | 700 | 300 | 38 | 30 | 132 | 0 | 1200 | 0.0507 | 1.1 | 21.71 |
| 13 | 700 | 300 | 38 | 90 | 0 | 72 | 1200 | 0.0507 | 1.2 | 23.68 |
| 14 | 700 | 300 | 38 | 84 | 0 | 78 | 1200 | 0.0507 | 1.3 | 25.66 |
| 15 | 700 | 300 | 38 | 78 | 0 | 84 | 1200 | 0.0507 | 1.4 | 27.63 |
| 16 | 700 | 300 | 38 | 72 | 0 | 90 | 1200 | 0.0507 | 1.5 | 29.61 |
| 17 | 700 | 300 | 38 | 66 | 0 | 96 | 1200 | 0.0507 | 1.6 | 31.58 |
| 18 | 700 | 300 | 38 | 60 | 0 | 102 | 1200 | 0.0507 | 1.7 | 33.55 |
| 19 | 700 | 300 | 38 | 54 | 0 | 108 | 1200 | 0.0507 | 1.8 | 35.53 |
| 20 | 700 | 300 | 38 | 48 | 0 | 114 | 1200 | 0.0507 | 1.9 | 37.5 |
| 21 | 700 | 300 | 38 | 42 | 0 | 120 | 1200 | 0.0507 | 2.0 | 39.47 |
| 22 | 700 | 300 | 38 | 36 | 0 | 126 | 1200 | 0.0507 | 2.1 | 41.45 |

**Table S2.** Quantities of stock solutions used for the variation of the DOC concentration without any cosurfactants. Stock solutions were prepared with following concentrations: 20%wt/V PEG, 20% wt/V dextran, 1% wt/V DOC, 5% wt/V DOC, all in $D_2O$. The SWCNT solution had a starting DOC concentration of 1%wt/V, as verified by absorption spectroscopy. The total volume is kept constant, assuming no change upon mixing.

| | PEG 20% (µL) | Dextran 20% (µL) | SWCNT (1% DOC) (µL) | $D_2O$ (µL) | DOC 1% (µL) | DOC 5% (µL) | total volume (µL) | DOC conc. (% wt/V) |
|---|---|---|---|---|---|---|---|---|
| **1** | 700 | 300 | 10 | 190 | 0 | 0 | 1200 | 0.00833 |
| **2** | 700 | 300 | 15 | 185 | 0 | 0 | 1200 | 0.0125 |
| **3** | 700 | 300 | 20 | 180 | 0 | 0 | 1200 | 0.01667 |
| **4** | 700 | 300 | 25 | 175 | 0 | 0 | 1200 | 0.0208 |
| **5** | 700 | 300 | 30 | 170 | 0 | 0 | 1200 | 0.025 |
| **6** | 700 | 300 | 30 | 160 | 10 | 0 | 1200 | 0.0333 |
| **7** | 700 | 300 | 30 | 150 | 20 | 0 | 1200 | 0.0417 |
| **8** | 700 | 300 | 30 | 140 | 30 | 0 | 1200 | 0.05 |
| **9** | 700 | 300 | 30 | 130 | 40 | 0 | 1200 | 0.0583 |
| **10** | 700 | 300 | 30 | 120 | 50 | 0 | 1200 | 0.0667 |
| **11** | 700 | 300 | 30 | 110 | 60 | 0 | 1200 | 0.075 |
| **12** | 700 | 300 | 30 | 100 | 70 | 0 | 1200 | 0.0833 |
| **13** | 700 | 300 | 30 | 90 | 80 | 0 | 1200 | 0.0917 |
| **14** | 700 | 300 | 30 | 80 | 90 | 0 | 1200 | 0.1 |
| **15** | 700 | 300 | 30 | 50 | 120 | 0 | 1200 | 0.125 |
| **16** | 700 | 300 | 30 | 30 | 140 | 0 | 1200 | 0.142 |
| **17** | 700 | 300 | 30 | 10 | 160 | 0 | 1200 | 0.158 |
| **18** | 700 | 300 | 30 | 35 | 120 | 15 | 1200 | 0.1875 |
| **19** | 700 | 300 | 30 | 20 | 120 | 30 | 1200 | 0.25 |

**Table S3.** Quantities of stock solutions used for the variation of the SDBS concentration at a fixed DOC concentration of 0.05%. Stock solutions were prepared with following concentrations: 20%wt/V PEG, 20% wt/V dextran, 1, 5 and 10% wt/V SDBS, 1% wt/V DOC, all in $D_2O$. The SWCNT solution had a starting DOC concentration of 1%wt/V, as verified by absorption spectroscopy. The total volume is kept constant, assuming no change upon mixing.

| | **PEG 20% (µL)** | **Dextran 20% (µL)** | **SWCNT (1% DOC) (µL)** | **DOC 1% (µL)** | **SDBS 1% (µL)** | **SDBS 5% (µL)** | **SDBS 10% (µL)** | **$D_2O$ (µL)** | **total volume (µL)** | **DOC conc. (% wt/V)** | **SDBS conc. (% wt/V)** |
|---|---|---|---|---|---|---|---|---|---|---|---|
| **1** | 700 | 300 | 35 | 25 | 0 | 0 | 0 | 140 | 1200 | 0.05 | 0 |
| **2** | 700 | 300 | 35 | 25 | 15 | 0 | 0 | 125 | 1200 | 0.05 | 0.0125 |
| **3** | 700 | 300 | 35 | 25 | 30 | 0 | 0 | 110 | 1200 | 0.05 | 0.025 |
| **4** | 700 | 300 | 35 | 25 | 45 | 0 | 0 | 95 | 1200 | 0.05 | 0.0375 |
| **5** | 700 | 300 | 35 | 25 | 60 | 0 | 0 | 80 | 1200 | 0.05 | 0.05 |
| **6** | 700 | 300 | 35 | 25 | 75 | 0 | 0 | 65 | 1200 | 0.05 | 0.0625 |
| **7** | 700 | 300 | 35 | 25 | 90 | 0 | 0 | 50 | 1200 | 0.05 | 0.075 |
| **8** | 700 | 300 | 35 | 25 | 105 | 0 | 0 | 35 | 1200 | 0.05 | 0.0875 |
| **9** | 700 | 300 | 35 | 25 | 0 | 24 | 0 | 116 | 1200 | 0.05 | 0.1 |
| **10** | 700 | 300 | 35 | 25 | 0 | 36 | 0 | 104 | 1200 | 0.05 | 0.15 |
| **11** | 700 | 300 | 35 | 25 | 0 | 48 | 0 | 92 | 1200 | 0.05 | 0.2 |
| **12** | 700 | 300 | 35 | 25 | 0 | 60 | 0 | 80 | 1200 | 0.05 | 0.25 |
| **13** | 700 | 300 | 35 | 25 | 0 | 0 | 36 | 104 | 1200 | 0.05 | 0.3 |
| **14** | 700 | 300 | 35 | 25 | 0 | 0 | 42 | 98 | 1200 | 0.05 | 0.35 |
| **15** | 700 | 300 | 35 | 25 | 0 | 0 | 48 | 92 | 1200 | 0.05 | 0.4 |
| **16** | 700 | 300 | 35 | 25 | 0 | 0 | 54 | 86 | 1200 | 0.05 | 0.45 |
| **17** | 700 | 300 | 35 | 25 | 0 | 0 | 60 | 80 | 1200 | 0.05 | 0.5 |
| **18** | 700 | 300 | 35 | 25 | 0 | 0 | 66 | 74 | 1200 | 0.05 | 0.55 |
| **19** | 700 | 300 | 35 | 25 | 0 | 0 | 72 | 68 | 1200 | 0.05 | 0.6 |
| **20** | 700 | 300 | 35 | 25 | 0 | 0 | 78 | 62 | 1200 | 0.05 | 0.65 |
| **21** | 700 | 300 | 35 | 25 | 0 | 0 | 90 | 50 | 1200 | 0.05 | 0.75 |
| **22** | 700 | 300 | 35 | 25 | 0 | 0 | 102 | 38 | 1200 | 0.05 | 0.85 |

**Table S4.** Quantities of stock solutions used for the variation of the SDBS concentration at a fixed DOC concentration of 0.1%. Stock solutions were prepared with following concentrations: 20%wt/V PEG, 20% wt/V dextran, 10, 15 and 20% wt/V SDBS, 5% wt/V DOC, all in $D_2O$. The SWCNT solution had a starting DOC concentration of 1%wt/V, as verified by absorption spectroscopy. The total volume is kept constant, assuming no change upon mixing.

| | PEG 20% (µL) | Dextran 20% (µL) | SWCNT (1% DOC) (µL) | DOC 5% (µL) | SDBS 10% (µL) | SDBS 15% (µL) | SDBS 20% (µL) | $D_2O$ (µL) | total volume (µL) | DOC conc. (% wt/V) | SDBS conc. (% wt/V) |
|---|---|---|---|---|---|---|---|---|---|---|---|
| **1** | 700 | 300 | 35 | 17 | 0 | 0 | 0 | 148 | 1200 | 0.1 | 0 |
| **2** | 700 | 300 | 35 | 17 | 12 | 0 | 0 | 136 | 1200 | 0.1 | 0.1 |
| **3** | 700 | 300 | 35 | 17 | 24 | 0 | 0 | 124 | 1200 | 0.1 | 0.2 |
| **4** | 700 | 300 | 35 | 17 | 36 | 0 | 0 | 112 | 1200 | 0.1 | 0.3 |
| **5** | 700 | 300 | 35 | 17 | 48 | 0 | 0 | 100 | 1200 | 0.1 | 0.4 |
| **6** | 700 | 300 | 35 | 17 | 60 | 0 | 0 | 88 | 1200 | 0.1 | 0.5 |
| **7** | 700 | 300 | 35 | 17 | 72 | 0 | 0 | 76 | 1200 | 0.1 | 0.6 |
| **8** | 700 | 300 | 35 | 17 | 84 | 0 | 0 | 64 | 1200 | 0.1 | 0.7 |
| **9** | 700 | 300 | 35 | 17 | 96 | 0 | 0 | 52 | 1200 | 0.1 | 0.8 |
| **10** | 700 | 300 | 35 | 17 | 108 | 0 | 0 | 40 | 1200 | 0.1 | 0.9 |
| **11** | 700 | 300 | 35 | 17 | 120 | 0 | 0 | 28 | 1200 | 0.1 | 1 |
| **12** | 700 | 300 | 35 | 17 | 132 | 0 | 0 | 16 | 1200 | 0.1 | 1.1 |
| **13** | 700 | 300 | 35 | 17 | 0 | 96 | 0 | 52 | 1200 | 0.1 | 1.2 |
| **14** | 700 | 300 | 35 | 17 | 0 | 104 | 0 | 44 | 1200 | 0.1 | 1.3 |
| **15** | 700 | 300 | 35 | 17 | 0 | 112 | 0 | 36 | 1200 | 0.1 | 1.4 |
| **16** | 700 | 300 | 35 | 17 | 0 | 120 | 0 | 28 | 1200 | 0.1 | 1.5 |
| **17** | 700 | 300 | 35 | 17 | 0 | 128 | 0 | 20 | 1200 | 0.1 | 1.6 |
| **18** | 700 | 300 | 35 | 17 | 0 | 136 | 0 | 12 | 1200 | 0.1 | 1.7 |
| **19** | 700 | 300 | 35 | 17 | 0 | 0 | 108 | 40 | 1200 | 0.1 | 1.8 |
| **20** | 700 | 300 | 35 | 17 | 0 | 0 | 114 | 34 | 1200 | 0.1 | 1.9 |
| **21** | 700 | 300 | 35 | 17 | 0 | 0 | 120 | 28 | 1200 | 0.1 | 2 |
| **22** | 700 | 300 | 35 | 17 | 0 | 0 | 126 | 22 | 1200 | 0.1 | 2.1 |
| **23** | 700 | 300 | 35 | 17 | 0 | 0 | 132 | 16 | 1200 | 0.1 | 2.2 |
| **24** | 700 | 300 | 35 | 17 | 0 | 0 | 138 | 10 | 1200 | 0.1 | 2.3 |
| **25** | 700 | 300 | 35 | 17 | 0 | 0 | 148 | 0 | 1200 | 0.1 | 2.47 |

## Section 2. Measuring the composition of the phases by absorption spectroscopy

To assess the composition of each phase, the absorption spectra of the individual components, *i.e.* PEG, dextran, DOC and SDS or SDBS in $D_2O$, were measured separately with $D_2O$ as a baseline. Each of these components have characteristic absorption peaks in the IR range of the spectrum, allowing to obtain the relative concentration of these components by manually fitting the long wavelength part of the spectra with a linear combination of individual component spectra. To this end, the linear combination is subtracted from the mixed phase spectrum, and the coefficients (proportional to the concentrations) of the linear combination are manually adjusted until a flat difference spectrum is obtained. The coefficients are optimized manually because the noise around 2000nm due to the strong $D_2O$ absorption band complicates an automatic fitting procedure.

As an example, Figure S1 shows such a subtraction. These fits were used for two purposes: (I) monitoring that the phase separation does not change (*i.e.* PEG and dextran concentrations in top and bottom phase) by adding *e.g.* more surfactants so that the absolute intensities of SWCNTs can be directly compared between different samples and (II) monitoring the re-distribution of surfactants in top and bottom phase, after phase separation since this does not occur evenly between the phases and is important for the second separation step. From this, we noticed that in the bottom phase, typically much less SDS is present as assumed from equal surfactant distribution among both phases, up to a factor of 6 difference.

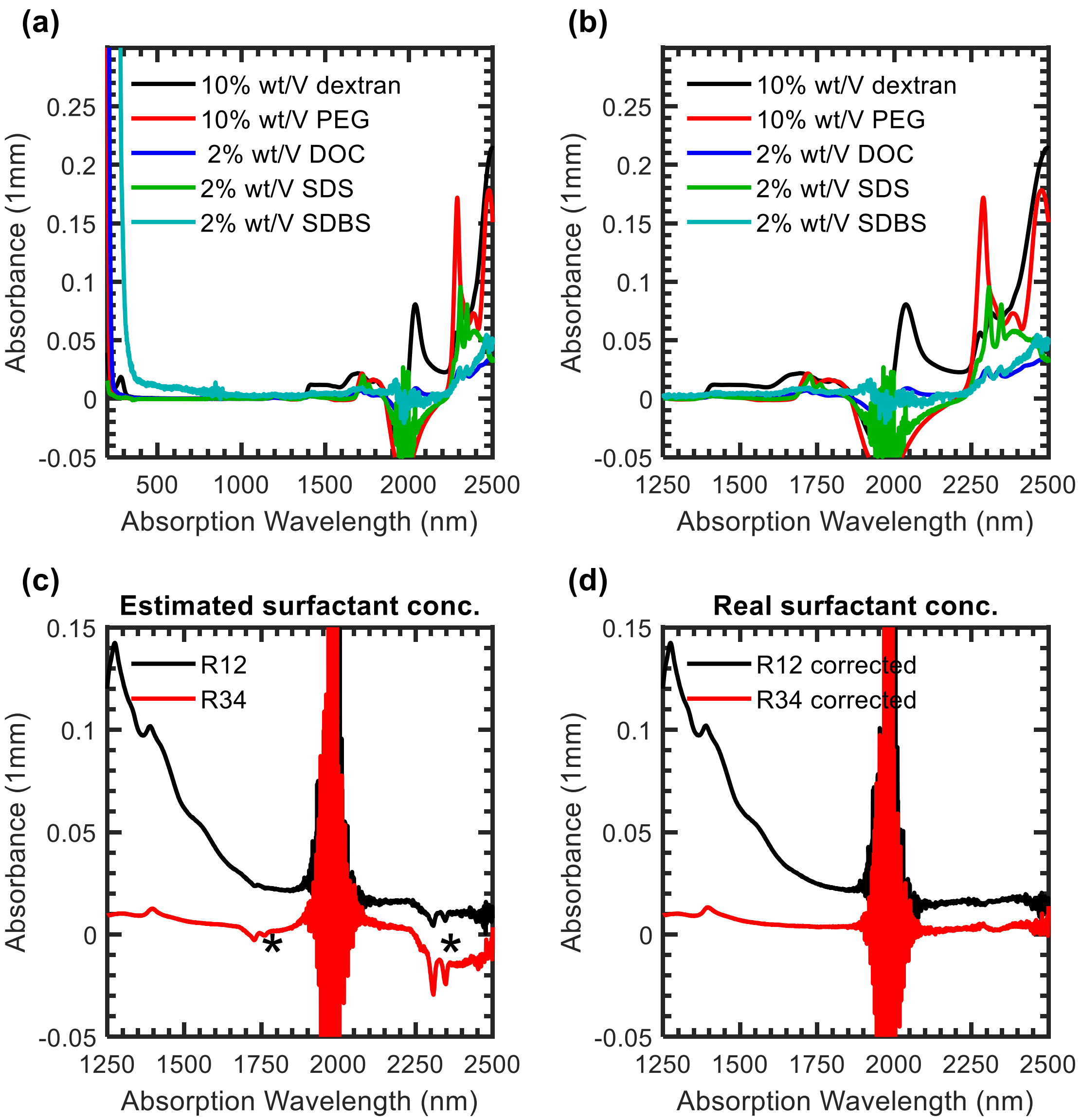

Figure S1: (a) Absorption spectra of the different components (10% wt/V dextran, 10% wt/V PEG, 2% wt/V DOC, 2% wt/V SDS and 2% wt/V SDBS in $D_2O$) measured in a quartz cell with optical path length of 1 mm. These spectra are used as basis spectra for the absorption spectra of the bottom and top phases. (b) The same spectra as in panel (1) but zoomed in on the IR wavelength range to show the distinct absorption peaks for the different components in this range, allowing to assess their specific concentrations. (c) absorption spectra of the bottom phases of the sample with SDS/DOC ratio 11.84 (black) and 33.55 (red), after subtracting a linear combination of the basis spectra, assuming equal distribution of surfactants across both phases. This clearly shows that there is too much SDS subtracted (indicated by the * in the Figure). (d) similar as in (c) but with corrected SDS surfactant concentration. For R12, instead of 0.6 % wt/V SDS, only 0.1 % wt/V SDS was subtracted, for R34, instead of 1.7% wt/V SDS, only 0.36% wt/V SDS was subtracted. Since prior to the absorption measurement, the samples are diluted by a factor of 2 with 4% DOC/$D_2O$, the exact concentration of DOC is difficult to obtain, and an equal distribution is assumed also in panel (c).

**Section 3. PLE maps of bottom and top phases for the SDS/DOC variation**

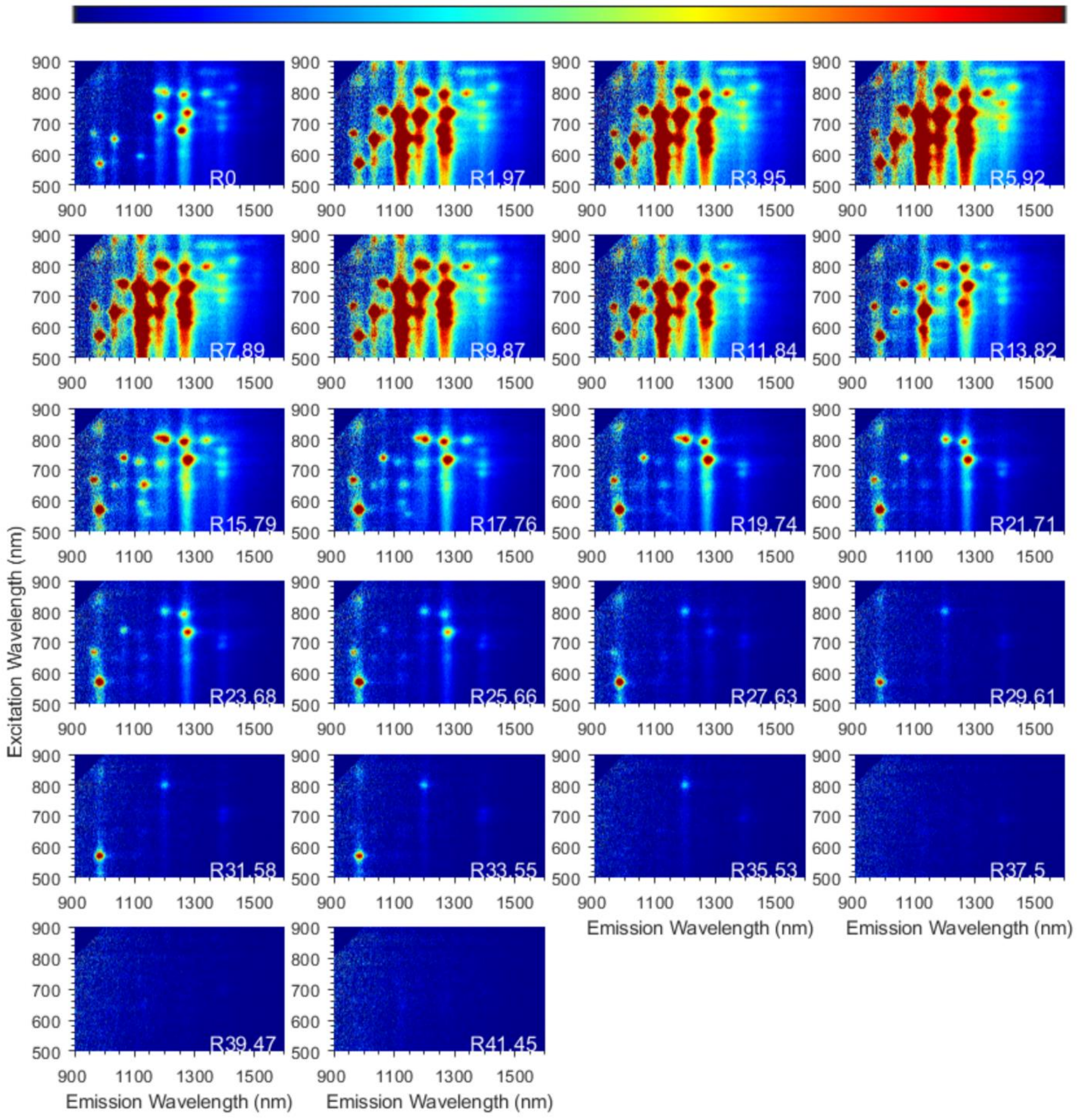

Figure S2: Overview of PLE maps of the bottom phases for the SDS/DOC variation. Intensities can be directly compared and SDS/DOC ratios (R#) are plotted in white on the corresponding PLE maps.

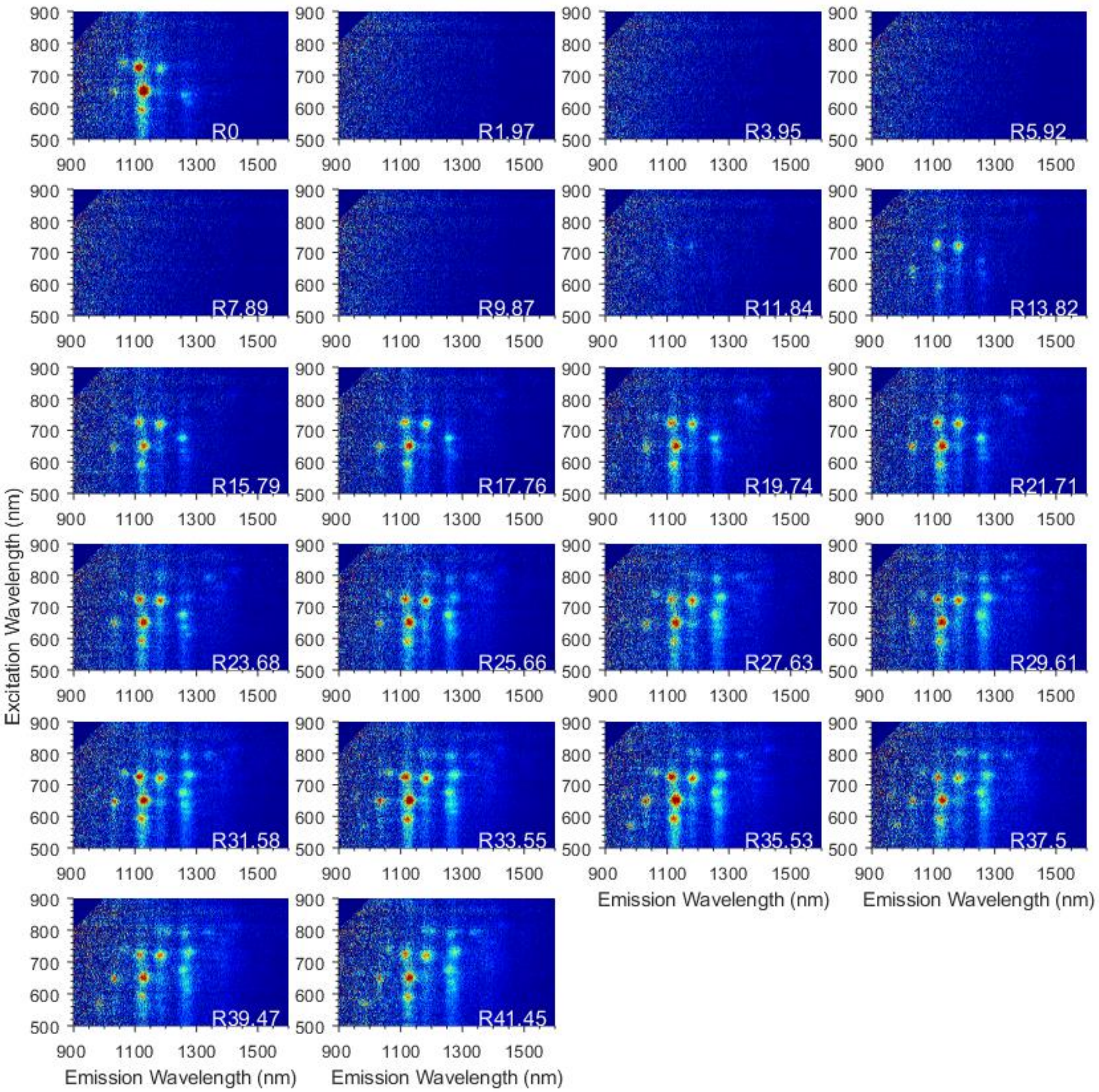

Figure S3: Overview of PLE maps of the top phases from the SDS/DOC variation. Intensities can be directly compared and SDS/DOC ratios (R#) are plotted in white on the corresponding PLE maps. The overall intensities in these PLE maps are smaller than for the bottom phases, due to the fact that the top phase has a much larger volume than the bottom phase after separation. Color coding is the same as in Figure S2.

**Section 4. Simultaneous 2D PLE fits to obtain the PL intensities for each chirality (fit examples)**

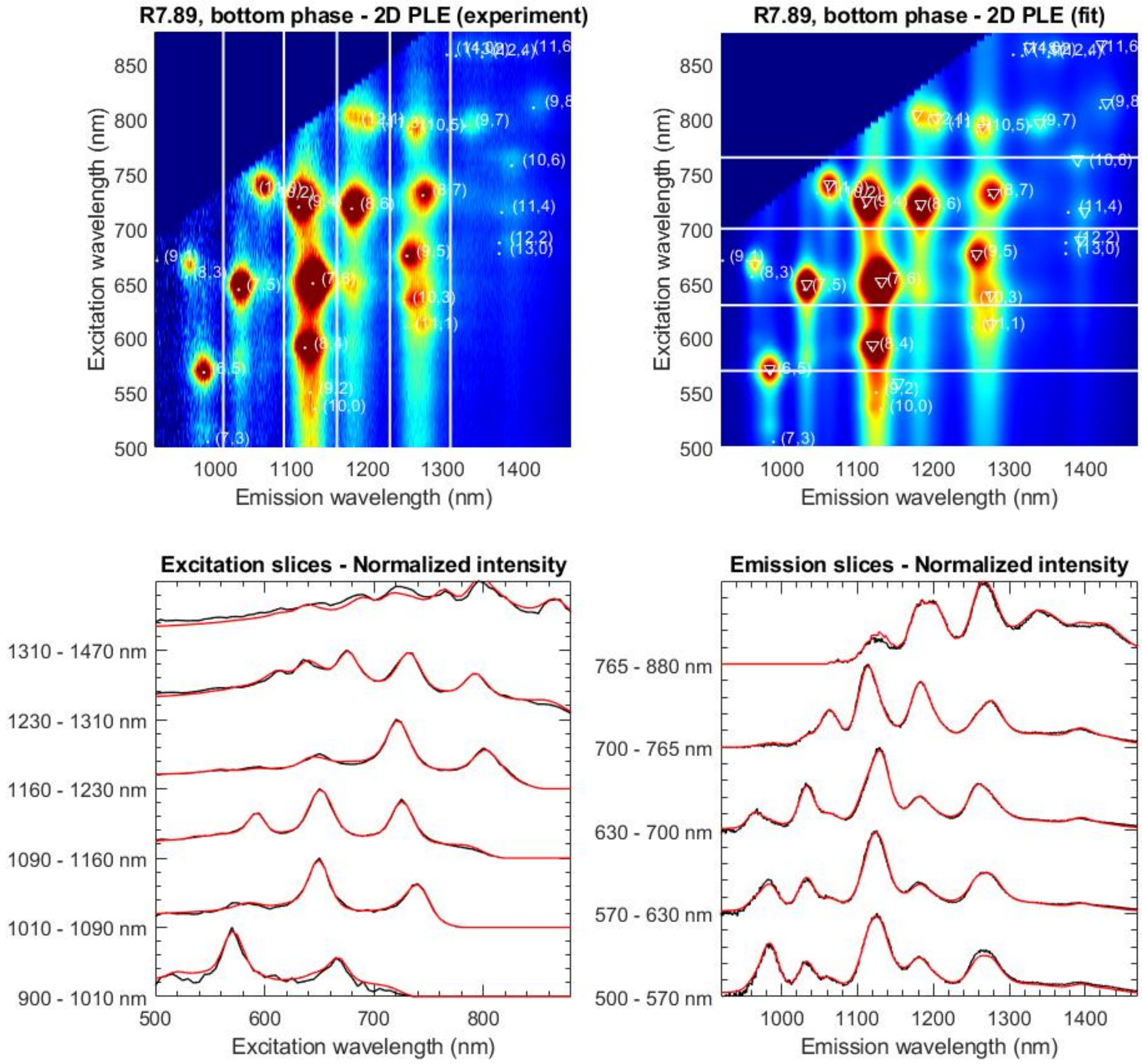

Figure S4a: 2D PLE experimental data and fits for the bottom phase of SDS/DOC ratio R7.89 at a fixed concentration of 0.0507% DOC. The top left panel represents the experimental data, while the top right panel represents the 2D fit of the experimental data. The bottom two panels then represent excitation and emission slices of the experimental data (black) and fits (red), obtained by integrating over specific emission or excitation ranges, respectively, as indicated in the panels and also highlighted in the PLE maps by the white lines.

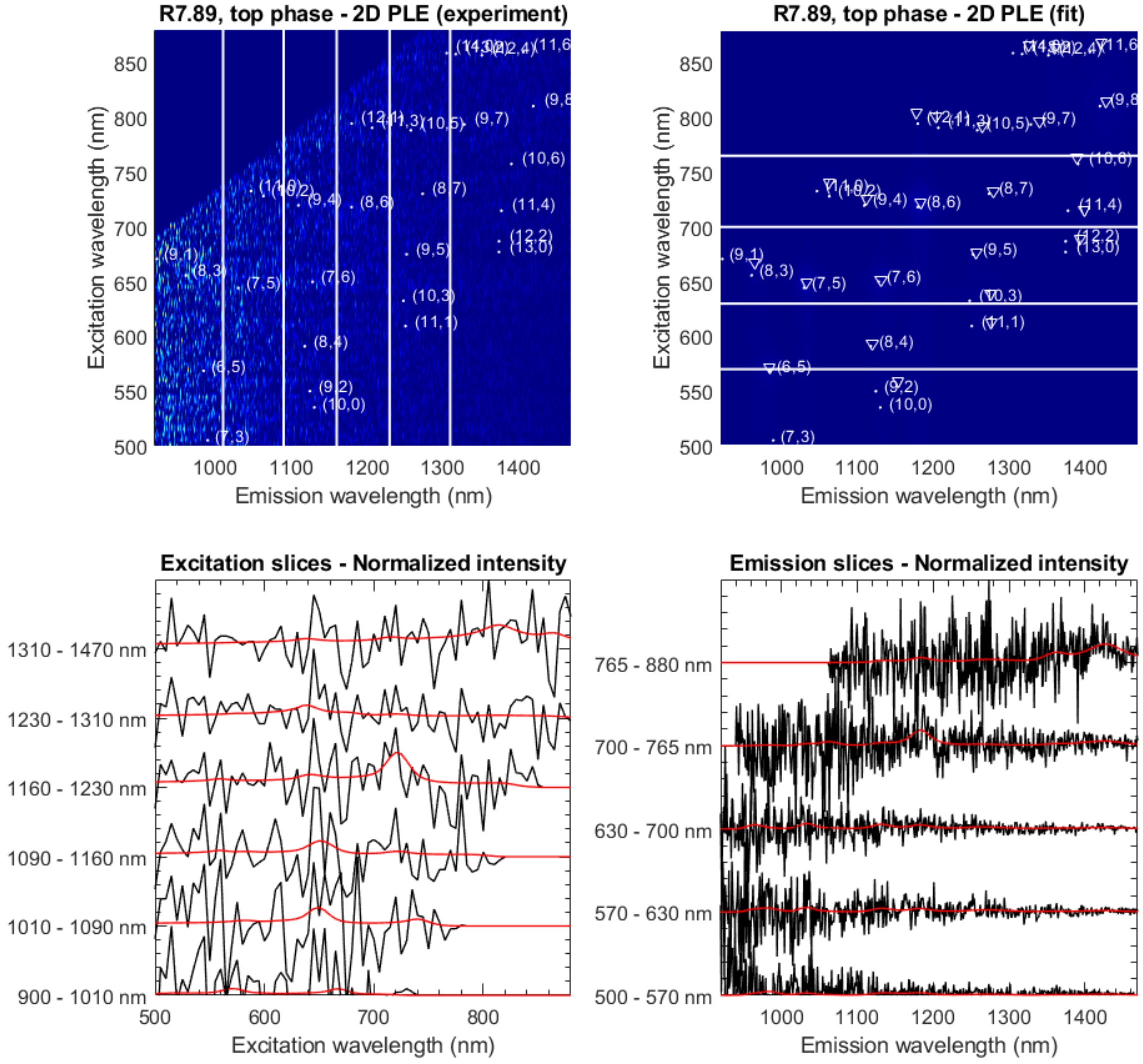

Figure S4b: 2D PLE experimental data and fits for the top phase of SDS/DOC ratio R7.89 at a fixed concentration of 0.0507% DOC. (same color coding as in Figure S4a).

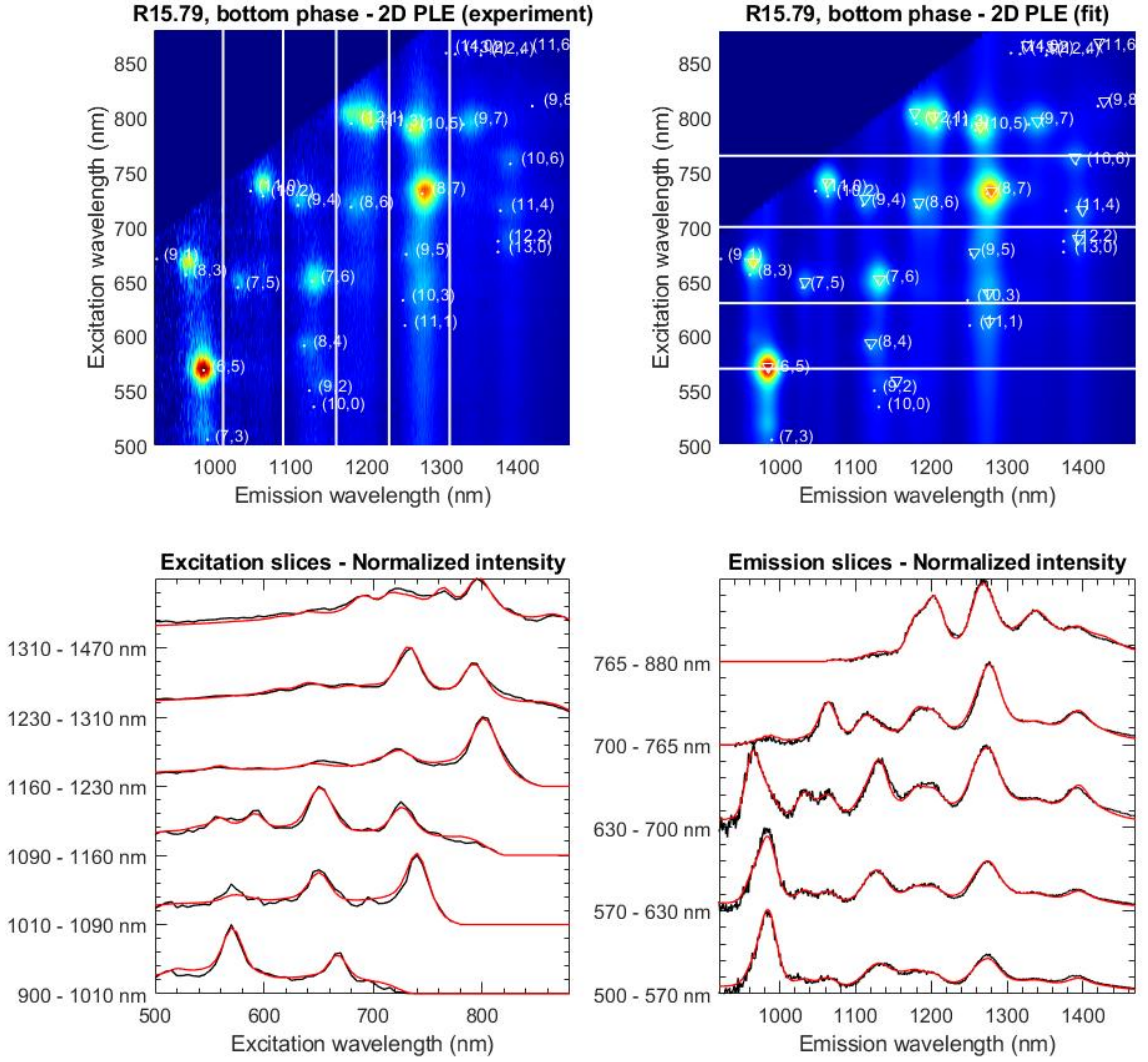

Figure S4c: 2D PLE experimental data and fits for the bottom phase of SDS/DOC ratio R15.79 at a fixed concentration of 0.0507% DOC. (same color coding as in Figure S4a).

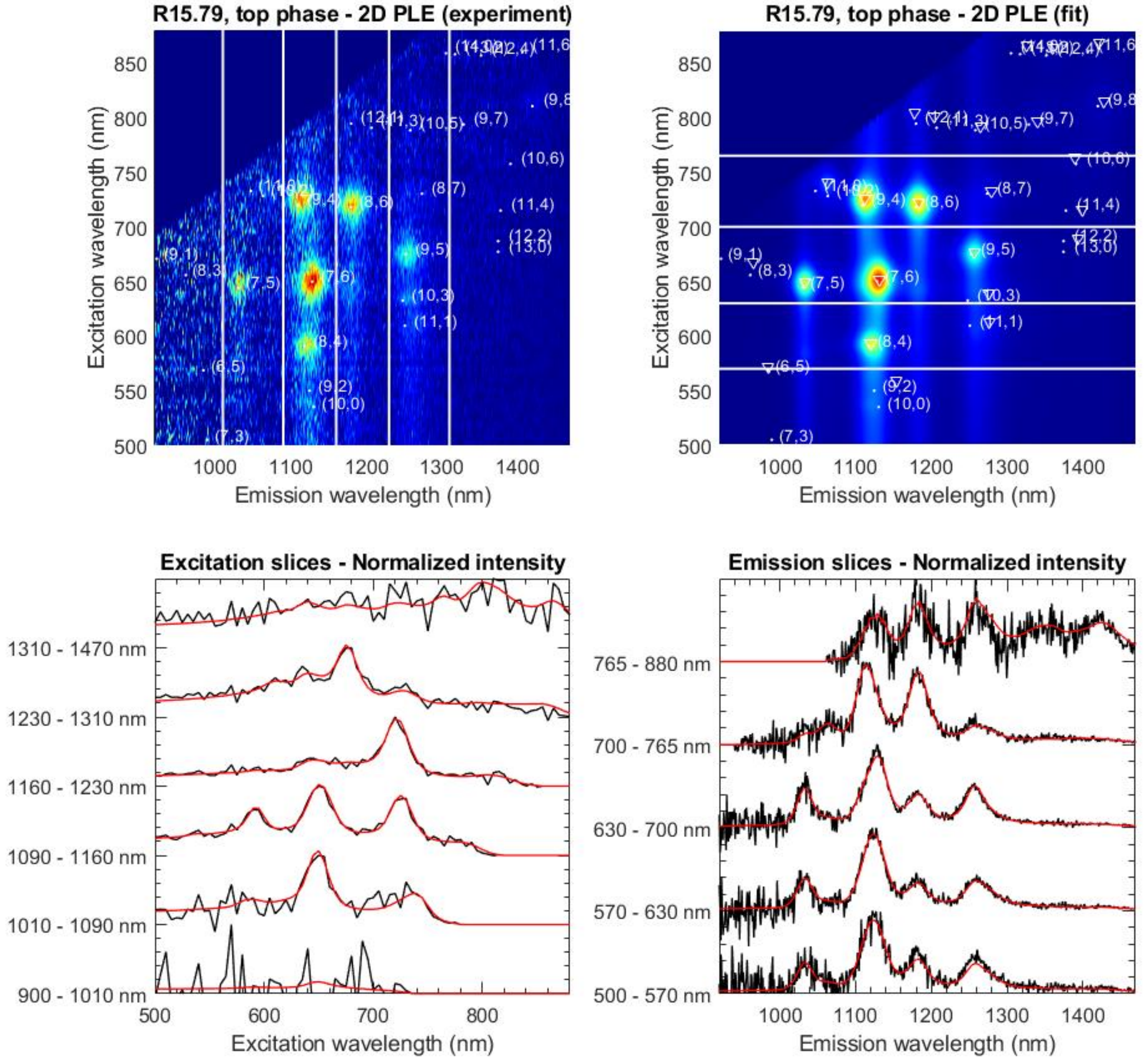

Figure S4d: 2D PLE experimental data and fits for the top phase of SDS/DOC ratio R15.79 at a fixed concentration of 0.0507% DOC. (same color coding as in Figure S4a).

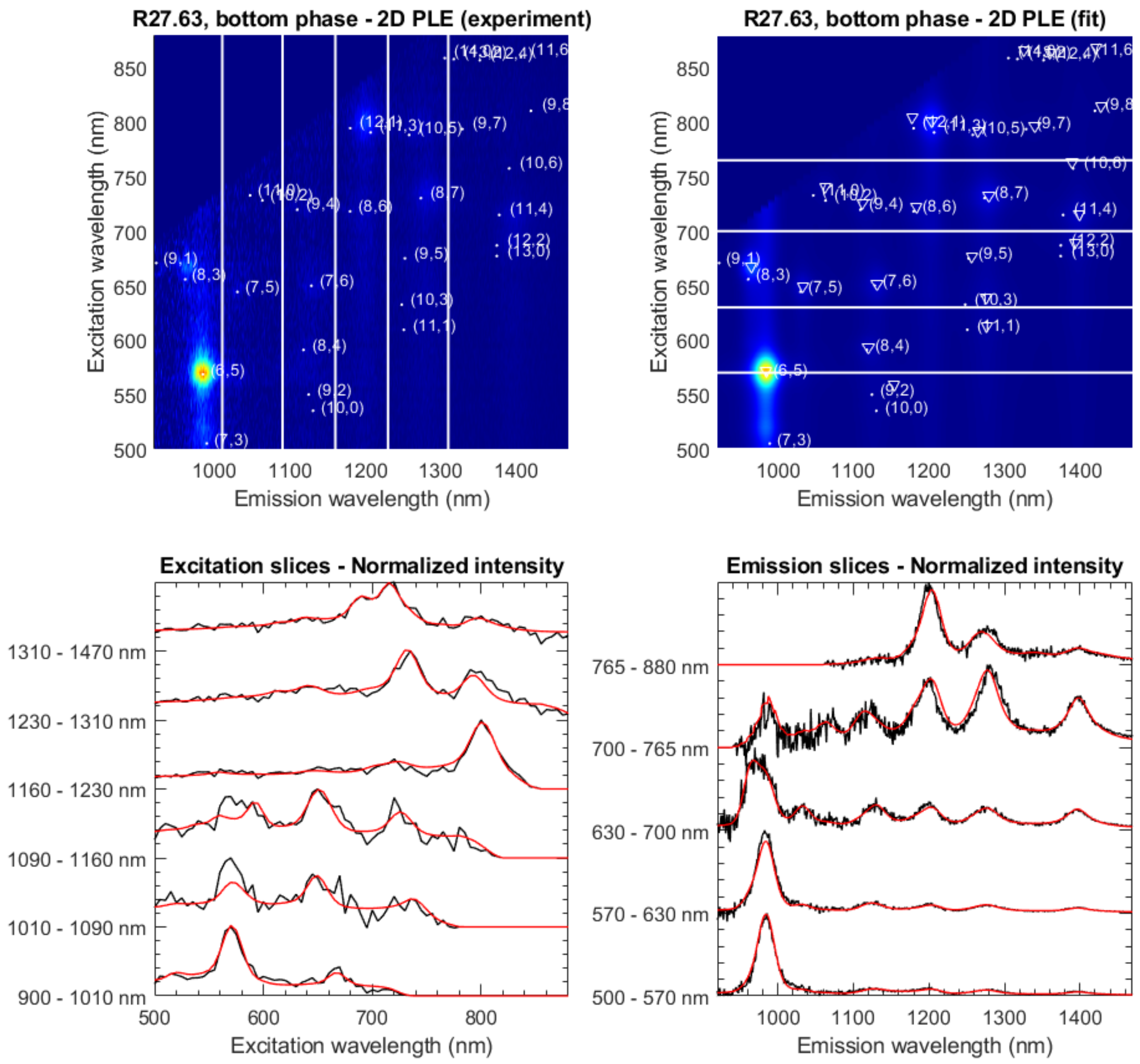

Figure S4e: 2D PLE experimental data and fits for the bottom phase of SDS/DOC ratio R27.63 at a fixed concentration of 0.0507% DOC. (same color coding as in Figure S4a).

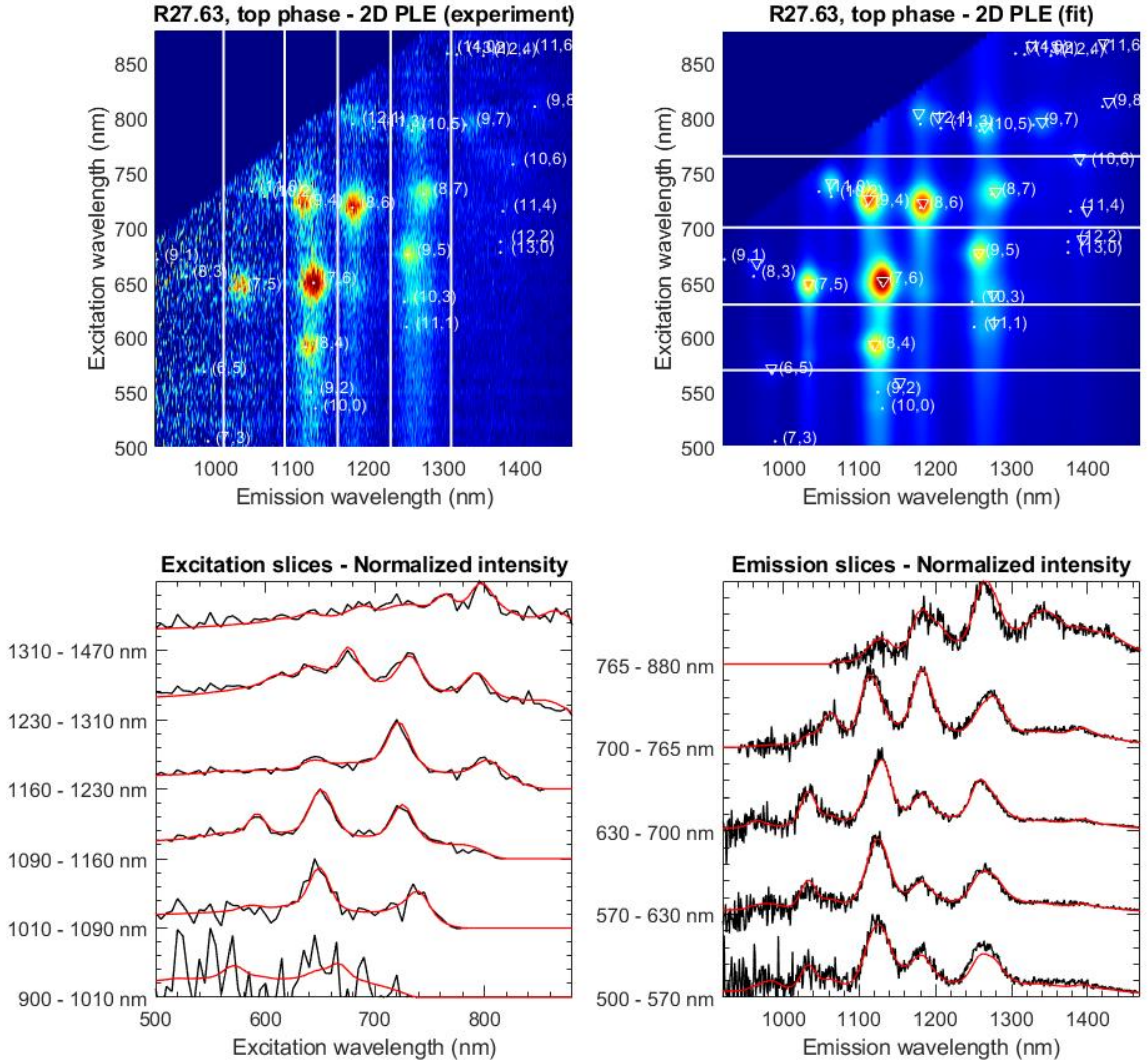

Figure S4f: 2D PLE experimental data and fits for the top phase of SDS/DOC ratio R27.63 at a fixed concentration of 0.0507% DOC. (same color coding as in Figure S4a).

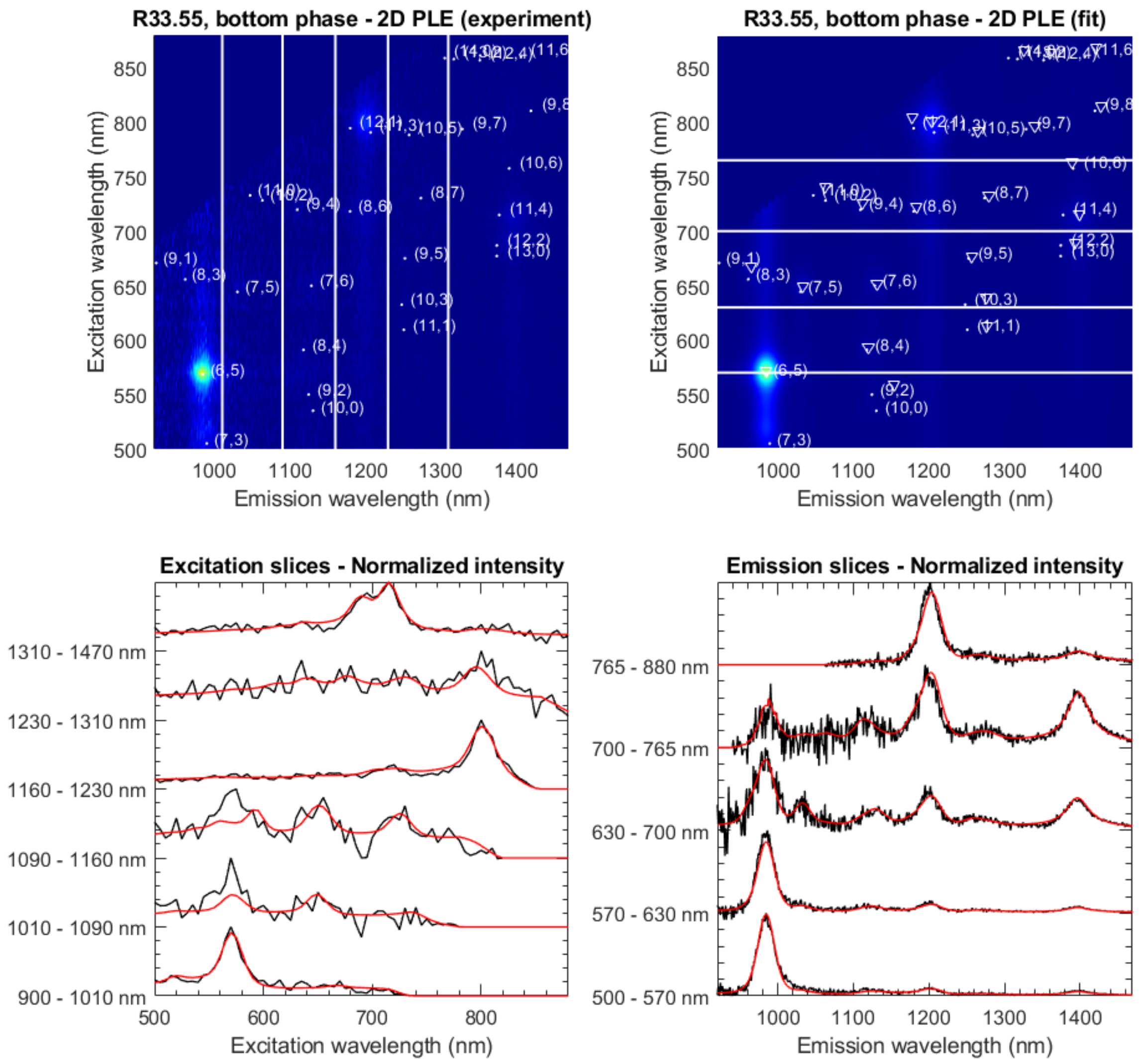

Figure S4g: 2D PLE experimental data and fits for the bottom phase of SDS/DOC ratio R33.55 at a fixed concentration of 0.0507% DOC. (same color coding as in Figure S4a).

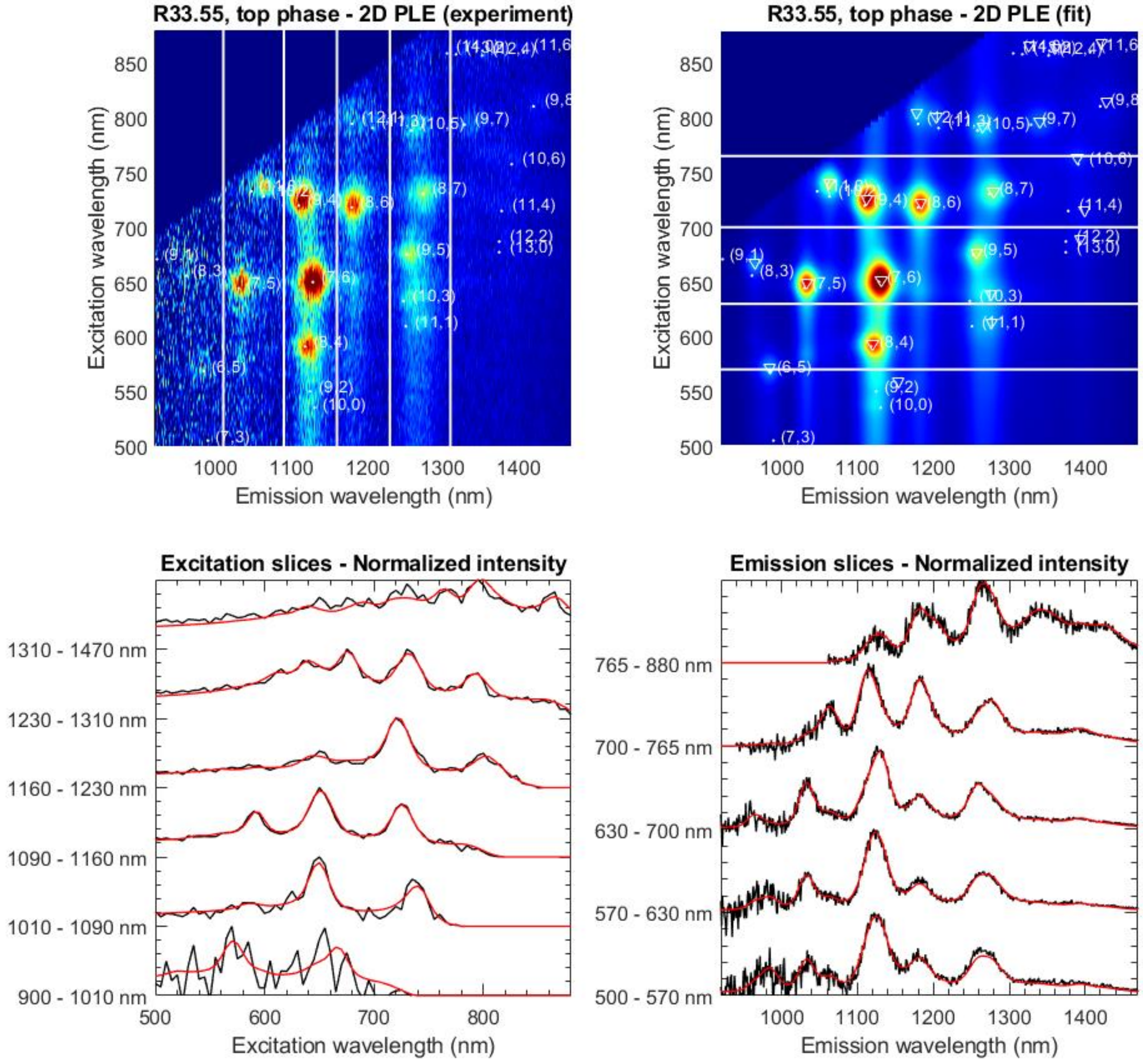

Figure S4h: 2D PLE experimental data and fits for the top phase of SDS/DOC ratio R33.55 at a fixed concentration of 0.0507% DOC. (same color coding as in Figure S4a).

## Section 5. Resonant Raman spectra and fits at different excitation wavelengths

**Table S5:** Excitation wavelengths and corresponding measured SWCNT chiralities in RRS.

| Laser | Excitation Wavelength | Chiralities |
|---|---|---|
| **Ti:Sapphire** | 824 nm | (5,4) |
| | 785 nm | (9,7), (10,5), (11,3), (12,1) |
| | 725 nm | (11,4), (8,7), (8,6), (9,4), (10,2) |
| | 710 nm | (5,3), (9,1) |
| **$Ar^+$** | 502 nm | (7,7), (8,5), (9,3) |
| | 457 nm | (12,2), (6,6), (7,4) |
| **$Kr^+$** | 647 nm | (10,3), (7,6), (7,5), (8,3), (14,2), (13,4), (12,6), (16,1) |
| **Rh6G Dye Laser** | 570 nm | (10,4), (11,2), (6,5), (6,4), (7,2) |

Figures S5 – S19 represent all Raman spectra obtained for the SDS/DOC variation with fixed DOC concentration.

5.1: Raman spectra at 824 nm: (5,4)

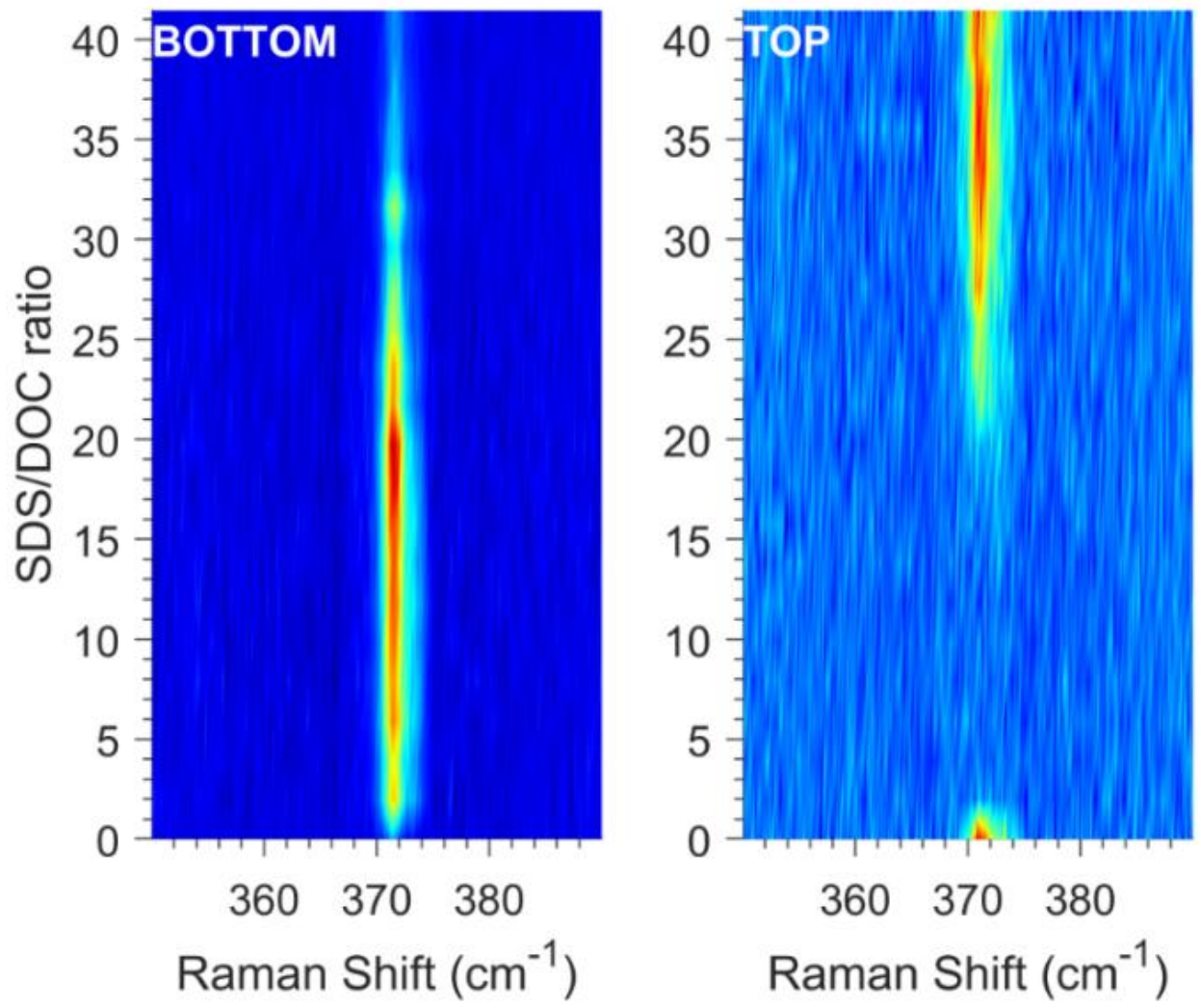

Figure S5: Intensity colormap of the RRS spectra obtained at 824 nm, as a function of increasing SDS/DOC ratio, showing the (5,4) SWCNT.

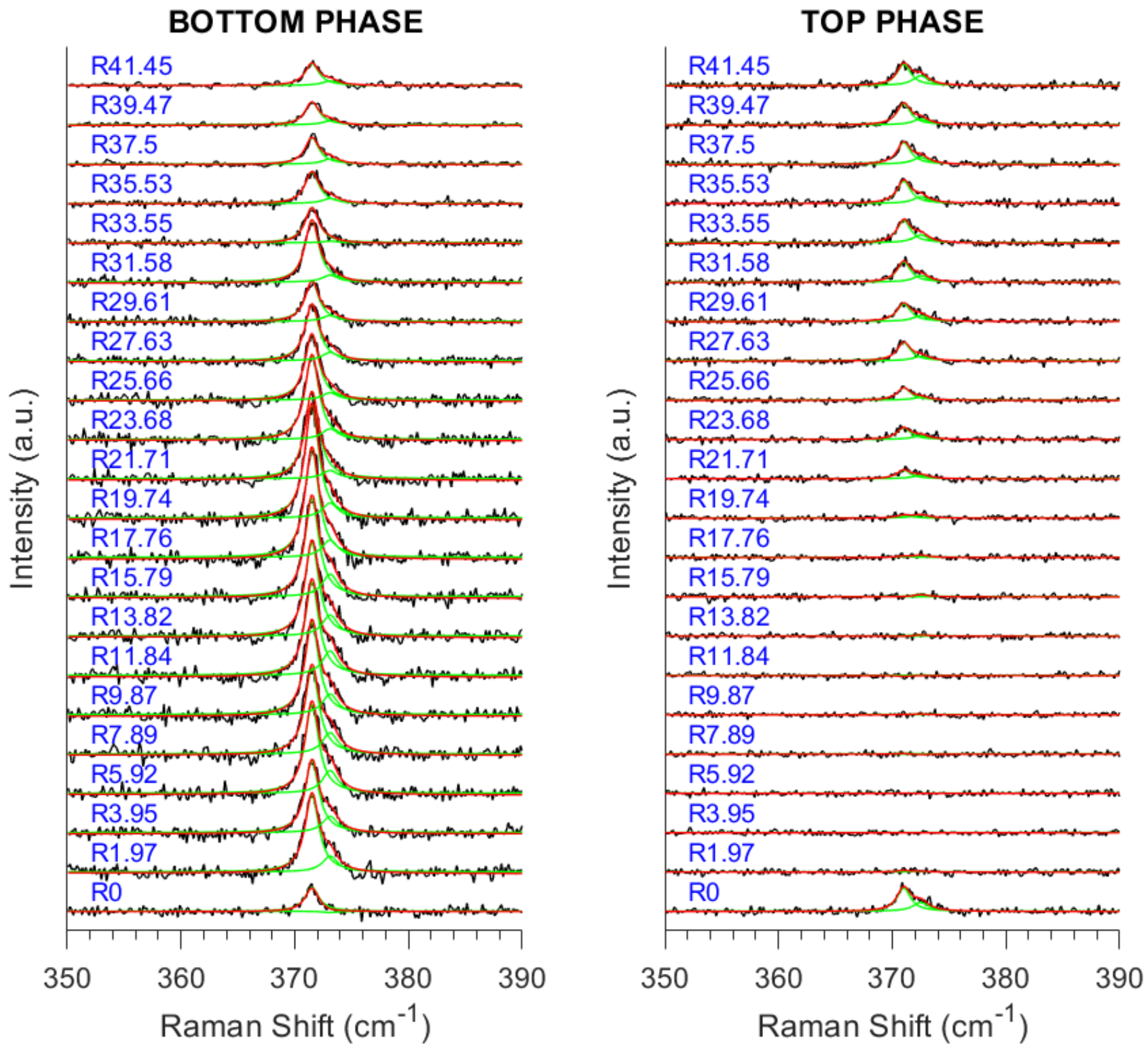

Figure S6: Individual Raman spectra at 824 nm of the different bottom and top phases (black) and fits (red) composed out of RBMs of both empty (371.5 $cm^{-1}$) and water-filled (373 $cm^{-1}$) (5,4) SWCNTs (individual Lorentzians shown in green). The R# denotes the SDS/DOC ratio.

5.2: Raman spectra at 785 nm: (9,7), (10,5), (11,3) and (12,1)

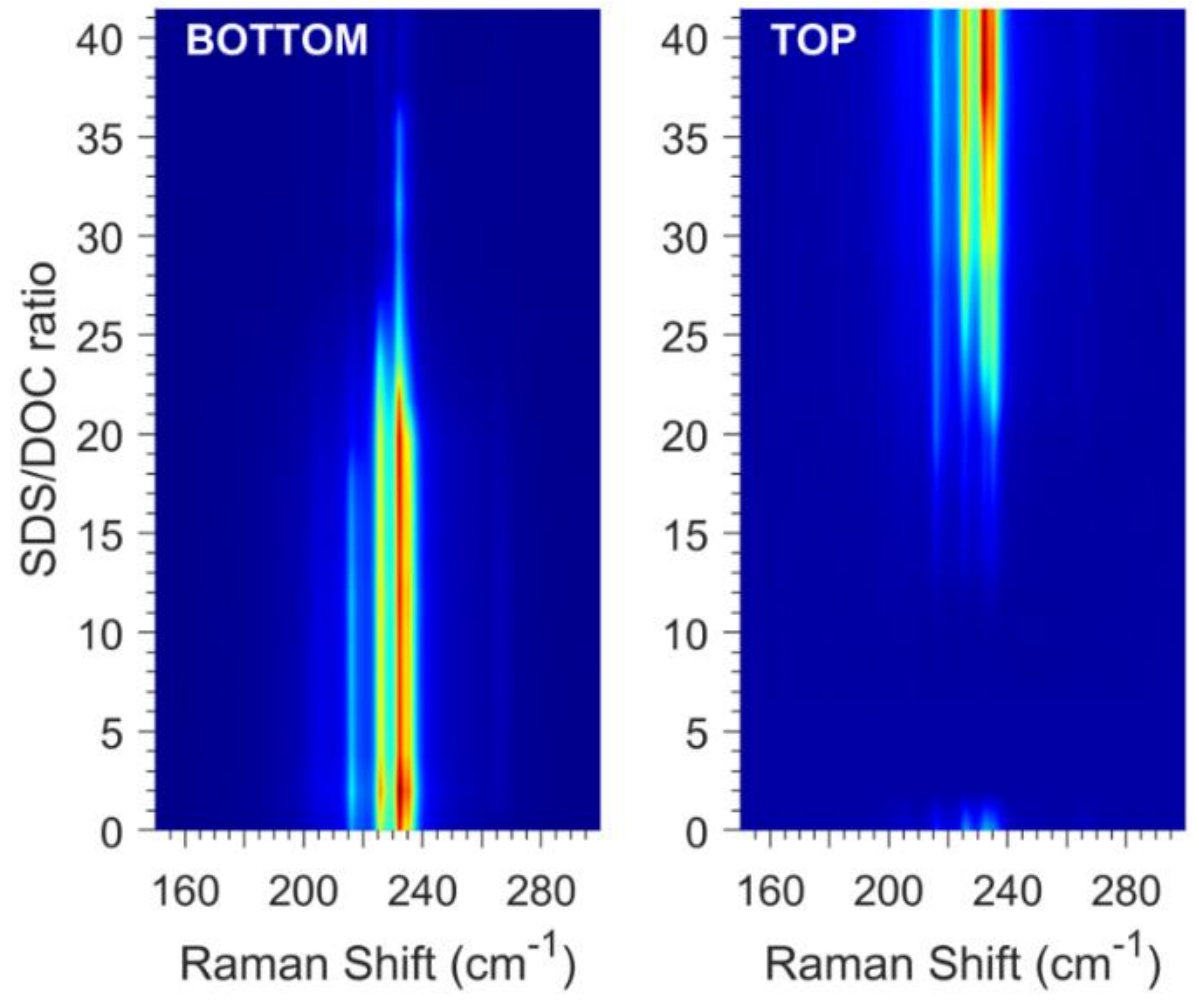

Figure S7: Intensity colormap of the RRS spectra obtained at 785 nm, as a function of increasing SDS/DOC ratio, showing the (9,7), (10,5), (11,3) and (12,1) SWCNTs

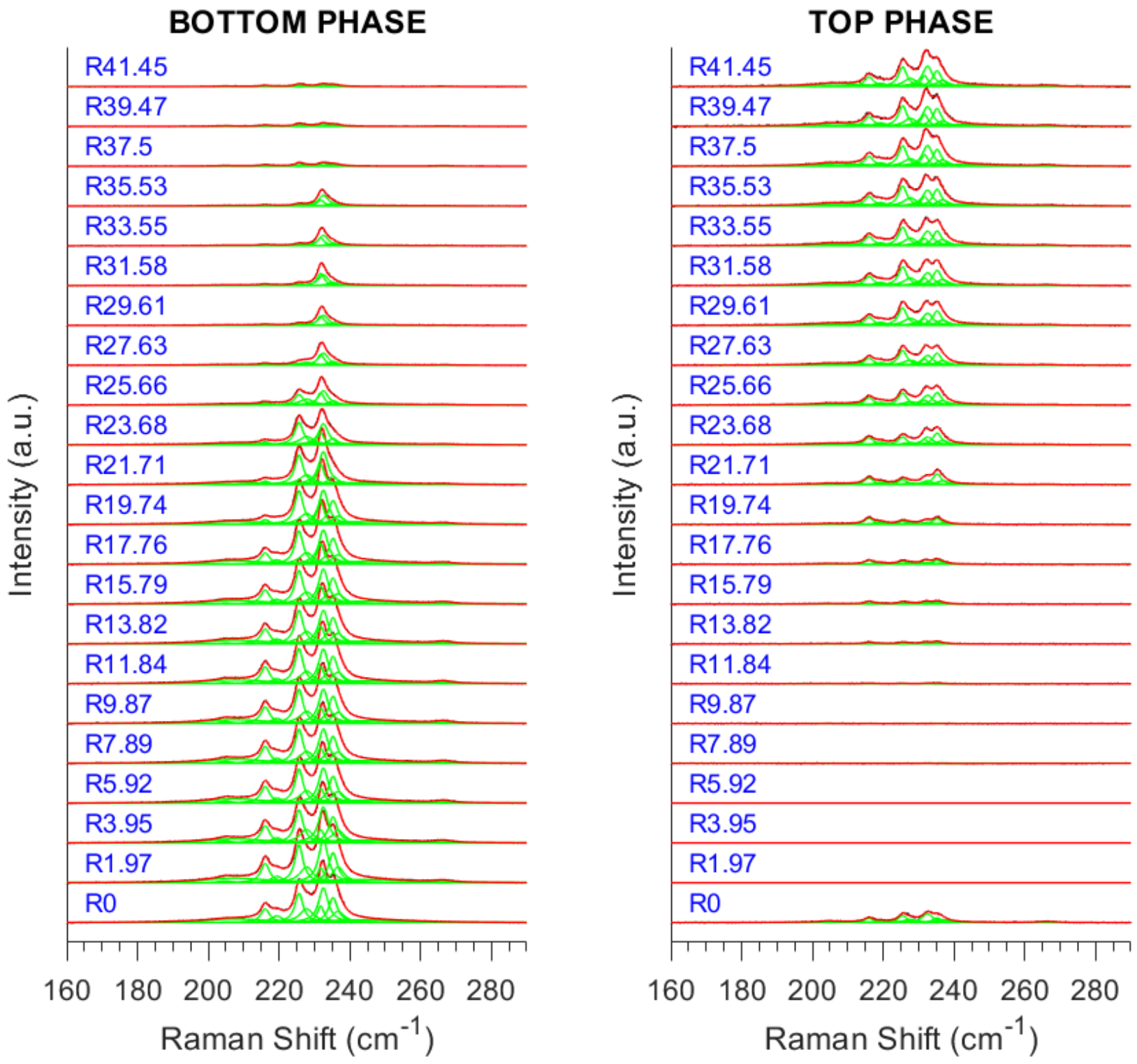

Figure S8: RRS spectra (black) and fits (red) obtained at 785 nm composed out of individual Lorentzians (green) corresponding to the RBMs of both empty and water-filled (9,7), (10,5), (11,3) and (12,1) SWCNTs. The R# denotes the SDS/DOC ratio.

5.3: Raman spectra at 725 nm: (11,4), (8,7), (8,6), (9,4) and (10,2)

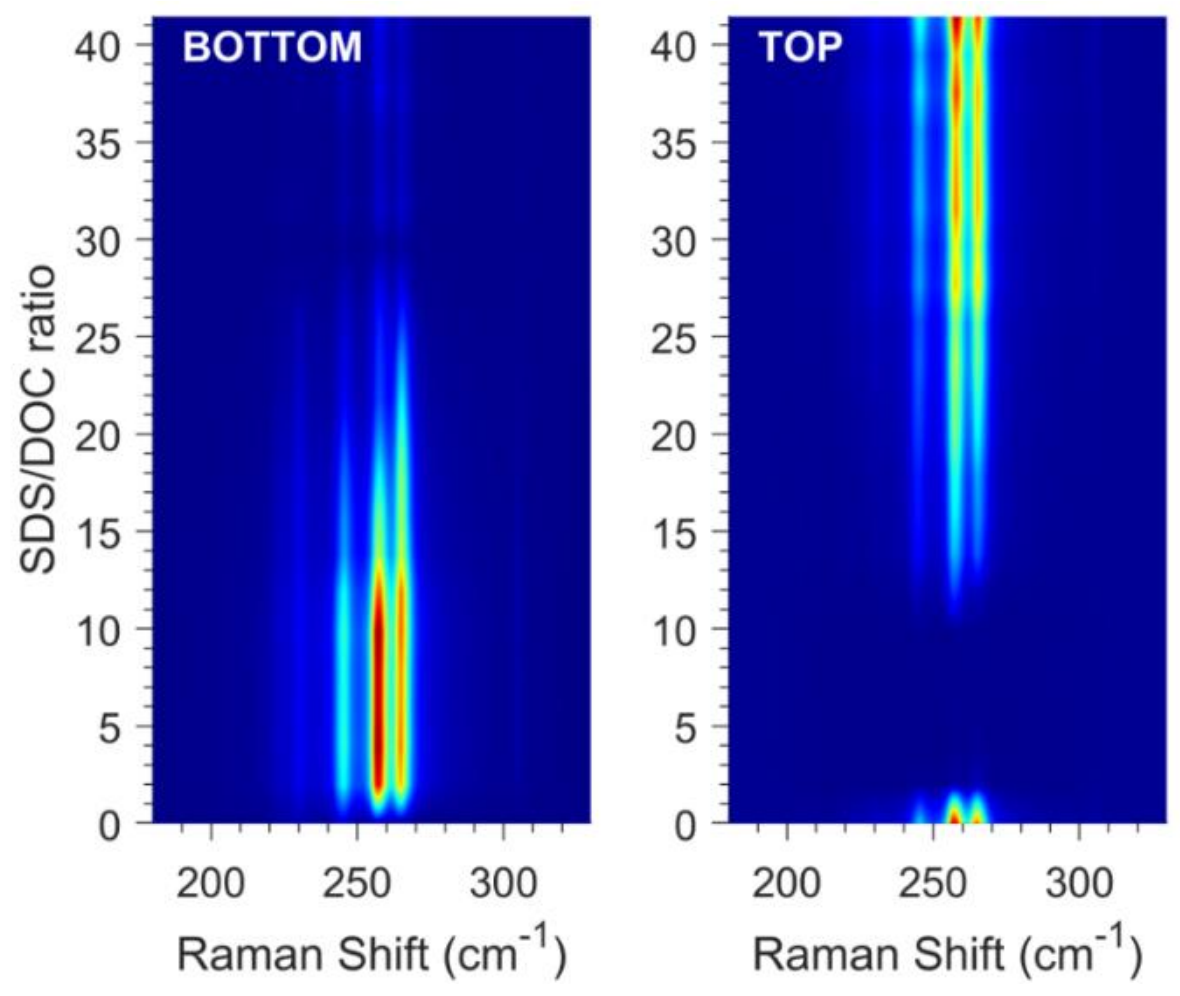

Figure S9: Intensity colormap of the RRS spectra obtained at 725 nm, as a function of increasing SDS/DOC ratio, showing the (11,4), (8,7), (8,6) (9,4) and (10,2) SWCNTs.

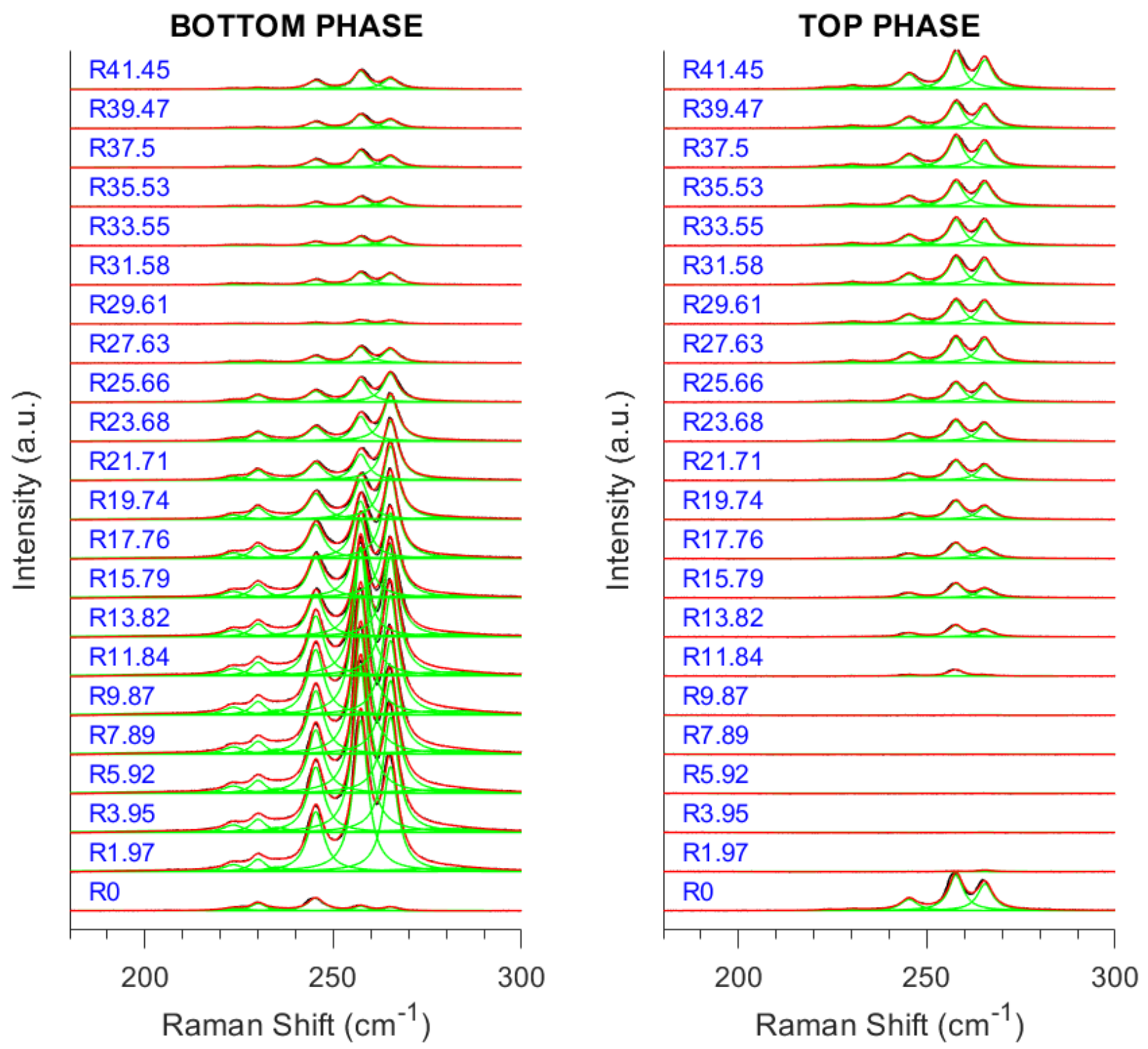

Figure S10: RRS spectra (black) and fits (red) obtained at 725 nm composed out of individual Lorentzians (green) corresponding to the RBMs of (11,4), (8,7), (8,6), (9,4) and (10,2) SWCNTs. The R# denotes the SDS/DOC ratio.

5.4: Raman spectra at 710 nm: (5,3) and (9,1)

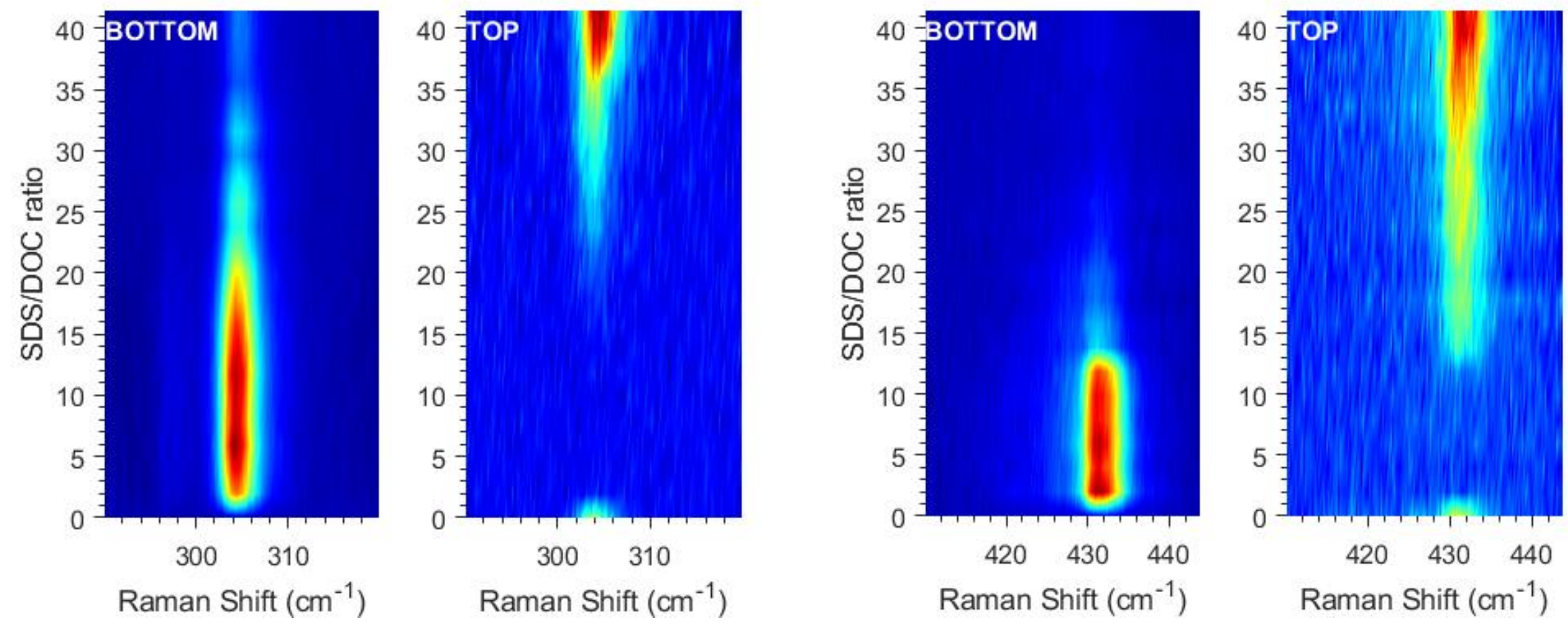

Figure S11: Intensity colormap of the RRS spectra obtained at 710 nm, as a function of increasing SDS/DOC ratio, showing the (9,1) (left 2 panels) and (5,3) (right 2 panels) SWCNTs.

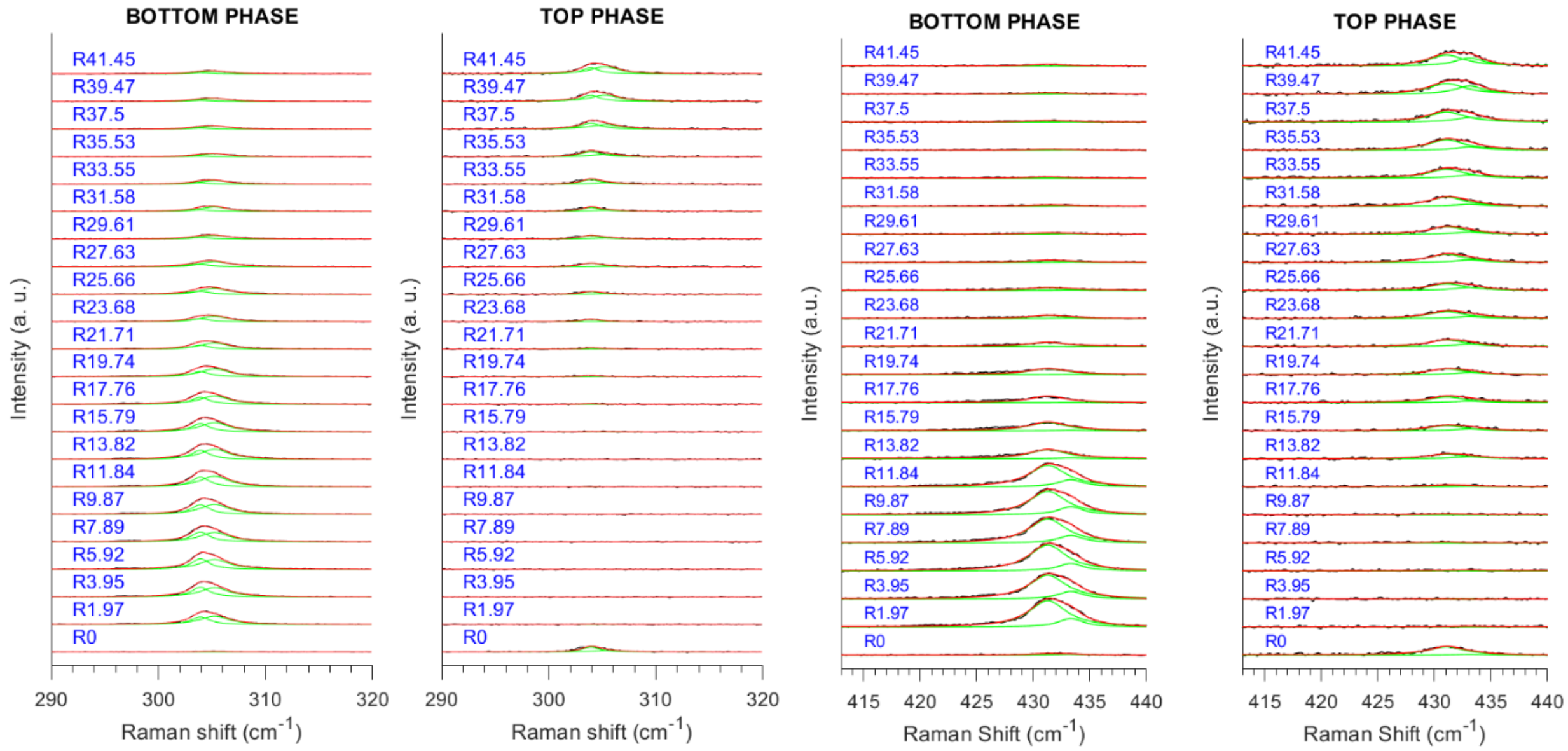

Figure S12: RRS spectra (black) and fits (red) obtained at 710 nm composed of individual Lorentzians (green) corresponding to the RBMs of empty and water-filled (9,1) (left 2 panels) and (5,3) (right 2 panels) SWCNTs. The R# denotes the SDS/DOC ratio.

5.5: Raman spectra at 502 nm: (7,7), (8,5) and (9,3)

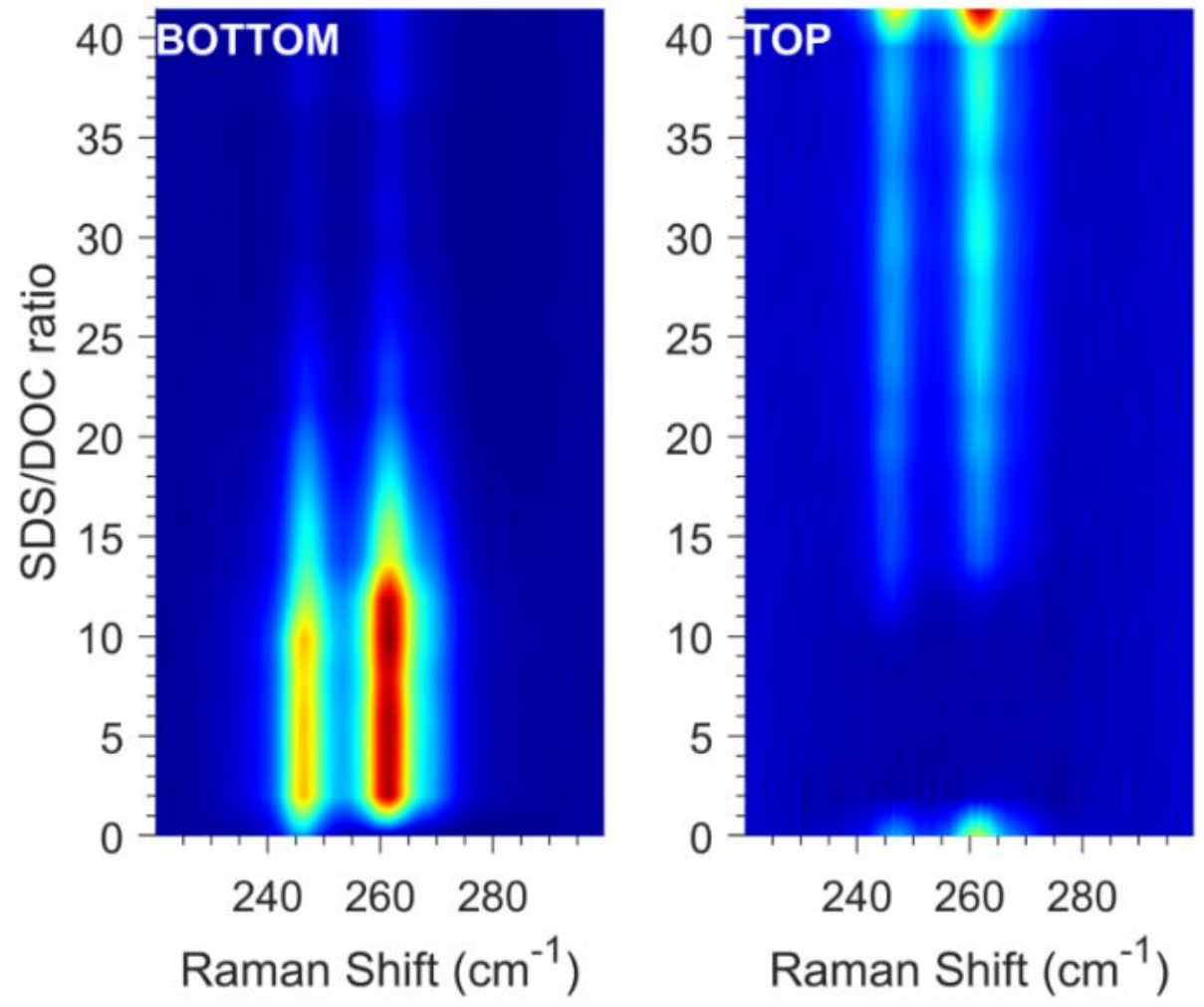

Figure S13: Intensity colormap of the RRS spectra obtained at 502 nm, as a function of increasing SDS/DOC ratio, showing the (7,7), (8,5) and (9,3) SWCNTs

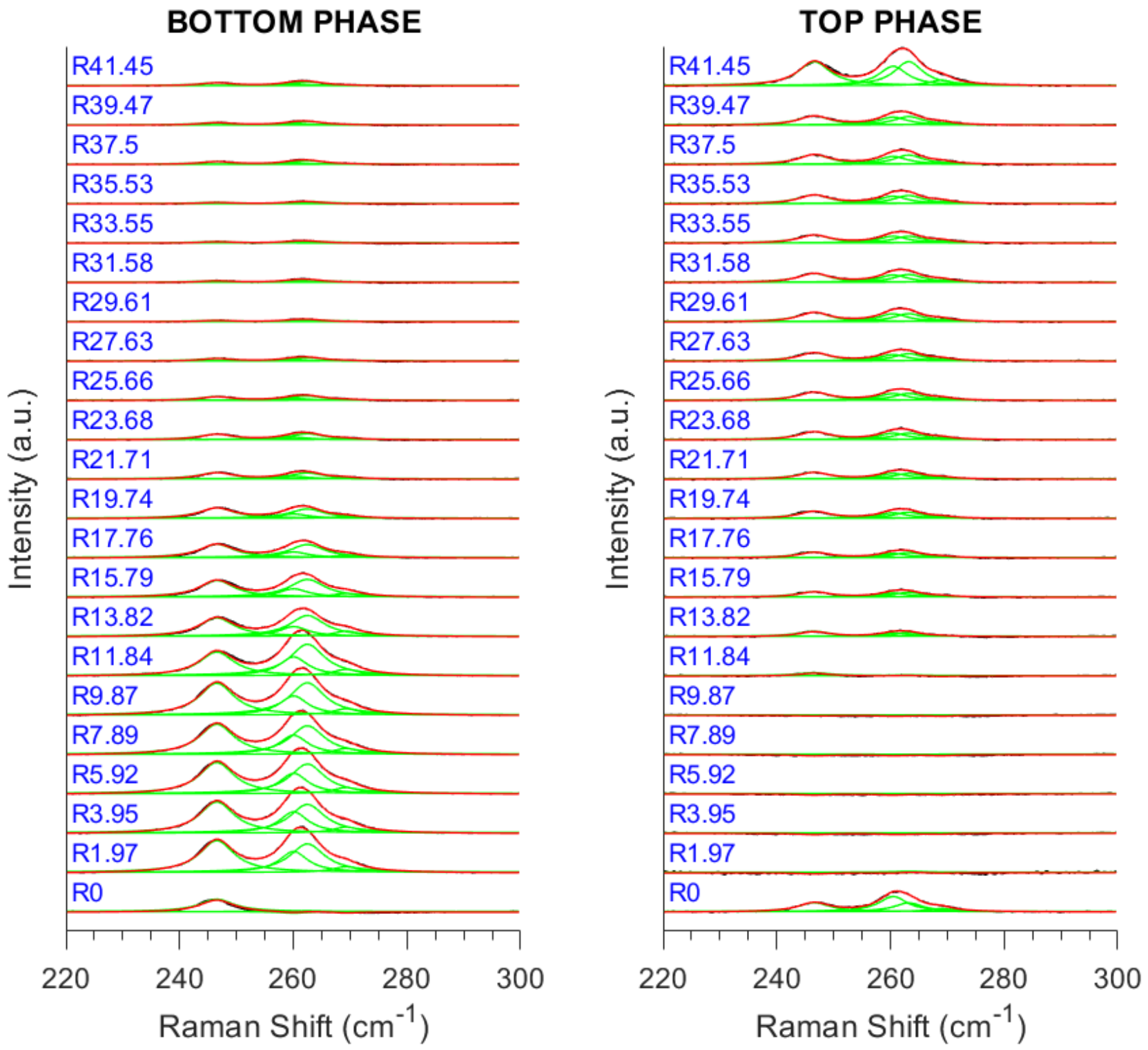

Figure S14: RRS spectra (black) and fits (red) obtained at 502 nm composed out of individual Lorentzians (green) corresponding to the RBMs of (7,7), empty and filled (8,5) and (9,3) SWCNTs. The R# denotes the SDS/DOC ratio.

5.6: Raman spectra at 457 nm: (12,2), (6,6) and (7,4)

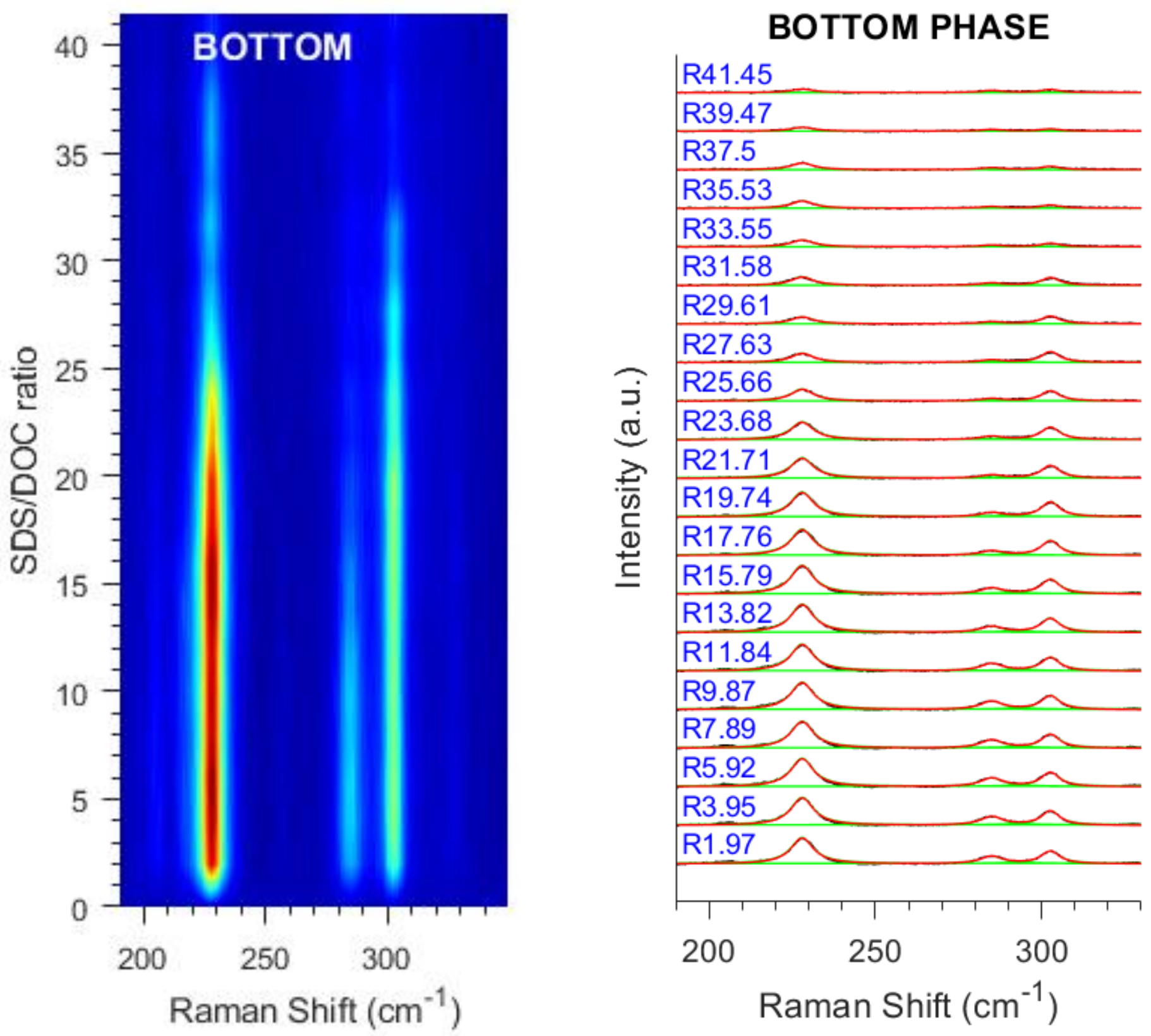

Figure S15: Intensity colormap of the RRS spectra obtained at 457 nm for the bottom phases, as well as the fitted RRS spectra, as a function of increasing SDS/DOC ratio, showing the (12,2), (6,6) and (7,4) SWCNTs.

5.7: Raman spectra at 647 nm: (10,3), (7,6), (7,5), (8,3), (14,2), (13,4), (12,6) and (16,1)

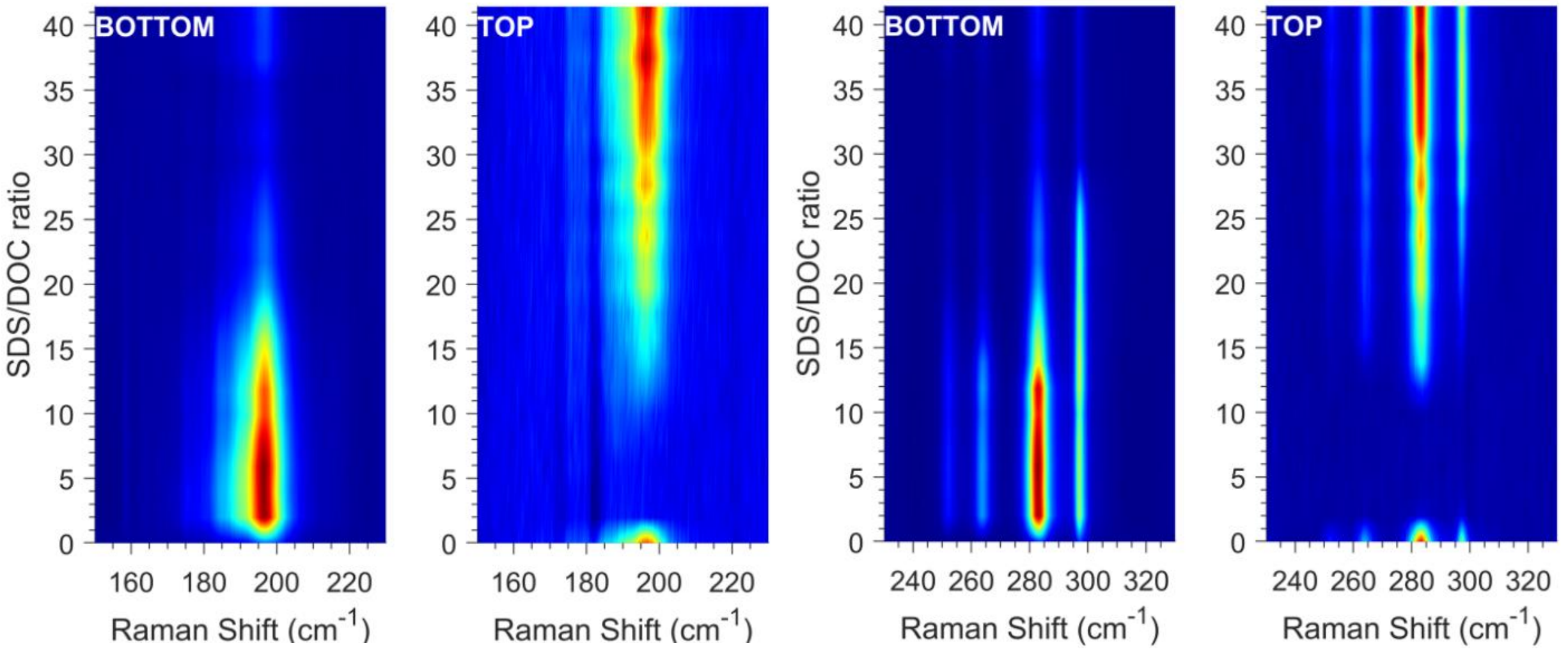

Figure S16: Intensity colormap of the RRS spectra obtained at 647 nm, as a function of increasing SDS/DOC ratio, showing the (left 2 panels) (16,1), (14,2), (13,4), (12,6), and (right 2 panels) (10,3), (7,6), (7,5) and (8,3) SWCNTs.

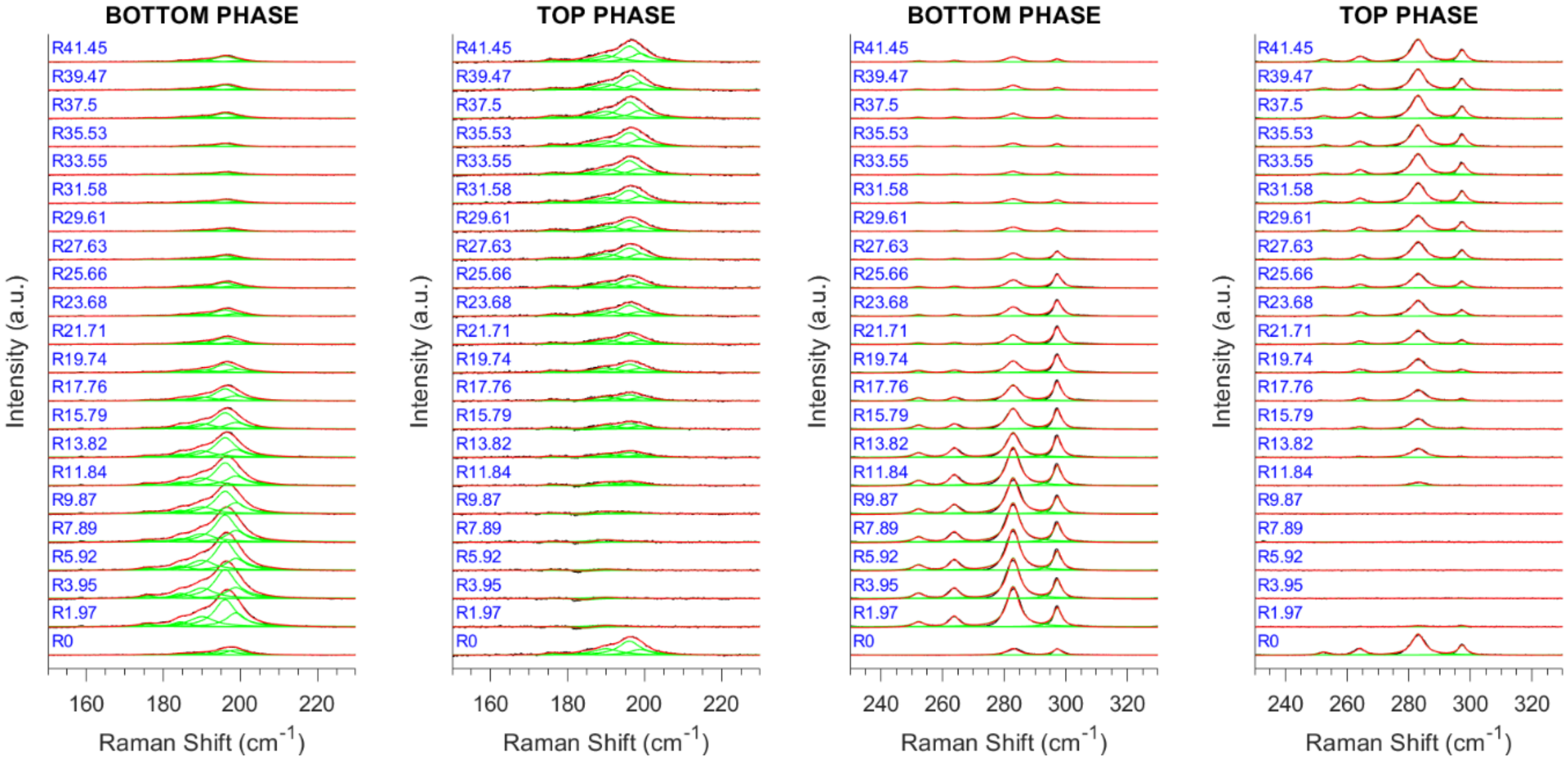

Figure S17: RRS spectra (black) and fits (red) obtained at 647 nm composed out of individual Lorentzians (green) corresponding to the RBMs (left 2 panels) (16,1), (14,2), (13,4), (12,6), and (right 2 panels) (10,3), (7,6), (7,5) and (8,3) SWCNTs. The R# denotes the SDS/DOC ratio.

5.8: Raman spectra at 570 nm: (10,4), (11,2), (6,5), (6,4) and (7,2)

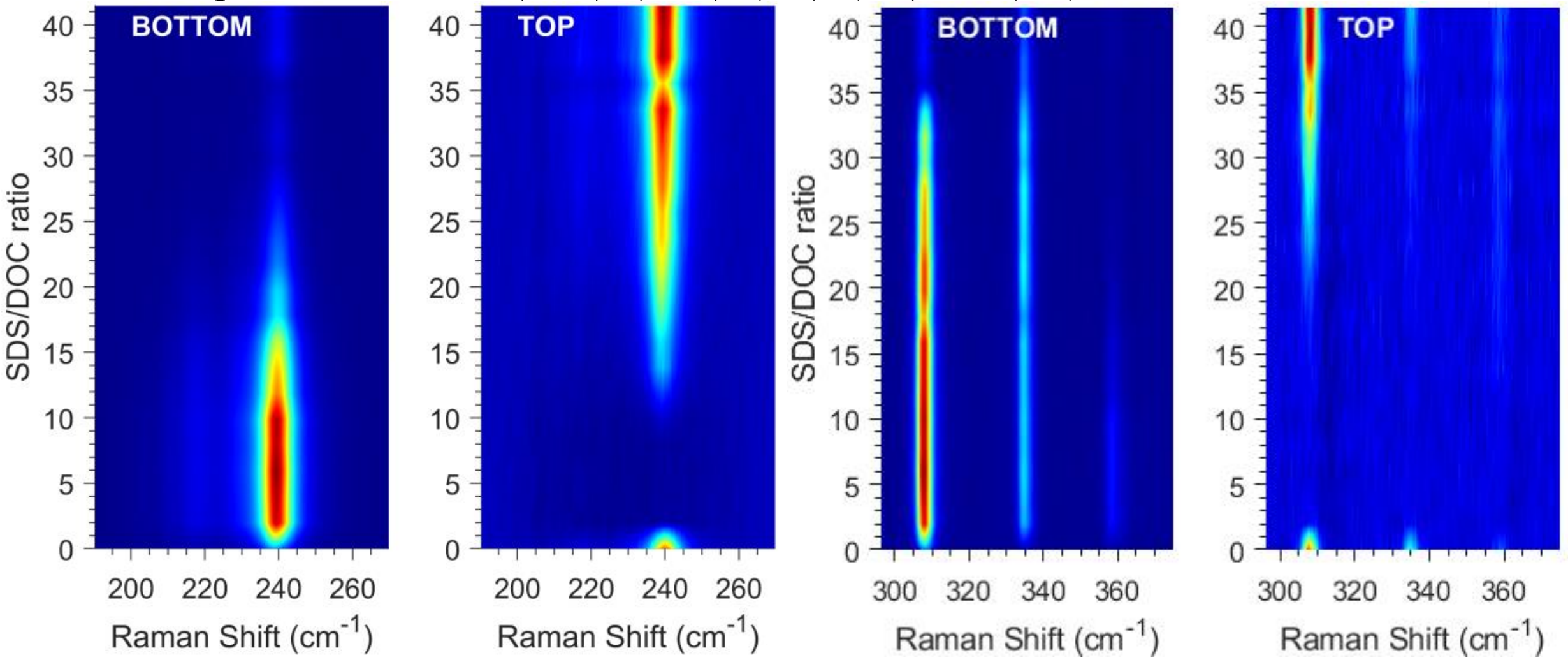

Figure S18: Intensity colormap of the RRS spectra obtained at 570 nm, as a function of increasing SDS/DOC ratio, showing the (10,4) and (11,2) (left 2 panels) and (6,5), (6,4) and (7,2) SWCNT (right 2 panels).

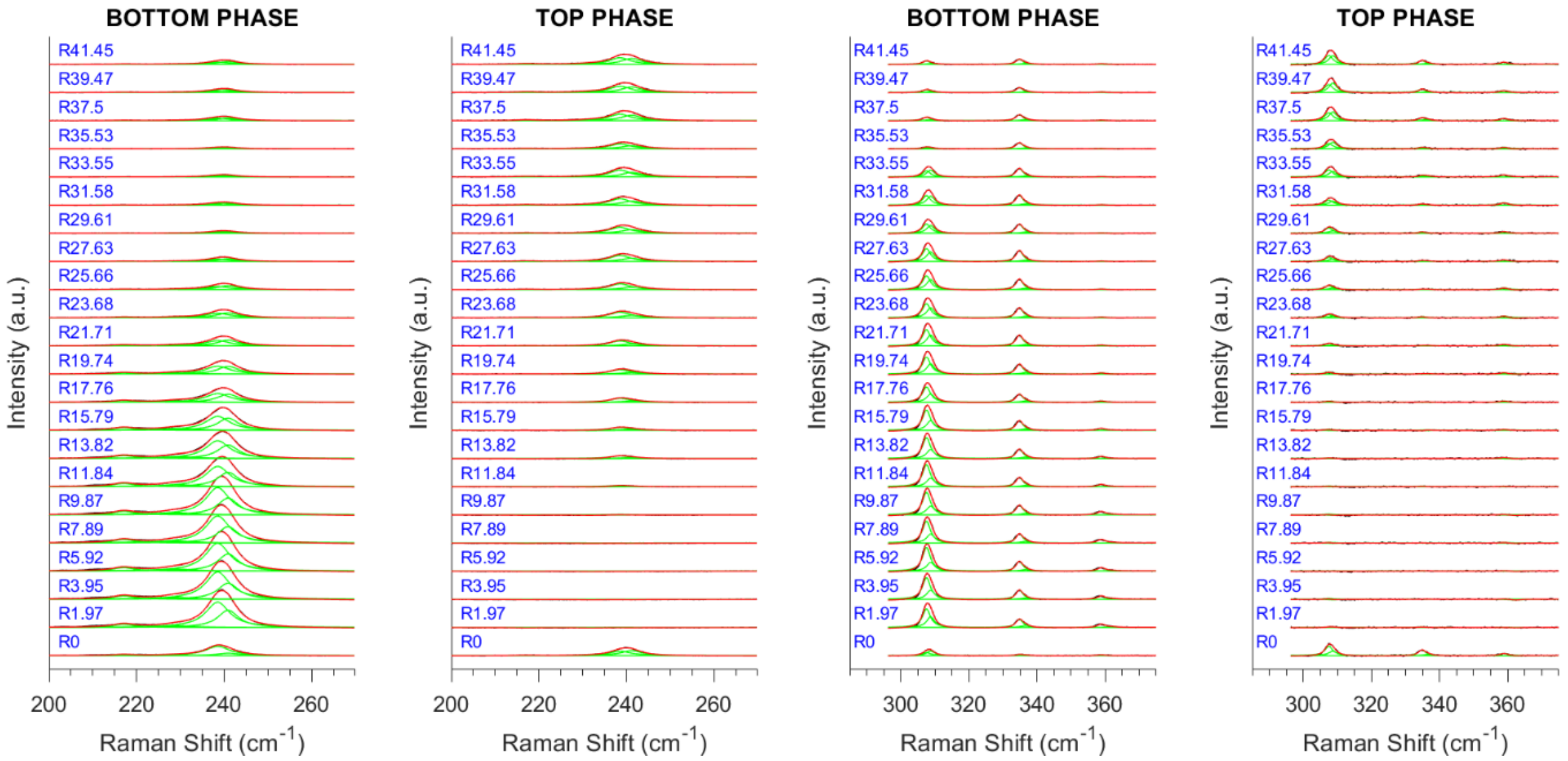

Figure S19: RRS spectra (black) and fits (red) obtained at 570 nm composed out of individual Lorentzians (green) corresponding to the RBMs of (10,4) and (11,2) (left 2 panels) and empty and water-filled (6,5), (6,4) and (7,2) SWCNT (right 2 panels). The R# denotes the SDS/DOC ratio.

**Section 6. Fits of partition coefficient curves for SDS/DOC (0.0507% wt/V DOC).**

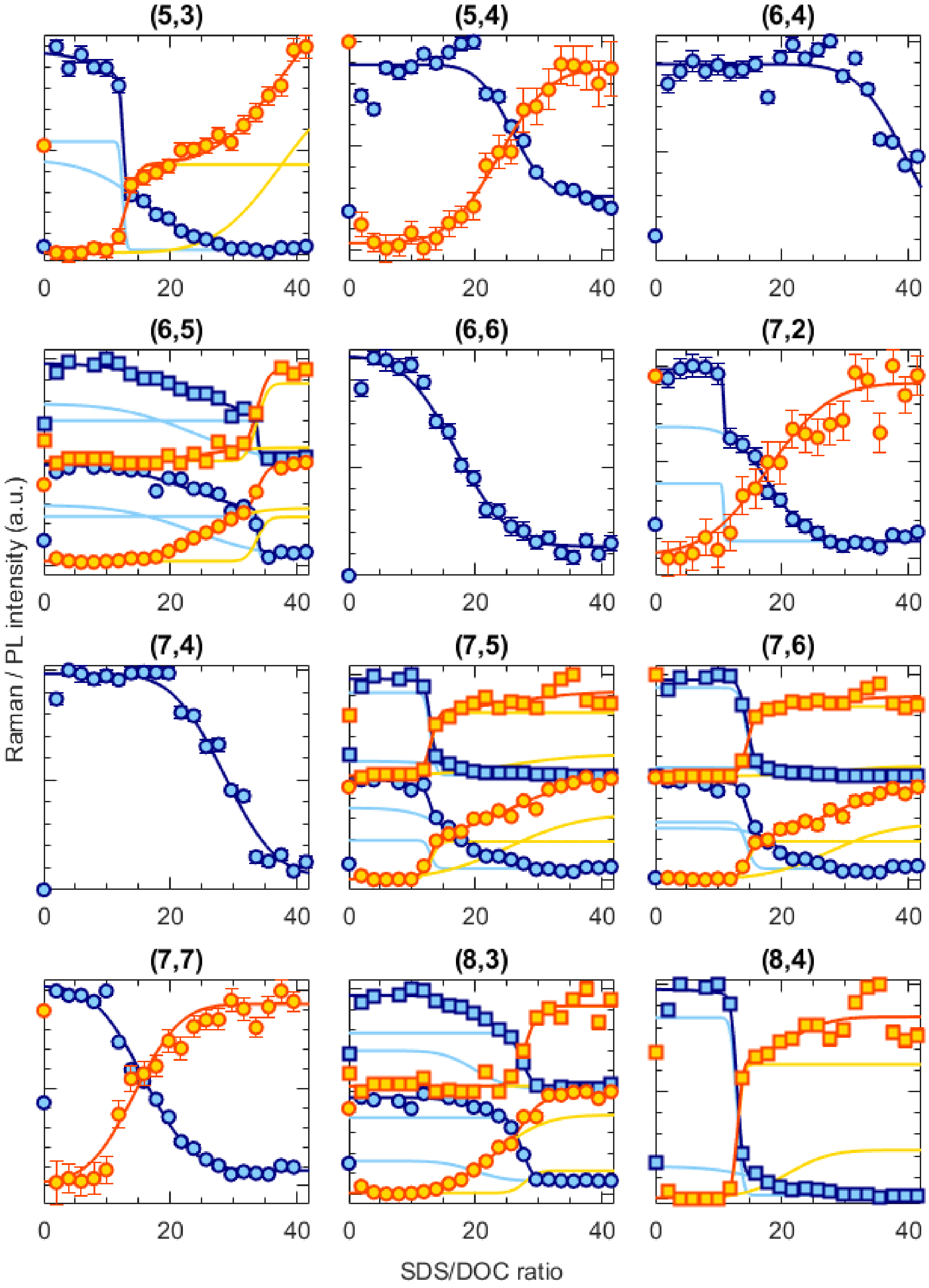

Figure S20: continues on next pages

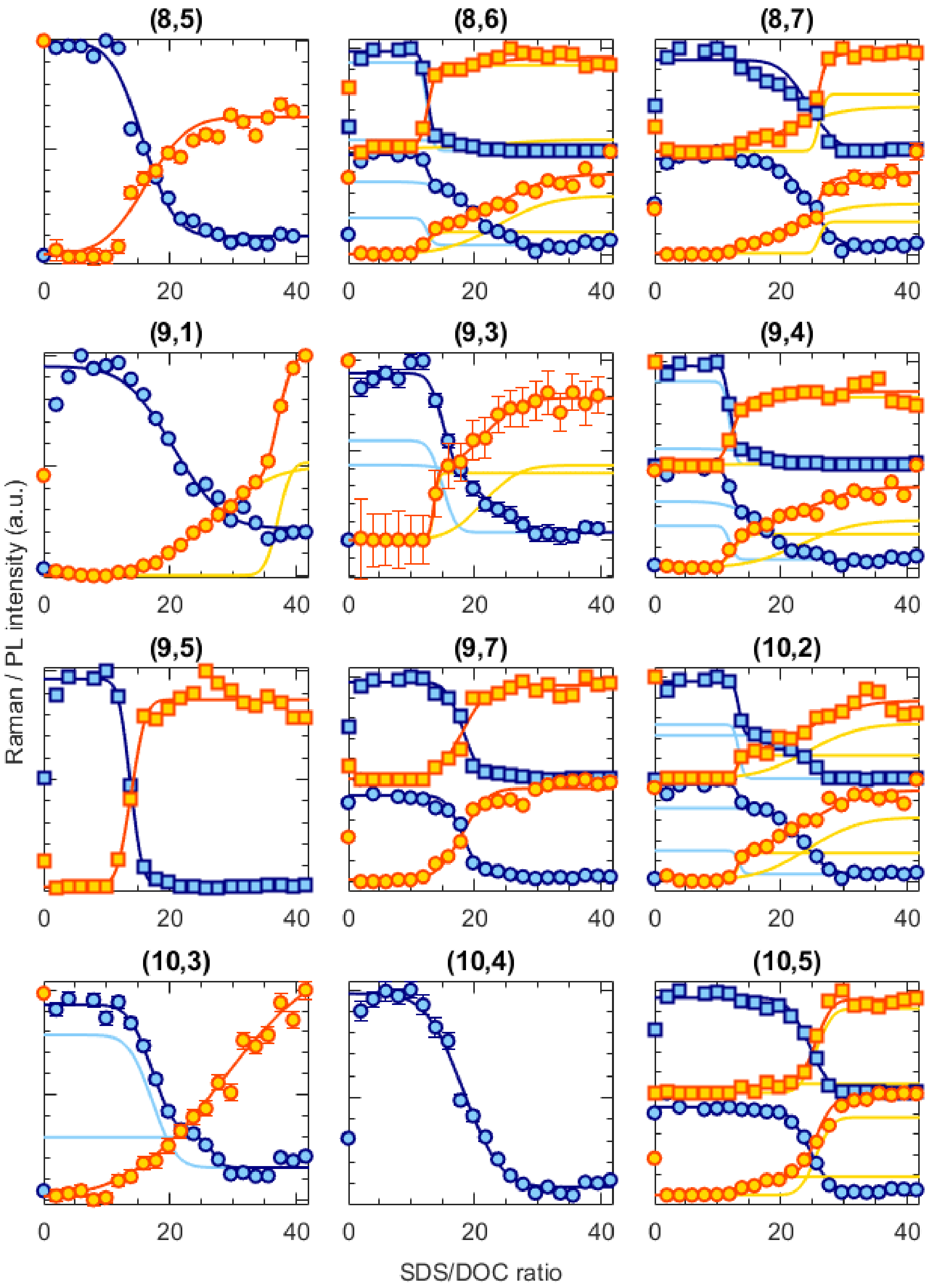

Figure S20: continues on next page

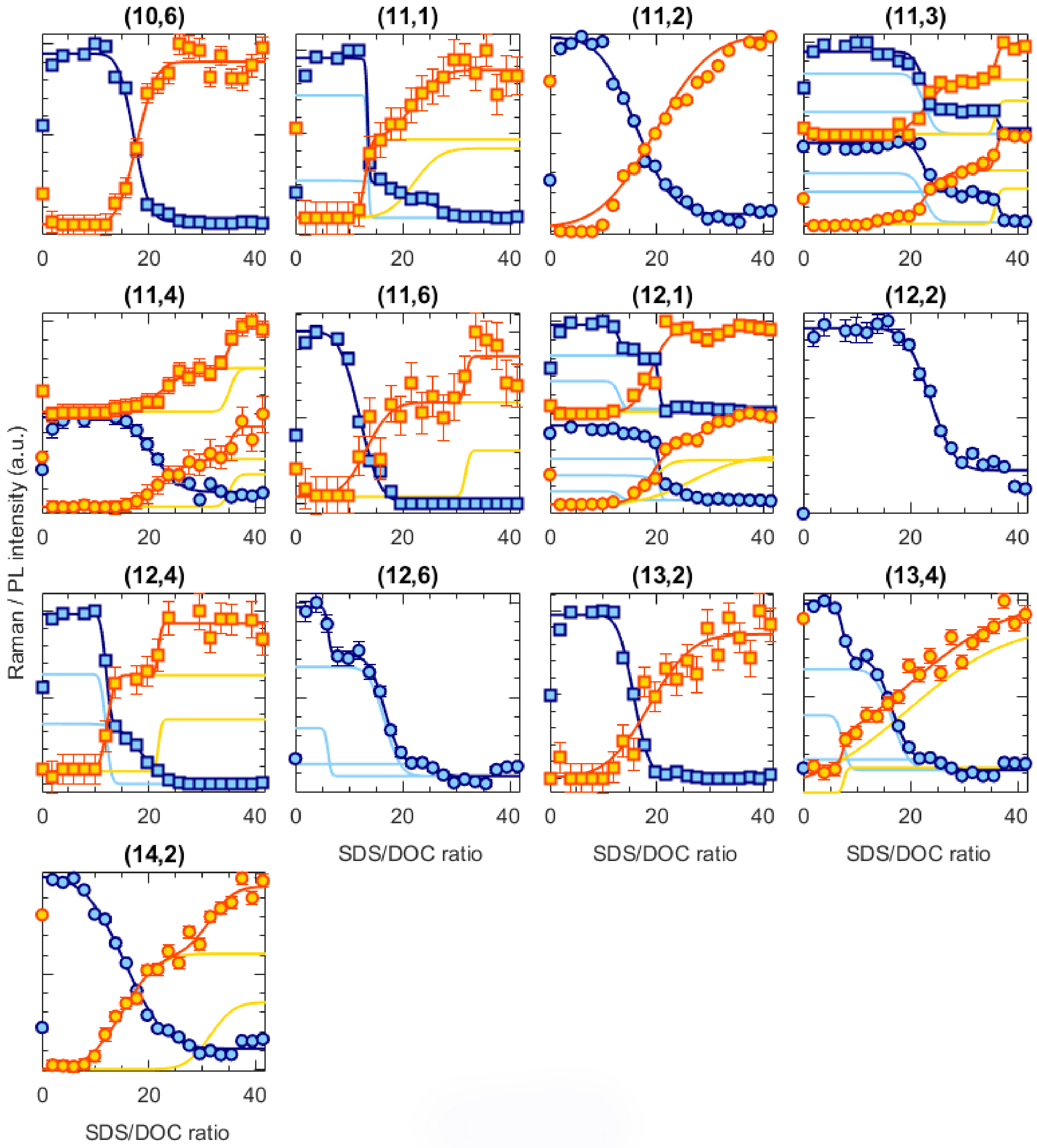

Figure S20: Normalized PL (squares) and Raman (circles) intensities (*i.e.*, $K_{(n,m)}^{top}$ and $K_{(n,m)}^{bottom}$) for SWCNT chiralities as a function of increasing SDS/DOC ratio. The fit components for the bottom and top phases are shown with blue and yellow solid lines, respectively; their sums are shown in darker blue and orange, respectively. The transition curves for some chiralities in the top phase are missing due to absence of corresponding measurements or reliable data.

## Section 7. Overview figures for other surfactant combinations

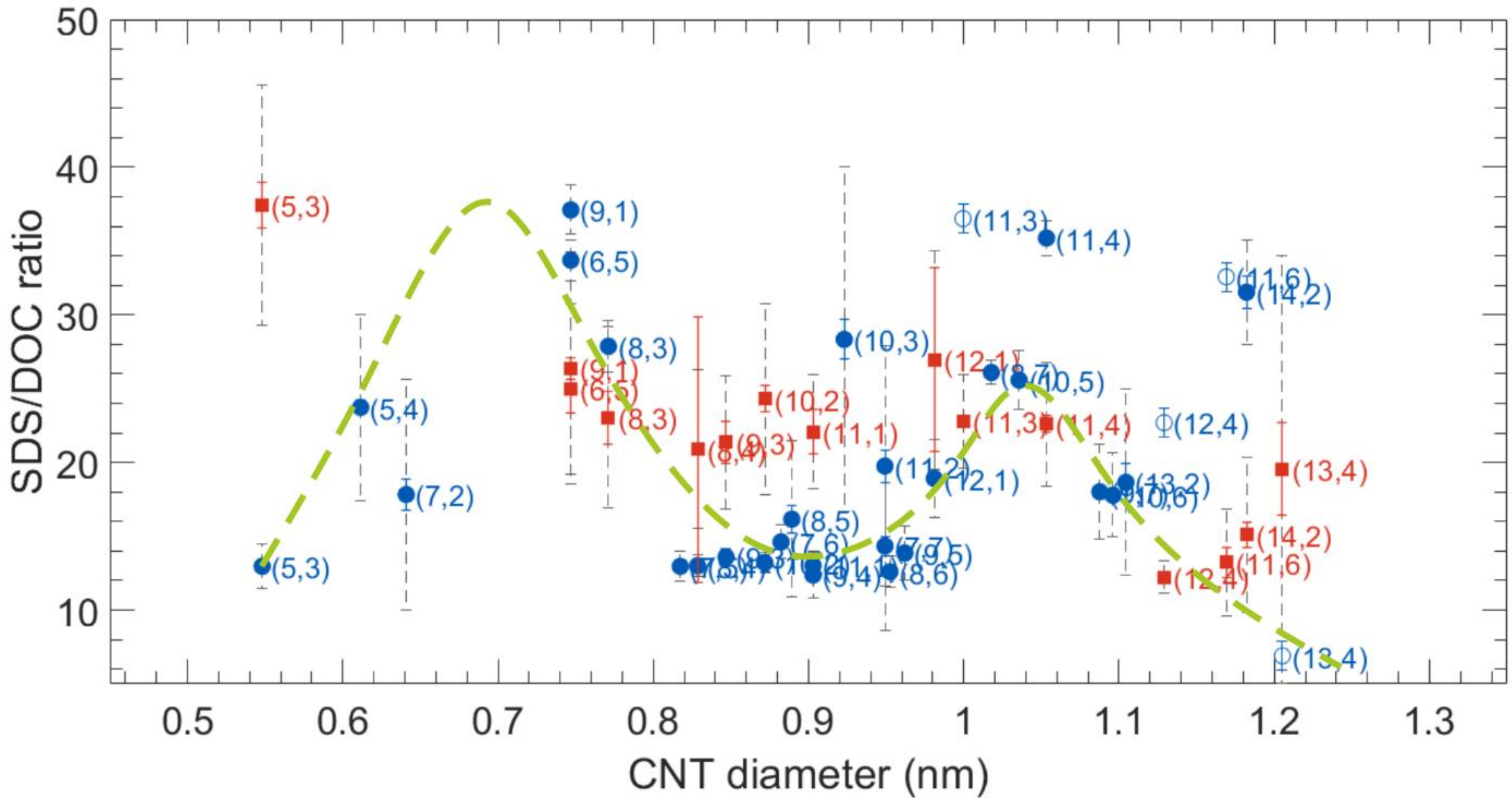

Figure S21: Transition points from **bottom to top phase** with increasing SDS concentration as determined from the **top phase data** for SWCNTs in SDS/DOC (0.0507% wt/V DOC) as a function of SWCNT diameter obtained by fitting the PL and RRS intensities of the bottom phase with a complementary error function. Blue circles correspond to the transitions that are either the only one (for all the chiralities that show one transition) or the steepest one (for all the chiralities that have two- or three-step transition curves). The second transition, if present, is shown with red squares. Blue and red solid-line error bars are 1$\sigma$ errors of the peak position fit. Dashed grey error bars are the transition linewidths obtained from the fit procedure (defined as FWHM of the corresponding Gaussians). If available, RRS and PL partition coefficient curves were fitted simultaneously for the same phase to obtain the best values for the transition points. If it was difficult to determine parameters of a transition (position, width) by means of fitting, *e.g.*, in case no experimental data point was present in the transition and the fit gave vast errors due to an underdetermined fit, we estimated the width of the transition as the widest visible boundaries of transition parameters (corresponding points are shown with open symbols). The green dashed line shows a trend towards periodic modulation.

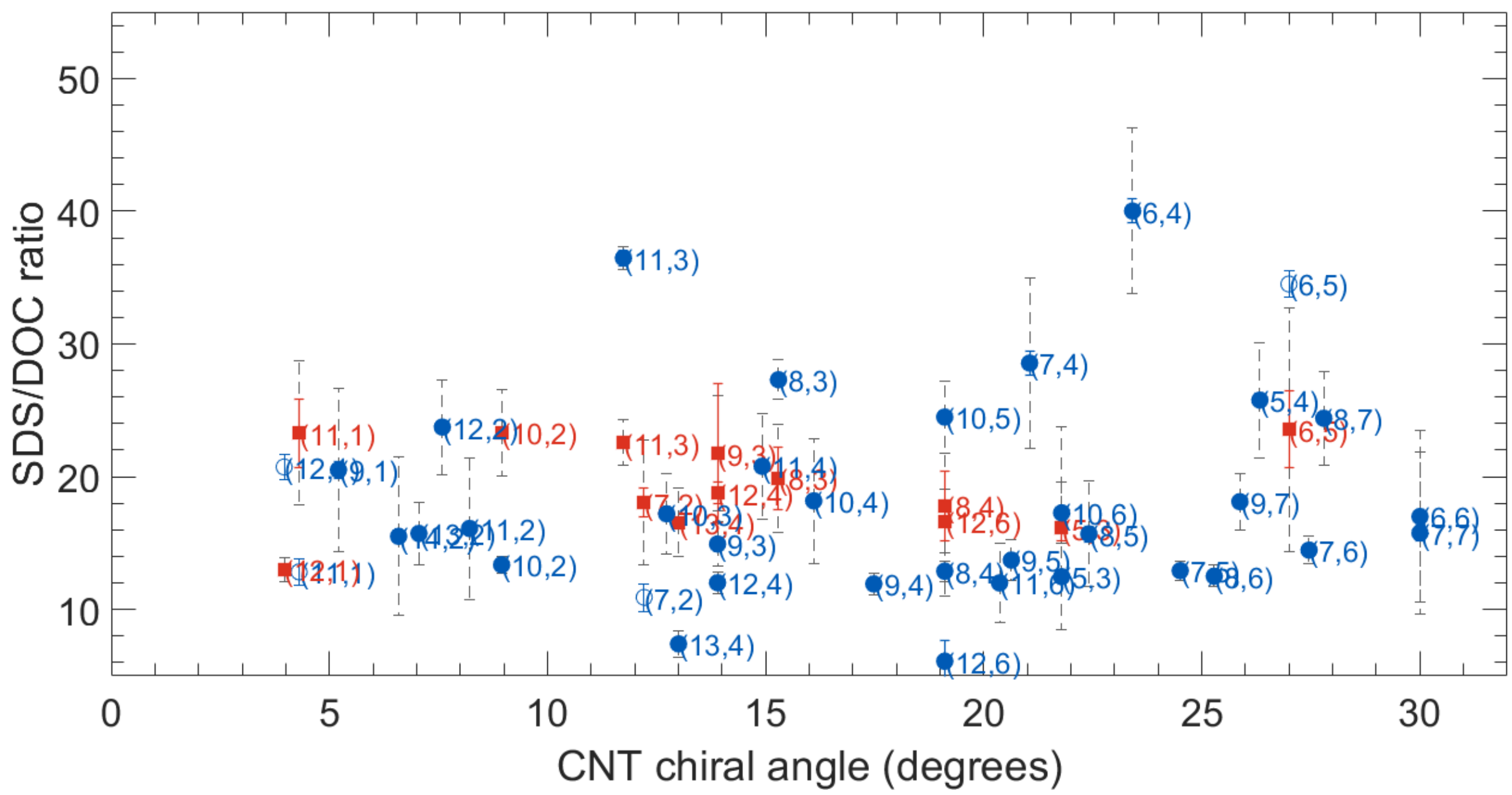

Figure S22: Transition points from **bottom to top phase** with increasing SDS concentration as determined from the **bottom phase data** for SWCNTs in SDS/DOC (0.0507% wt/V DOC) as a function of SWCNT chiral angle obtained by fitting the PL and RRS intensities of the bottom phase with an inverse error function, showing no obvious dependence on chiral angle. (Same color coding as in the caption of Figure S21)

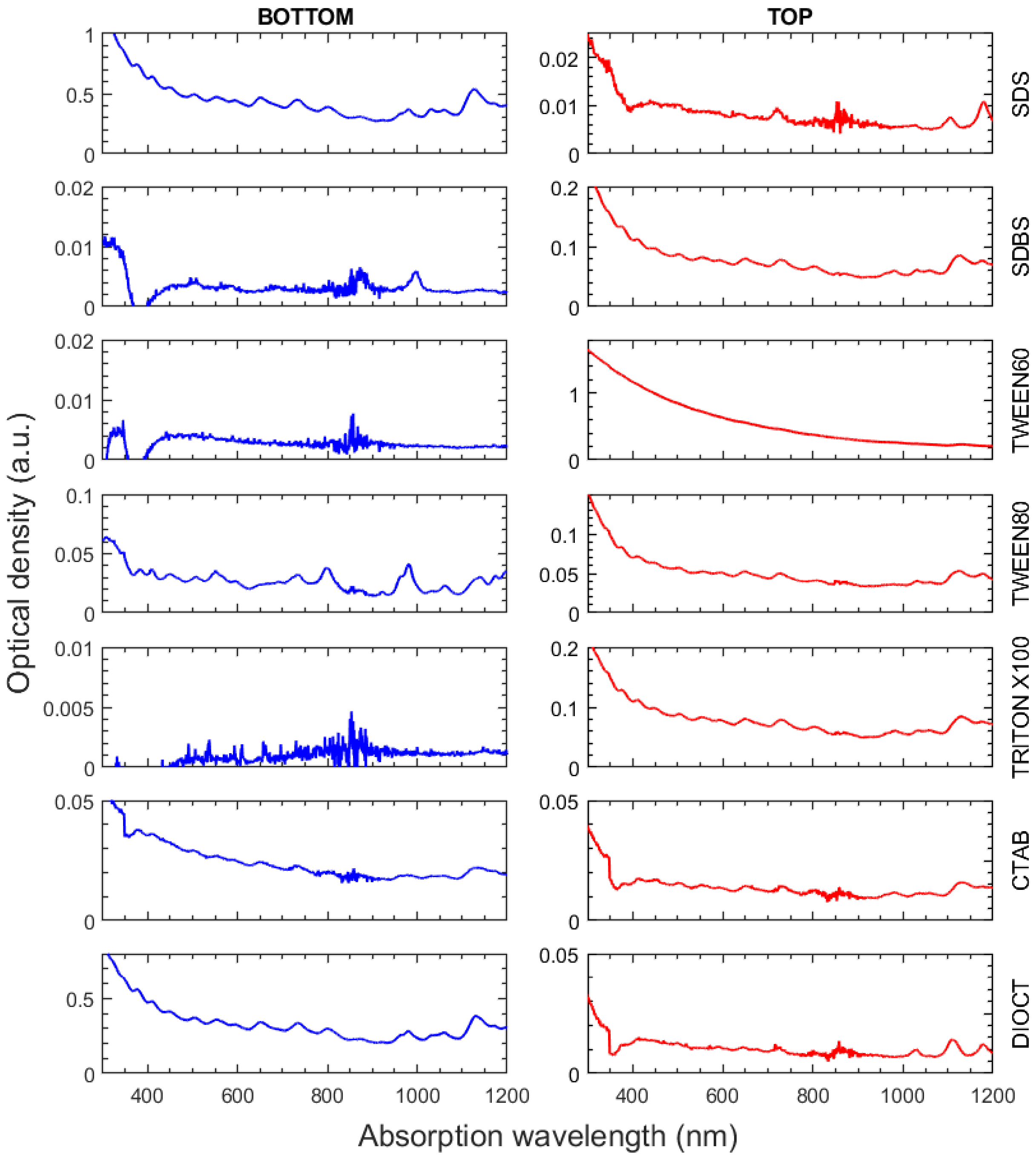

Figure S23: Absorption spectra of the bottom (in blue) and top (in red) phases for the different cosurfactants (SDS, SDBS, TWEEN60, TWEEN80, TRITON X100, CTAB, DIOCT). Separations were performed starting from the same SWCNT solution in 1% wt/V DOC/$D_2O$ and at a DOC concentration of 0.05% wt/V and a cosurfactant/DOC ratio of 10.

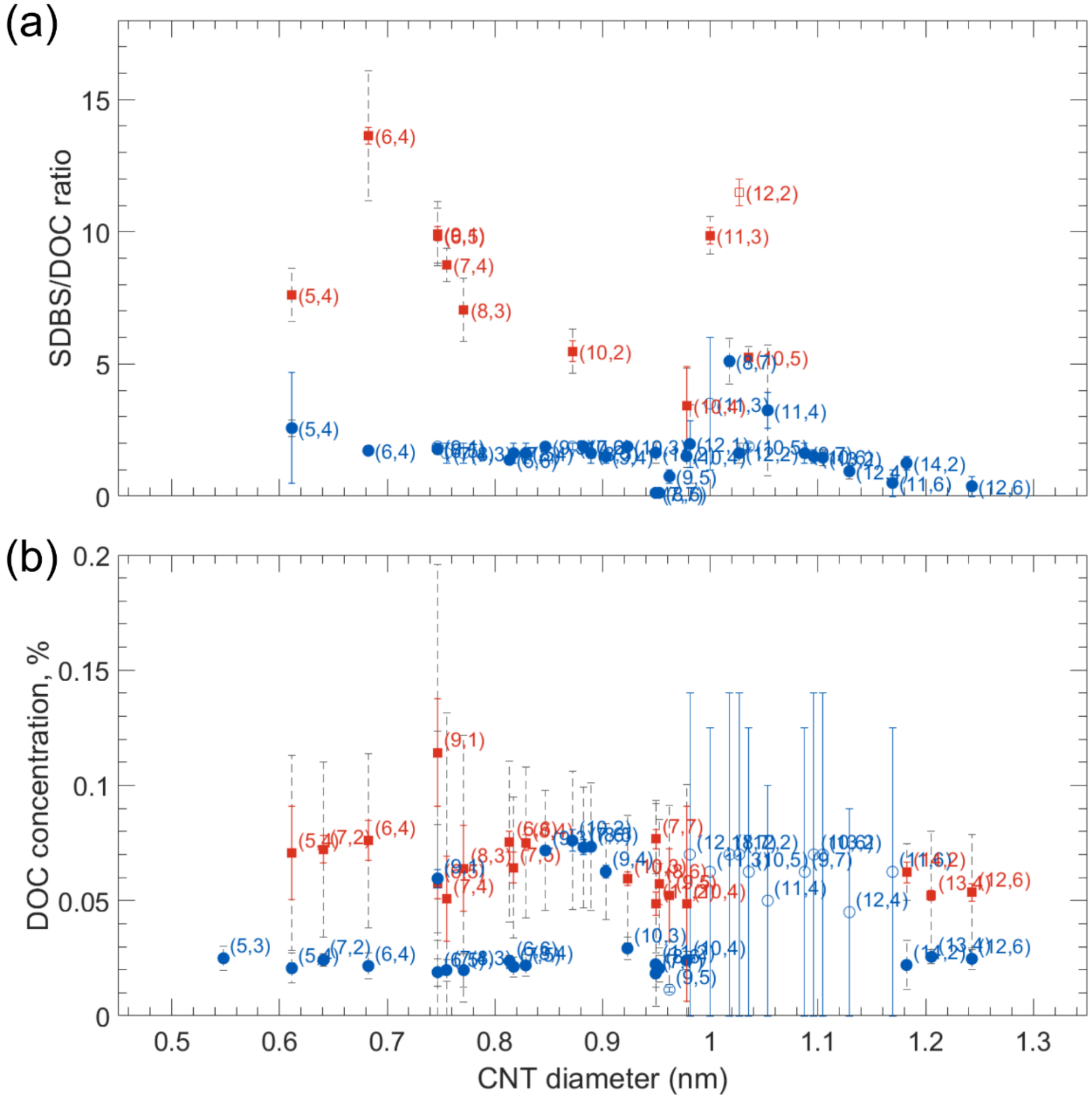

Figure S24: Transition points (a) from **bottom to top phase** with increasing SDBS concentration as determined from the **bottom phase data** for SWCNTs in SBDS/DOC (0.05% wt/V DOC) and (b) from **top to bottom phase** when increasing the DOC concentration as determined from the **bottom phase data** for SWCNTs in only DOC, both as a function of SWCNT diameter obtained by fitting the PL and RRS intensities of the bottom phase with a (complementary) error function. (Similar color coding as in the caption of Figure S21)

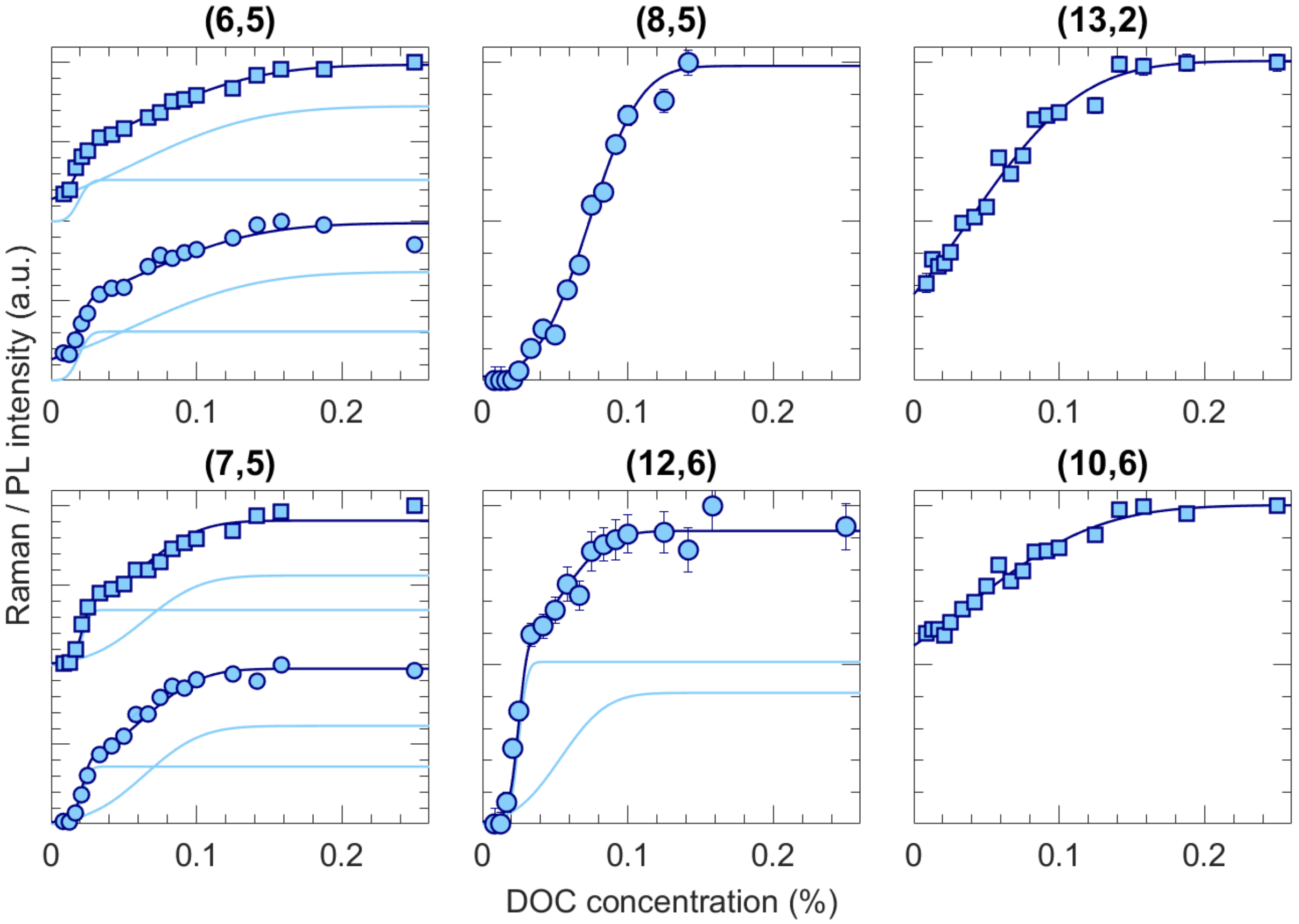

Figure S25: Normalized PL (squares) and Raman (circles) intensities (*i.e.*, $K_{(n,m)}^{top}$ and $K_{(n,m)}^{bottom}$) for a selected set of SWCNT chiralities as a function of DOC concentration (bottom phases, when only DOC is used). The fit components are shown with blue solid lines, their sums are shown in dark blue.

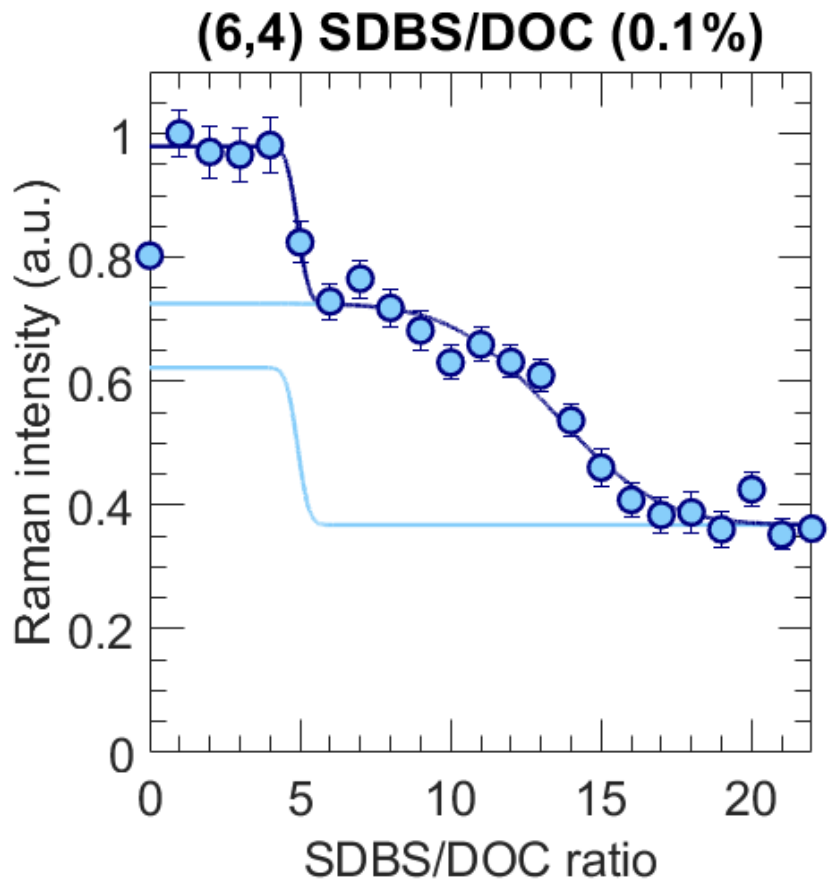

Figure S26: Normalized Raman (circles) intensities (*i.e.*, $K_{(n,m)}^{top}$ and $K_{(n,m)}^{bottom}$) for (6,4) SWCNTs as a function of increasing SDBS/DOC ratio (bottom phase), 0.1% wt/V DOC. The individual fit components are shown with blue solid lines; their sum (overall fit) is shown in darker blue.

## Section 8. Tables of transition points for all surfactant combinations

**Table S6:** Transitions points from **bottom to top phase** with increasing SDS concentration, as determined from the **bottom phase data** for SWCNTs in SDS/DOC (at a fixed concentration of 0.0507% wt/V DOC). Color representation of the transitions corresponds to Figure 6(a) of the main text.

| Chirality | Diameter, nm | First transition (position), SDS/DOC ratio | First transition (position error), SDS/DOC ratio | First transition (width), SDS/DOC ratio | First transition (width error), SDS/DOC ratio | Second transition (position), SDS/DOC ratio | Second transition (position error), SDS/DOC ratio | Second transition (width), SDS/DOC ratio | Second transition (width error), SDS/DOC ratio |
|---|---|---|---|---|---|---|---|---|---|
| **(5,4)** | 0.611 | 25.772 | 0.503 | 4.339 | 0.728 | | | | |
| **(6,4)** | 0.683 | 34.214 | 0.511 | 1.899 | 0.686 | | | | |
| **(6,6)** | 0.814 | 17.005 | 0.448 | 6.486 | 0.595 | | | | |
| **(7,4)** | 0.755 | 28.558 | 0.895 | 6.459 | 1.499 | | | | |
| **(7,5)** | 0.817 | 12.926 | 0.134 | 0.706 | 0.103 | | | | |
| **(7,6)** | 0.882 | 14.473 | 0.146 | 1.027 | 0.181 | | | | |
| **(7,7)** | 0.949 | 15.763 | 0.573 | 6.069 | 0.764 | | | | |
| **(8,5)** | 0.889 | 15.700 | 0.578 | 3.979 | 0.774 | | | | |
| **(8,6)** | 0.952 | 12.510 | 0.094 | 0.818 | 0.097 | | | | |
| **(8,7)** | 1.018 | 24.405 | 0.358 | 3.544 | 0.483 | | | | |
| **(9,1)** | 0.747 | 20.509 | 0.603 | 6.119 | 0.843 | | | | |
| **(9,4)** | 0.903 | 11.932 | 0.081 | 0.797 | 0.450 | | | | |
| **(9,5)** | 0.962 | 13.722 | 0.054 | 1.505 | 0.077 | | | | |
| **(9,7)** | 1.088 | 18.122 | 0.180 | 2.146 | 0.245 | | | | |
| **(10,3)** | 0.923 | 17.209 | 0.379 | 2.998 | 0.495 | | | | |
| **(10,4)** | 0.978 | 18.185 | 0.259 | 4.699 | 0.349 | | | | |
| **(10,5)** | 1.036 | 24.496 | 0.186 | 2.685 | 0.253 | | | | |
| **(10,6)** | 1.096 | 17.283 | 0.207 | 2.318 | 0.280 | | | | |
| **(11,2)** | 0.949 | 16.103 | 0.481 | 5.347 | 0.645 | | | | |
| **(11,4)** | 1.053 | 20.787 | 0.365 | 3.983 | 0.506 | | | | |
| **(11,6)** | 1.169 | 12.006 | 0.360 | 3.016 | 0.445 | | | | |
| **(12,2)** | 1.027 | 23.739 | 0.280 | 3.583 | 0.378 | | | | |
| **(13,2)** | 1.104 | 15.729 | 0.205 | 2.331 | 0.278 | | | | |
| **(14,2)** | 1.182 | 15.513 | 0.299 | 5.945 | 0.421 | | | | |
| **(7,2)** | 0.641 | 9.868…11.96 | | <2.092 | | 18.056 | 1.062 | 4.702 | 0.897 |
| **(11,1)** | 0.903 | 11.84…13.82 | | <0.99 | | 23.286 | 2.583 | 5.429 | 3.030 |
| **(6,5)** | 0.747 | 23.563 | 2.890 | 9.182 | 3.232 | 33.5…35.53 | | <1.015 | |
| **(12,1)** | 0.981 | 12.999 | 0.421 | 0.916 | 0.386 | 19.76…21.71 | | <0.975 | |
| **(5,3)** | 0.548 | 12.473 | 0.374 | 0.436 | 0.389 | 16.148 | 1.012 | 7.629 | 0.910 |
| **(8,3)** | 0.771 | 19.889 | 2.344 | 4.047 | 2.173 | 27.305 | 0.262 | 1.509 | 0.352 |
| **(8,4)** | 0.829 | 12.887 | 0.059 | 0.757 | 0.052 | 17.810 | 2.606 | 6.754 | 1.843 |
| **(9,3)** | 0.847 | 14.935 | 0.521 | 1.698 | 0.774 | 21.744 | 5.249 | 4.328 | 3.563 |
| **(10,2)** | 0.872 | 13.367 | 0.228 | 0.658 | 0.316 | 23.343 | 0.194 | 3.259 | 0.256 |
| **(11,3)** | 1.000 | 22.556 | 0.381 | 1.715 | 0.521 | 36.478 | 0.582 | 0.879 | 0.507 |
| **(12,4)** | 1.129 | 12.018 | 0.127 | 0.796 | 0.420 | 18.756 | 0.808 | 3.452 | 0.664 |
| **(12,6)** | 1.243 | 6.108 | 1.583 | 0.487 | 4.411 | 16.663 | 0.182 | 2.361 | 0.185 |
| **(13,4)** | 1.205 | 7.410 | 0.143 | 0.982 | 0.192 | 16.571 | 0.149 | 2.572 | 0.212 |

**Table S7:** Transitions points from **bottom to top phase** with increasing SDS concentration, as determined from the **top phase data** for SWCNTs in SDS/DOC (at a fixed concentration of 0.0507% wt/V DOC). Color representation of the transitions corresponds to Figure S21.

| Chirality | Diameter, nm | First transition (position), SDS/DOC ratio | First transition (position error), SDS/DOC ratio | First transition (width), SDS/DOC ratio | First transition (width error), SDS/DOC ratio | Second transition (position), SDS/DOC ratio | Second transition (position error), SDS/DOC ratio | Second transition (width), SDS/DOC ratio | Second transition (width error), SDS/DOC ratio |
|---|---|---|---|---|---|---|---|---|---|
| **(5,4)** | 0.611 | 23.735 | 0.502 | 6.300 | 0.709 | | | | |
| **(7,2)** | 0.641 | 17.828 | 1.062 | 7.796 | 1.380 | | | | |
| **(7,5)** | 0.817 | 12.962 | 0.320 | 1.035 | 0.354 | | | | |
| **(7,6)** | 0.882 | 14.618 | 0.194 | 1.128 | 0.236 | | | | |
| **(7,7)** | 0.949 | 14.340 | 0.633 | 5.747 | 0.855 | | | | |
| **(8,5)** | 0.889 | 16.157 | 0.921 | 5.298 | 1.250 | | | | |
| **(8,6)** | 0.952 | 12.598 | 0.190 | 1.077 | 0.216 | | | | |
| **(8,7)** | 1.018 | 26.090 | 0.245 | 0.812 | 0.404 | | | | |
| **(9,4)** | 0.903 | 12.368 | 0.293 | 1.551 | 0.431 | | | | |
| **(9,5)** | 0.962 | 13.866 | 0.223 | 1.805 | 0.307 | | | | |
| **(9,7)** | 1.088 | 18.013 | 0.359 | 3.192 | 0.479 | | | | |
| **(10,3)** | 0.923 | 28.335 | 1.357 | 11.678 | 1.643 | | | | |
| **(10,5)** | 1.036 | 25.571 | 0.289 | 2.016 | 0.354 | | | | |
| **(10,6)** | 1.096 | 17.781 | 0.346 | 2.849 | 0.469 | | | | |
| **(11,2)** | 0.949 | 19.754 | 1.103 | 8.130 | 1.324 | | | | |
| **(13,2)** | 1.104 | 18.654 | 1.259 | 6.317 | 1.745 | | | | |
| **(6,4)** | 0.683 | >30 | | | | | | | |
| **(13,4)** | 1.205 | 5.9…7.9 | | <1 | | 19.552 | 3.145 | 14.494 | 2.590 |
| **(11,3)** | 1.000 | 22.786 | 0.405 | 3.142 | 0.519 | 35.53…37.5 | | <0.985 | |
| **(11,6)** | 1.169 | 13.230 | 0.995 | 3.643 | 1.347 | 31.59…33.55 | | <0.98 | |
| **(12,4)** | 1.129 | 12.224 | 0.202 | 1.120 | 0.306 | 21.71…23.68 | | <0.985 | |
| **(5,3)** | 0.548 | 12.968 | 0.376 | 1.534 | 0.412 | 37.426 | 1.541 | 8.103 | 0.871 |
| **(6,5)** | 0.747 | 24.994 | 1.643 | 5.794 | 1.468 | 33.698 | 0.130 | 1.365 | 0.342 |
| **(8,3)** | 0.771 | 23.051 | 1.798 | 6.155 | 1.165 | 27.871 | 0.299 | 1.723 | 0.413 |
| **(8,4)** | 0.829 | 12.998 | 0.496 | 0.753 | 0.551 | 20.874 | 9.002 | 5.370 | 7.516 |
| **(9,1)** | 0.747 | 26.390 | 0.722 | 7.858 | 0.494 | 37.106 | 0.072 | 1.670 | 0.116 |
| **(9,3)** | 0.847 | 13.596 | 0.567 | 0.594 | 0.828 | 21.365 | 1.436 | 4.503 | 1.401 |
| **(10,2)** | 0.872 | 13.214 | 0.575 | 0.692 | 0.680 | 24.314 | 0.914 | 6.460 | 1.557 |
| **(11,1)** | 0.903 | 13.025 | 0.426 | 0.945 | 0.393 | 22.086 | 1.523 | 3.860 | 1.773 |
| **(11,4)** | 1.053 | 22.615 | 0.547 | 4.189 | 0.704 | 35.205 | 0.272 | 1.168 | 0.431 |
| **(12,1)** | 0.981 | 18.915 | 0.454 | 2.657 | 0.657 | 26.937 | 6.219 | 7.401 | 3.565 |
| **(14,2)** | 1.182 | 15.095 | 0.870 | 5.222 | 1.195 | 31.532 | 1.116 | 3.574 | 1.566 |

**Table S8:** Transitions points from **bottom to top phase** with increasing SDBS concentration, as determined from the **bottom phases data** for SWCNTs in SDBS/DOC (at a fixed concentration of 0.1% wt/V DOC). Color representation of the transitions corresponds to Figure 6(b) of the main text.

| Chirality | Diameter, nm | First transition (position), SDBS/DOC ratio | First transition (position error), SDBS/DOC ratio | First transition (width), SDBS/DOC ratio | First transition (width error), SDBS/DOC ratio | Second transition (position), SDBS/DOC ratio | Second transition (position error), SDBS/DOC ratio | Second transition (width), SDBS/DOC ratio | Second transition (width error), SDBS/DOC ratio |
|---|---|---|---|---|---|---|---|---|---|
| **(5,3)** | 0.548 | 4.354 | 0.031 | 0.348 | 0.027 | | | | |
| **(6,6)** | 0.814 | 4.517 | 0.066 | 0.433 | 0.054 | | | | |
| **(7,2)** | 0.641 | 3.203 | 0.063 | 0.784 | 0.084 | | | | |
| **(7,5)** | 0.817 | 4.778 | 0.020 | 0.579 | 0.029 | | | | |
| **(7,6)** | 0.882 | 5.062 | 0.032 | 0.619 | 0.053 | | | | |
| **(7,7)** | 0.949 | 3.900 | 0.022 | 0.748 | 0.032 | | | | |
| **(8,3)** | 0.771 | 10.101 | 0.446 | 3.173 | 0.562 | | | | |
| **(8,4)** | 0.829 | 4.603 | 0.023 | 0.386 | 0.019 | | | | |
| **(8,5)** | 0.889 | 4.474 | 0.026 | 0.744 | 0.033 | | | | |
| **(8,6)** | 0.952 | 4.255 | 0.051 | 0.337 | 0.063 | | | | |
| **(9,2)** | 0.795 | 6.813 | 0.118 | 1.982 | 0.158 | | | | |
| **(9,3)** | 0.847 | 4.788 | 0.239 | 1.316 | 0.293 | | | | |
| **(9,4)** | 0.903 | 4.166 | 0.033 | 0.478 | 0.060 | | | | |
| **(9,5)** | 0.962 | 5.169 | 0.018 | 0.401 | 0.037 | | | | |
| **(9,7)** | 1.088 | 5.419 | 0.048 | 0.975 | 0.063 | | | | |
| **(10,3)** | 0.923 | 4.227 | 0.329 | 1.846 | 0.367 | | | | |
| **(10,4)** | 0.978 | 5.587 | 0.149 | 1.824 | 0.195 | | | | |
| **(11,2)** | 0.949 | 4.394 | 0.072 | 1.381 | 0.093 | | | | |
| **(11,4)** | 1.053 | 8.128 | 0.182 | 1.978 | 0.247 | | | | |
| **(11,6)** | 1.169 | 2.081 | 0.001 | 0.617 | 0.001 | | | | |
| **(12,4)** | 1.129 | 3.670 | 0.056 | 0.372 | 0.055 | | | | |
| **(13,2)** | 1.104 | 5.006 | 0.021 | 0.612 | 0.037 | | | | |
| **(14,2)** | 1.182 | 2.709 | 0.197 | 1.776 | 0.158 | | | | |
| **(10,6)** | 1.096 | 5…6 | | <0.5 | | | | | |
| **(5,4)** | 0.611 | 6.774 | 1.150 | 4.675 | 1.178 | 13.204 | 0.429 | 0.341 | 0.700 |
| **(6,5)** | 0.747 | 5.477 | 0.712 | 3.208 | 1.082 | 14.633 | 0.148 | 1.091 | 0.187 |
| **(7,4)** | 0.755 | 6.837 | 0.569 | 2.334 | 0.887 | 13.543 | 0.144 | 0.584 | 0.152 |
| **(9,1)** | 0.747 | 4.272 | 0.314 | 1.808 | 0.331 | 14.358 | 0.263 | 1.342 | 0.355 |
| **(10,2)** | 0.872 | 4.989 | 0.072 | 0.563 | 0.148 | 8.581 | 0.234 | 1.073 | 0.280 |
| **(11,3)** | 1.000 | 8.212 | 0.128 | 1.429 | 0.172 | 14.478 | 0.169 | 0.966 | 0.221 |
| **(12,2)** | 1.027 | 7.750 | 0.263 | 2.148 | 0.376 | 15.319 | 0.624 | 1.412 | 0.784 |
| **(6,4)** | 0.683 | 4…6 | | <1 | | 13.481 | 0.377 | 2.762 | 0.536 |
| **(12,1)** | 0.981 | 4…5 | | <0.5 | | 7.373 | 0.100 | 0.273 | 0.069 |
| **(13,4)** | 1.205 | 0…1 | | <0.5 | | 4.042 | 0.428 | 1.156 | 0.424 |
| **(8,7)** | 1.018 | 6.771 | 0.407 | 2.033 | 0.358 | 9…10 | | <0.5 | |
| **(10,5)** | 1.036 | 7.526 | 0.471 | 1.307 | 0.364 | 9…10 | | <0.5 | |

**Table S9:** Transitions points from **bottom to top phase** with increasing SDBS concentration, as determined from the **bottom phases data** for SWCNTs in SDBS/DOC (at a fixed concentration of 0.05% wt/V DOC). Color representation of the transitions corresponds to Figure S24a.

| Chirality | Diameter, nm | First transition (position), SDBS/DOC ratio | First transition (position error), SDBS/DOC ratio | First transition (width), SDBS/DOC ratio | First transition (width error), SDBS/DOC ratio | Second transition (position), SDBS/DOC ratio | Second transition (position error), SDBS/DOC ratio | Second transition (width), SDBS/DOC ratio | Second transition (width error), SDBS/DOC ratio |
|---|---|---|---|---|---|---|---|---|---|
| **(8,7)** | 1.018 | 5.106 | 0.189 | 0.871 | 0.260 | | | | |
| **(11,2)** | 0.949 | 1.647 | 0.051 | 0.397 | 0.118 | | | | |
| **(11,4)** | 1.053 | 3.247 | 0.679 | 2.486 | 0.716 | | | | |
| **(12,1)** | 0.981 | 1.971 | 0.882 | 1.493 | 0.626 | | | | |
| **(12,4)** | 1.129 | 0.942 | 0.012 | 0.296 | 0.017 | | | | |
| **(13,2)** | 1.104 | 1.459 | 0.016 | 0.306 | 0.016 | | | | |
| **(6,6)** | 0.814 | 1.25…1.5 | | <0.125 | | | | | |
| **(7,5)** | 0.817 | 1.25…2 | | <0.375 | | | | | |
| **(7,6)** | 0.882 | 1.75…2 | | <0.125 | | | | | |
| **(7,7)** | 0.949 | 0…0.25 | | <0.125 | | | | | |
| **(8,4)** | 0.829 | 1.25…2 | | <0.375 | | | | | |
| **(8,5)** | 0.889 | 1.25…2 | | <0.375 | | | | | |
| **(8,6)** | 0.952 | 0…0.25 | | <0.125 | | | | | |
| **(9,3)** | 0.847 | 1.75…2 | | <0.125 | | | | | |
| **(9,4)** | 0.903 | 1.25…1.75 | | <0.25 | | | | | |
| **(9,5)** | 0.962 | 0.5…1 | | <0.25 | | | | | |
| **(9,7)** | 1.088 | 1.25…2 | | <0.375 | | | | | |
| **(10,3)** | 0.923 | 1.75…2 | | <0.125 | | | | | |
| **(10,6)** | 1.096 | 1.25…1.75 | | <0.25 | | | | | |
| **(11,6)** | 1.169 | 0…1 | | <0.5 | | | | | |
| **(12,6)** | 1.243 | 0…0.75 | | <0.375 | | | | | |
| **(14,2)** | 1.182 | 1…1.5 | | <0.25 | | | | | |
| **(5,4)** | 0.611 | 2.576 | 2.093 | 0.310 | 1.458 | 7.619 | 0.156 | 1.005 | 0.200 |
| **(6,4)** | 0.683 | 1.722 | 0.139 | 0.172 | 0.201 | 13.629 | 0.305 | 2.453 | 0.415 |
| **(6,5)** | 0.747 | 1.774 | 0.091 | 0.192 | 0.143 | 9.852 | 0.128 | 1.055 | 0.176 |
| **(10,4)** | 0.978 | 1.512 | 0.076 | 0.415 | 0.147 | 3.418 | 1.491 | 1.407 | 0.892 |
| **(7,4)** | 0.755 | 1.25…2 | | <0.375 | | 8.746 | 0.166 | 0.642 | 0.227 |
| **(8,3)** | 0.771 | 1.25…2 | | <0.375 | | 7.053 | 0.120 | 1.201 | 0.160 |
| **(9,1)** | 0.747 | 1.75…2 | | <0.125 | | 9.925 | 0.260 | 1.207 | 0.348 |
| **(10,2)** | 0.872 | 1.75…2 | | <0.125 | | 5.480 | 0.395 | 0.832 | 0.516 |
| **(10,5)** | 1.036 | 1.75…2 | | <0.125 | | 5.248 | 0.130 | 0.407 | 0.176 |
| **(11,3)** | 1.000 | 1…6 | | <2.5 | | 9.854 | 0.317 | 0.708 | 0.511 |
| **(12,2)** | 1.027 | 1.25…2 | | <0.375 | | 11…12 | | <0.5 | |

**Table S10:** Transitions from **top to bottom phase** with increasing DOC concentration as obtained from the **bottom phase data** for the SWCNTs in DOC alone. Color representation of the transitions corresponds to Figure S24b.

| Chirality | Diameter, nm | First transition (position), % DOC | First transition (position error), % DOC | First transition (width), % DOC | First transition (width error), % DOC | Second transition (position), % DOC | Second transition (position error), % DOC | Second transition (width), % DOC | Second transition (width error), % DOC |
|---|---|---|---|---|---|---|---|---|---|
| **(5,3)** | 0.548 | 0.02497 | 0.00071 | 0.00517 | 0.00110 | | | | |
| **(7,6)** | 0.882 | 0.07312 | 0.00327 | 0.02629 | 0.00409 | | | | |
| **(8,5)** | 0.889 | 0.07337 | 0.00199 | 0.02775 | 0.00296 | | | | |
| **(9,3)** | 0.847 | 0.07177 | 0.00126 | 0.02591 | 0.00147 | | | | |
| **(9,4)** | 0.903 | 0.06257 | 0.00268 | 0.02087 | 0.00330 | | | | |
| **(10,2)** | 0.872 | 0.07606 | 0.00464 | 0.02993 | 0.00556 | | | | |
| **(8,7)** | 1.018 | <0.14 | | <0.07 | | | | | |
| **(9,7)** | 1.088 | <0.125 | | <0.0625 | | | | | |
| **(10,5)** | 1.036 | <0.125 | | <0.0625 | | | | | |
| **(10,6)** | 1.096 | <0.14 | | <0.07 | | | | | |
| **(11,3)** | 1.000 | <0.125 | | <0.0625 | | | | | |
| **(11,4)** | 1.053 | <0.1 | | <0.05 | | | | | |
| **(11,6)** | 1.169 | <0.125 | | <0.0625 | | | | | |
| **(12,1)** | 0.981 | <0.14 | | <0.07 | | | | | |
| **(12,2)** | 1.027 | <0.14 | | <0.07 | | | | | |
| **(12,4)** | 1.129 | <0.09 | | <0.045 | | | | | |
| **(13,2)** | 1.104 | <0.14 | | <0.07 | | | | | |
| **(9,5)** | 0.962 | <0.013 | | <0.0065 | | 0.05244 | 0.02001 | 0.03894 | 0.02004 |
| **(5,4)** | 0.611 | 0.02066 | 0.00222 | 0.00651 | 0.00438 | 0.07074 | 0.02026 | 0.04218 | 0.01883 |
| **(6,4)** | 0.683 | 0.02164 | 0.00083 | 0.00551 | 0.00167 | 0.07605 | 0.00868 | 0.03778 | 0.00989 |
| **(6,5)** | 0.747 | 0.01894 | 0.00090 | 0.00599 | 0.00132 | 0.05724 | 0.00626 | 0.06619 | 0.00577 |
| **(6,6)** | 0.814 | 0.02374 | 0.00070 | 0.00265 | 0.00103 | 0.07559 | 0.00469 | 0.03479 | 0.00496 |
| **(7,2)** | 0.641 | 0.02423 | 0.00046 | 0.00282 | 0.00079 | 0.07225 | 0.00597 | 0.03793 | 0.00568 |
| **(7,4)** | 0.755 | 0.01984 | 0.00189 | 0.00466 | 0.00263 | 0.05073 | 0.01842 | 0.08073 | 0.02206 |
| **(7,5)** | 0.817 | 0.02130 | 0.00097 | 0.00436 | 0.00177 | 0.06437 | 0.00655 | 0.03046 | 0.00577 |
| **(7,7)** | 0.949 | 0.01838 | 0.00099 | 0.00578 | 0.00127 | 0.07681 | 0.00419 | 0.01523 | 0.00555 |
| **(8,3)** | 0.771 | 0.01983 | 0.00193 | 0.00719 | 0.00320 | 0.06400 | 0.01879 | 0.05786 | 0.01416 |
| **(8,4)** | 0.829 | 0.02193 | 0.00091 | 0.00465 | 0.00172 | 0.07518 | 0.00354 | 0.03271 | 0.00425 |
| **(8,6)** | 0.952 | 0.02053 | 0.00070 | 0.00586 | 0.00150 | 0.05723 | 0.00880 | 0.02791 | 0.00653 |
| **(9,1)** | 0.747 | 0.05959 | 0.00391 | 0.02347 | 0.00665 | 0.11427 | 0.02338 | 0.08163 | 0.01648 |
| **(10,3)** | 0.923 | 0.02928 | 0.00205 | 0.00494 | 0.00240 | 0.05943 | 0.00285 | 0.02765 | 0.00194 |
| **(10,4)** | 0.978 | 0.02401 | 0.00184 | 0.00394 | 0.00397 | 0.04870 | 0.04224 | 0.05151 | 0.02983 |
| **(11,2)** | 0.949 | 0.02237 | 0.00087 | 0.00504 | 0.00142 | 0.04873 | 0.00503 | 0.04474 | 0.00491 |
| **(12,6)** | 1.243 | 0.02473 | 0.00081 | 0.00474 | 0.00158 | 0.05362 | 0.00379 | 0.02504 | 0.00239 |
| **(13,4)** | 1.205 | 0.02560 | 0.00121 | 0.00302 | 0.00281 | 0.05209 | 0.00217 | 0.02807 | 0.00159 |
| **(14,2)** | 1.182 | 0.02206 | 0.00176 | 0.01059 | 0.00170 | 0.06234 | 0.00455 | 0.01243 | 0.00477 |

**Section 9. Details on the separation parameters for (6,5) SWCNTs in two steps.**

**Table S11:** Detailed volumes and concentrations for the separation of (6,5) SWCNTs with SDS/DOC (0.05% wt/V DOC) and SDBS/DOC (0.1% wt/V DOC) in two consecutive steps.

| SDS/DOC with 0.05% wt/V DOC | | | | | | |
|---|---|---|---|---|---|---|
| First step | | | | | | |
| PEG 20% wt/V | dextran 20% wt/V | SWCNT 1% wt/V DOC | SDS 15% wt/V | $D_2O$ | Total | SDS/DOC |
| 7 mL | 3 mL | 600 µL | 1160 µL | 240 µL | 12 mL | 29 |
| Mimicking top phase for step 2 (175 µL + 75 µL bottom phase) | | | | | | |
| PEG 20% wt/V | dextran 20% wt/V | | SDS 15% wt/V | DOC 1% wt/V | Total Volume | SDS/DOC |
| 1645 µL | 290 µL | | 350 µL | 122 µL | 2407 µL | 43.6 |
| SDBS/DOC with 0.1% wt/V DOC | | | | | | |
| First step | | | | | | |
| PEG 20% wt/V | dextran 20% wt/V | SWCNT 1% wt/V DOC | SDBS 12% wt/V | DOC 5% wt/V | Total Volume | SDBS/DOC |
| 7 mL | 3 mL | 600 µL | 1200 µL | 120 µL | 12 mL | 12 |
| Mimicking top phase for step 2 (175 µL + 75 µL bottom phase) | | | | | | |
| PEG 20% wt/V | dextran 20% wt/V | | SDBS 10% wt/V | DOC 5% wt/V | Total Volume | SDBS/DOC |
| 1645 µL | 290 µL | | 450 µL | 50 µL | 2435 µL | 18 |

## Section 10. Comparison with literature data

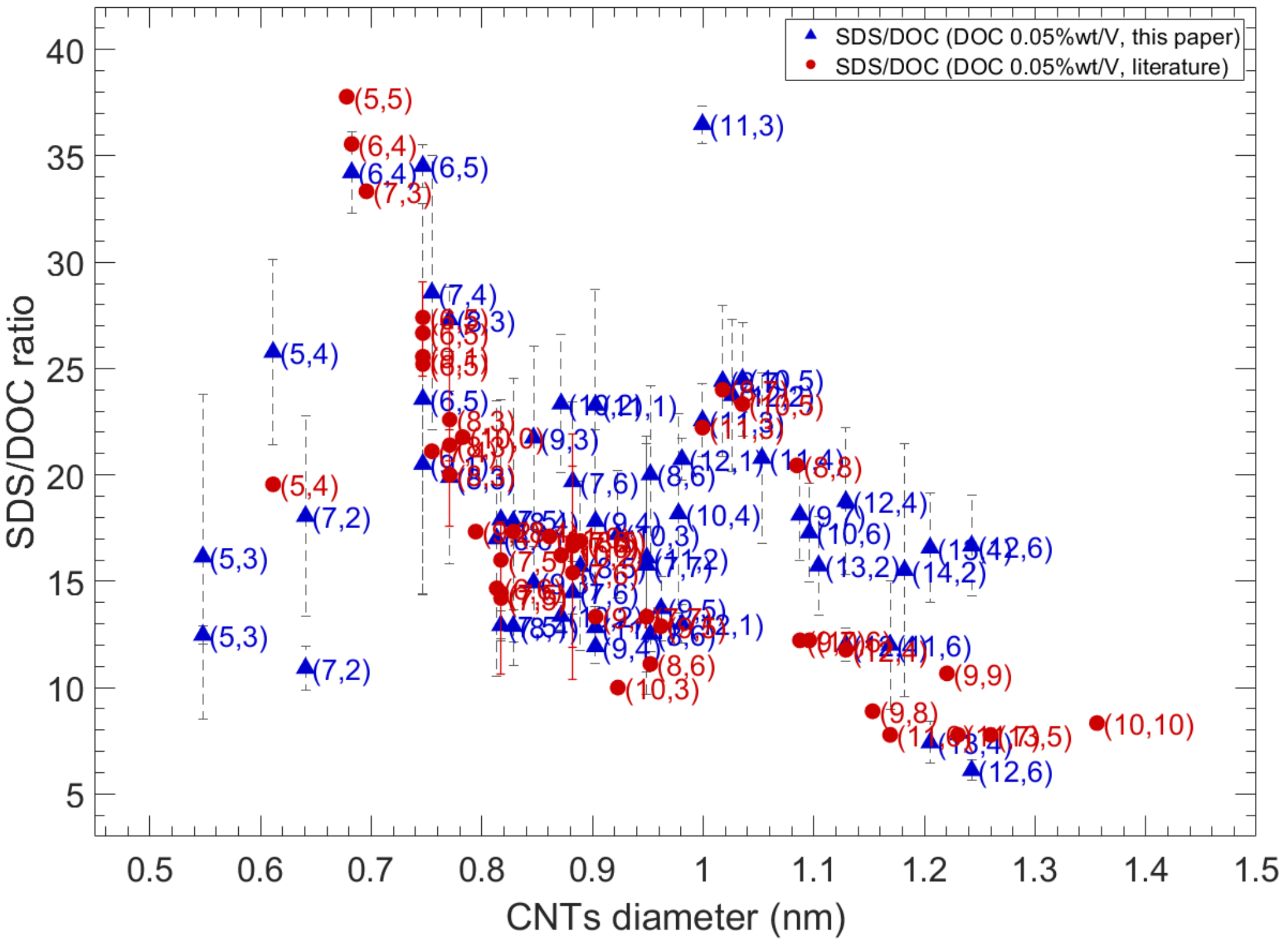

Figure S27: Comparison of transition points for SDS/DOC with 0.05% DOC determined in this work (blue triangles), and those obtained empirically from multistage ATP separations [J. A. Fagan *et al.*, ACS Nano 9, p. 5377 (2015)] or from surfactant-exchange NIR PL experiments [C. Sims *et al.*, Carbon 165, p. 196 (2020)]. The literature data agrees very well with our experimental data. The wider range of chiralities available from the present work reveals the non-monotonous, periodically modulated trend described in the main text.